\documentclass[12pt,preprint]{aastex}  

\newcommand{\lsim}{\, \lower2truept\hbox{${< \atop\hbox{\raise4truept\hbox{$\sim$}}}$}\,}
\newcommand{\gsim}{\, \lower2truept\hbox{${> \atop\hbox{\raise4truept\hbox{$\sim$}}}$}\,}
\newcommand{\nid}{\noindent}
\newcommand{\ch}{\colhead}
\newcommand{\oneskip}{\vskip\baselineskip}

\newcommand{\puncspace}{\ifmmode\,\else{\ifcat.\C{\if.\C\else\if,\C\else\if?\C\else%
\if:\C\else\if;\C\else\if-\C\else\if)\C\else\if/\C\else\if]\C\else\if'\C%
\else\space\fi\fi\fi\fi\fi\fi\fi\fi\fi\fi}%
\else\if\empty\C\else\if\space\C\else\space\fi\fi\fi}\fi}
\newcommand{\SP}{\let\\=\empty\futurelet\C\puncspace}

\begin{document}

\slugcomment{{\bf Accepted by ApJ: February 17, 2010}}

\title{Local Benchmarks for the Evolution of Major-Merger Galaxies
  --- Spitzer Observations of a K-Band Selected Sample}

\author{C. Kevin Xu \altaffilmark{1}, 
Donovan Domingue\altaffilmark{2},
Yi-Wen Cheng\altaffilmark{1,5,6}, 
Nanyao Lu\altaffilmark{1}, 
Jiasheng Huang\altaffilmark{3},
Yu Gao\altaffilmark{4},  
Joseph M. Mazzarella\altaffilmark{1},
Roc Cutri\altaffilmark{1},
Wei-Hsin Sun\altaffilmark{6},
Jason Surace\altaffilmark{7}
}

\altaffiltext{1}{Infrared Processing and Analysis Center, 
California Institute of Technology 100-22, Pasadena, CA 91125}
\altaffiltext{2}{Georgia College \& State University, CBX 82,
Milledgeville, GA 31061}
\altaffiltext{3}{Harvard-Smithsonian Center for Astrophysics, 60 Garden Street, Cambridge, MA 02138}
\altaffiltext{4}{Purple Mountain Observatory, Chinese Academy of Sciences, 2 West Beijing Road, Nanjing 210008, China}
\altaffiltext{5}{Institute of Astronomy, National Central University, Chung-Li 32054, Taiwan}
\altaffiltext{6}{Institute of Astrophysics and Leung Center for Cosmology and Particle Astrophysics, National Taiwan University, Taipei, Taiwan 10617}
\altaffiltext{7}{Spitzer Science Center, California Institute of Technology, Mail Stop 220-6, Pasadena, CA 91125}

\received{\today}

\begin{abstract}

We present Spitzer observations for a sample of close major-merger
galaxy pairs (KPAIR sample) selected from 2MASS/SDSS-DR3
cross-matches.  The goals are to study the star
formation activity in these galaxies and to set a local bench
mark for the cosmic evolution of close major mergers.  The Spitzer
KPAIR sample (27 pairs, 54 galaxies) includes all spectroscopically
confirmed spiral-spiral (S+S) and spiral-elliptical (S+E) pairs in a
parent sample that is complete for primaries brighter than K=12.5 mag,
projected separations of $\rm 5 \leq s \leq 20$h$^{-1}$ kpc, and mass
ratios $\leq 2.5$.  The Spitzer data, consisting of images in 7 bands
(3.6, 4.5, 5.8, 8, 24, 70, 160$\mu m$), show very diversified IR
emission properties.  Compared to single spiral galaxies in a control
sample, only spiral galaxies in S+S pairs show significantly enhanced
specific star formation rate (sSFR=SFR/M), whereas spiral galaxies in S+E
pairs do not.  Furthermore, the SFR enhancement of spiral galaxies in
S+S pairs is highly mass-dependent. Only those with $\rm M \gsim
10^{10.5} M_\sun$ show significant enhancement. Relatively low mass
($\rm M \sim 10^{10} M_\sun$) spirals in S+S pairs have about
the same SFR/M compared to their counterparts in the control sample,
while those with $10^{11} M_\sun$ have on average a $\sim 3$ times
higher SFR/M than single spirals.  There is evidence for a correlation
between the global star formation activities (but not the nuclear
activities) of the component galaxies in massive S+S major-merger
pairs (the ``Holmberg effect'').  There is no significant difference in the
SFR/M between the primaries and the secondaries, nor between
spirals of $\rm SEP <1$ and those of $\rm SEP \geq 1$, $\rm SEP$ being
the normalized separation parameter.  The contribution of KPAIR
galaxies to the cosmic SFR density in the local universe is only 1.7\%,
and amounts to $\rm
\rho^{.}_{KPAIR} = 2.54 \times 10^{-4} \; (M_\sun \; yr^{-1}\;
Mpc^{-3})$.
\end{abstract}


\section{Introduction}

Two questions about the evolution of galaxy mergers are
being intensely debated in the current literature: 
(1) Does merger rate have a strong or weak cosmic evolution? 
(2) Are mergers responsible for the strong
evolution of the cosmic star formation rate (SFR) 
since $z \sim 1$ (e.g., Lilly et al. 1996; 
Madau et al. 1998; Hopkins 2004)?
Answers to these questions have important implications to the
understanding of basic processes in galaxy formation/evolution,
such as mass growth and star formation.

Earlier studies of merger rate, using samples of close galaxy pairs
and morphologically disturbed systems in different redshift ranges,
yielded a very broad range of the evolutionary index $m$, $m = 0$ -- 6,
assuming the evolution has a power-law form $(1+z)^m$ (Zepf \& Koo
1989; Burkey et al. 1994; Carlberg et al. 1994; Yee \& Ellington 1995;
Woods et al. 1995; Patton et al. 1997; Wu \& Keel 1998; Brinchmann et
al. 1998; Le F\'evre et al. 2000; Carlberg et al. 2000; Patton et
al. 2002; Conselice et al. 2003; Lin et al. 2004; Bundy et al. 2004;
Lavery et al. 2004).
More recent results can be divided into two camps, the ``strong
evolution'' camp with $m \sim 3$ (Conselice 2006; Kampczyk et al. 2007;
Kartaltepe et al. 2007) and the ``weak evolution'' camp with $m \sim 0.5$
(Lin et al. 2008; Lotz et al. 2008).  The strong evolution scenario is
consistent with the cosmic time dependence of the merging rate of dark
matter halos (Lacey \& Cole 1993; Khochfar \& Burkert 2001). However,
more recent simulations including sub-halo structures
support the weak evolution scenario (Berrier et
al. 2006; Guo \& White 2008).

Since the discovery of a strong evolution of the cosmic SFR (Lilly et
al. 1996; Madau et al. 1998), many authors have argued that the
primary cause is a rapid decline of merger-induced star formation
(Driver et al. 1995; Glazebrook et al. 1995; Abraham et al. 1996;
Brinchmann et al. 1998, Le F\'evre et al. 2000; Conselice et
al. 2003). In particular, infrared (IR) surveys by both ISO (Elbaz et
al. 2002) and Spitzer (Le Floch et al. 2006) found that beyond $z \sim
0.7$--1.0, the cosmic star formation is contributed mostly by luminous
IR galaxies (LIRGs) with $L_{IR} \geq 10^{11} L_\sun$. In the local
universe most LIRGs are in merger systems, and they contribute only
a few percent of the integrated IR luminosity density (Sanders \&
Mirabel 1996). This seems to provide another argument for a strong
evolution of merger-induced starbursts as the primary driver of the
strong cosmic SFR evolution.  However, are LIRGs at $z \sim 1$ indeed
mostly in merger systems, as observed for their local
counterparts?  To this question there are both positive answers (Zheng
et al. 2004; Hammer et al. 2005; Bridge et al. 2007) and negative
answers (Flores et al. 1999; Bell et al. 2005; Melbourne et al. 2005;
Lotz et al. 2008).  Authors in the latter group found that most
LIRGs at $z \sim 1$ are normal late-type galaxies. According to them,
it is the secular evolution of normal late-type galaxies that is
mostly responsible for the cosmic SFR evolution, not the evolution of
mergers.

What are the reasons for these controversies?  The foremost among them
are sample selection effects.  There are two classical methods of
selecting interacting/merging galaxies. One is to find galaxies with
peculiar morphology (e.g. with tidal tails, double nuclei, or
distorted discs), and the other is to identify paired galaxies.  There are
systematic differences between interacting galaxies selected using
these two different methods. In
the so-called merger sequence (Toomre 1977), galaxies in close pairs
are usually mergers in the early stages when the two galaxies are
still separable. In contrast, peculiar galaxies are mostly found in
the later stages when the first collision between the two galaxies is
happening or passed.  However, this distinction is not clear-cut.
Mergers such as the Antennae (= Arp~244) can be identified using both
methods.

There are several known biases in the morphological selection method.
The most serious one is due to the mis-identifications of isolated
irregular galaxies or starburst galaxies as mergers (Lotz et
al. 2008).  This effect is particularly severe for samples selected in
the rest frame UV bands (and to some extend those selected in the rest
frame B-band), where the emission of young stars and dust extinction
can significantly affect the surface brightness distribution.  Another
bias is caused by missing low surface brightness merger features, such
as faint tidal tails and bridges, in observations that lack sufficient
sensitivity. This incompleteness becomes increasingly severe for high-z
surveys because of the cosmic dimming.

The most common bias affecting current pair selected samples is an
incompleteness known as ``missing the secondary'' (Xu et al. 2004).
For flux limited (= apparent magnitude limited) samples or luminosity
limited (= absolute magnitude limited = ``volume limited'') samples, a
paired galaxy brighter than the limit can be missed if its companion
is fainter than the limit. The amplitude of this incompleteness can
vary with the redshift, and cause significant bias in the results on
the evolution of merger rate.  For example, in many recent studies of
merger rate evolution, pair fractions of galaxies of $M \leq M_{lim}$
are compared, where $M_{lim}$ is an absolute magnitude limit in the
rest frame B or V band.  For these samples, the amplitude of the
incompleteness due to the ``missing the secondary'' bias is on the
order of $Q \sim 0.5 \phi (M_{lim}) \delta M/\int^{M_{lim}}_{-\inf} \phi
(M)\; dM$, where $ \phi $ is the luminosity function (LF) of the
paired galaxies (e.g. Xu et al. 2004; Domingue et al. 2009), and
$\delta M$ is the typical magnitude difference between the two
galaxies in a pair, which is $\sim 1$ mag for major mergers. When
$M_{lim}$ is fixed, the ratio $Q$ decreases with z if the LF has a
positive ``luminosity evolution'' (i.e. the ``knee'' of the LF becomes
brighter with increasing z), as being observed for the LF of field
galaxies (Wolf et al. 2003; Marshesini et al. 2007). This can
introduce an artificial ``evolution'' of the merger rate in studies
comparing pair fractions in samples at different redshifts that are
limited by the same $M_{lim}$ (e.g. Kartaltepe et al. 2007), and being
responsible for the high evolutionary index (m$\sim 3$) found in those
studies. On the other hand, all studies in the 'weak evolution' camp
invoke a correction for the ``passive luminosity evolution'' (PLE),
allowing the absolute value of $M_{lim}$ to increase with z
accordingly. This indeed reduces the effect of the incompleteness on
the evolutionary index of the merger rate.  However, as pointed out by
Kartaltepe et al. (2007), there is no strong empirical justification
for the PLE model.  If the true evolution of the luminosities of
interacting galaxies is different from PLE, then the bias is still
present. It is better to get rid of the incompleteness from the merger rate
studies in the first place.

Other biases for pair selections include: (1) Contamination due to
unphysical, projected pairs (``interlopers'').  This affects mostly
samples with incomplete redshifts or with only photometric redshifts.
(2) Incompleteness due to missing of pairs in which the two galaxies
are too close to be separated visually because of insufficient angular
resolution or obscuration by dust.
  
Being aware of the selection effects that lead to the conflicting
results, we set out to design a set of merger selection criteria that
minimize the biases mentioned above. Firstly, we opted for the pair
selection method instead of the morphological selection method because,
based purely on galaxy separation, pair selections are more objective.
However, this also confines our study to systems prior to the final
stages of the merging process. 
Secondly, we chose the rest frame K as the waveband in which our
samples are to be selected.  This is the band least affected by star
formation and dust extinction and most closely related to mass 
(Bell \& De
Jong 2001). Thirdly, we confine ourselves to close major-merger pairs 
with mass ratio $< 2.5$ and with projected separation in the range
$\rm 5 h^{-1} \leq s \leq 20 h^{-1}$ kpc ($\rm h = H_0/(100\; {\rm km~sec}^{-1} {\rm Mpc}^{-1})$).
Many studies have found that only major mergers with
separations comparable to the size of galaxies (i.e. $\lsim 20 h^{-1}$
kpc) show significant SFR enhancements (Xu \& Sulentic 1991; Barton et
al. 2000; Lambas et al. 2003; Nikolic et al. 2004; Alonso et al. 2004;
Woods et al. 2006; Barton et al. 2007; Ellison et al. 2008).
Other detailed selection criteria are presented in Section 2.
It should be emphasized that, when studying the evolution of
interacting galaxies, samples at different redshifts must be 
selected using identical criteria.  

In two earlier papers (Xu et al. 2004; Domingue et al. 2009), we
started from samples in the local universe, with the goal of setting
local benchmarks for merger rate evolution.  Xu et al. (2004),
using a sample of 19 close major-merger pairs selected from the
matched 2MASS/2dFGRS catalog (Cole et al. 2001), derived the K-band
luminosity function (KLF) and the differential pair fraction function
(DPFF) of local binary galaxies. This was followed up by a more
extended analysis (Domingue et al. 2009), exploiting a large sample of
173 close major-merger pairs selected from 2MASS/SDSS-DR6
cross-matches. Assuming the mass dependent merger time scale of Kitzbichler
\& White (2008), Domingue et al. (2009) found that
the differential merger rate increases with mass, 
and the merger rate v.s. mass relation is in good agreement
with what being found in the N-body simulations of Maller et al. (2006).

In this paper we report Spitzer imaging observations (7 bands at 3.6,
4.5, 5.8, 8, 24, 70, and 160$\mu m$) of 27 galaxy pairs in the local
universe, selected from cross matches of 2MASS and SDSS-DR3.  The
scientific goals are (1) studying the star formation activity in these
galaxies and (2) setting a local bench mark for the evolution of the
SFR in close major mergers. Assuming that only late type galaxies
contribute significantly to the total SFR, our analysis is
concentrated on spiral (S) galaxies in spiral-spiral (S+S) and
spiral-elliptical (S+E) pairs. The main focus of this paper is on the
enhancement (or the lack of it) of the SFR of paired galaxies. 
Most previous studies on this
subject are based on optical/UV observations susceptible to dust
extinction (Barton et al. 2000; Lambas et al. 2003; Nikolic et
al. 2004; Alonso et al. 2004; Woods et al. 2006; Woods \& Geller 2007;
Barton et al. 2007; Ellison et al. 2008). Early FIR studies 
based on IRAS observations (Kennicutt et al.
1987; Telesco et al. 1988; Xu \& Sulentic 1991) can not resolve the
pairs because of the coarse angular resolution of IRAS. The more
recent ISO study of Xu et al. (2001) and Spitzer study of Smith et
al. (2007), which resolved pairs into discrete regions 
(e.g.  nuclei of component galaxies, bridges and
tails, overlapping regions between the two disks, etc.), are confined
to pairs with strong interacting features, and therefore are biased to
interacting galaxies with strong star formation enhancement.
This work shall be neutral to these biases. Our studies on IR SED's (Domingue et
al. in preparation) and on the optical properties including new 
H$\alpha$ and H$\beta$ imaging observations (Cheng et al. in
preparation) will be published separately. Throughout this paper, we
adopt the $\Lambda$-cosmology with $\Omega_m=0.3$ and $\Omega_\Lambda
= 0.7$, and ${\rm H}_0= 75\; ({\rm km~sec}^{-1} {\rm Mpc}^{-1})$.

\section{The Sample}
The sample of galaxy pairs was selected from cross-matches between the Two
Micron All Sky Survey (2MASS) Extended Source Catalog (XSC; Jarrett et
al. 2000) and the galaxy catalog of Sloan Digital Sky Survey (SDSS)
Data Release 3 (DR3; Abazajian et al. 2005).  The selection procedure
is similar to that in Xu et al. (2004) and Domingue et
al. (2009). First, the following two criteria were set for
galaxies to be considered in the pair selection:
\begin{description}
\item{(1)} Galaxies should be brighter than $K_{s}$=13.5 mag, 
the completeness limit of the XSC (Jarrett et al. 2000). The default 
${\rm K_{20}}$ magnitude is used for the ${\rm K_s}$ band (2.16$\mu m$)
fluxes (Jarrett et al. 2000). The ${\rm K_{20}}$ magnitudes were
taken from Domingue et al. (2009), for which the photometric
error due to the blending of close neighbors was corrected. 
As in Xu et al. (2004), a uniform -0.2 mag correction was applied
to the ${\rm K_{20}}$ of galaxies when extrapolation to total K band
magnitude was necessary.
\item{(2)} Galaxies should have the redshift
completeness index c$_{z}$$>$0.5, where  c$_{z}$ is the ratio of the
number of galaxies with measured redshifts within 1\degr \ radius
from the center of the galaxy in question and the number of all
galaxies within the same radius (Xu et al. 2004). This confines
the pair selection to regions where the SDSS-DR3 has good spectroscopic 
coverage. 
\end{description}
The resulting sample has 50312 galaxies, of which 42847 have measured
redshifts (85\% redshift completeness). A comparison with the number counts
of 2MASS galaxies (Kochanek, et al. 2001) yielded an equivelent sky coverage
of $\Omega = 3000$ deg$^2$ ($\rm \Omega = N/CN$, where number counts $\rm CN$ 
is in units of deg$^{-2}$). 

Pair selection was then carried out with the following criteria:
\begin{description}
\item{(1)} The
${\rm K_s}$ magnitude of the primary is brighter than 12.5 mag.  A primary
is defined as the brighter component of a pair.  
\item{(2)} The ${\rm K_s}$
difference between the two components is less than 1 magnitude: 
$\rm \delta K_s \leq 1$ mag.  
\item{(3)} At least one of the components has spectroscopic redshift.
\item{(4)} The projected separation is in the range of 
${\rm 5 \leq s \leq 20 h^{-1}}$ kpc.  
\item{(5)} The velocity difference is less than 500 km sec$^{-1}$: 
$\rm \delta v \leq 500$ km sec$^{-1}$. 
\end{description}
Criteria (1) and (2) ensure that, by construction,
all galaxies in this pair sample are
brighter than ${\rm K_s}$=13.5, the completeness limit of 2MASS survey
(Jarrett et al. 2000). Therefore the sample {\it does not}
suffer from the ``missing the secondary''
bias that plagued many earlier merger studies (see Section 1).
This also restricts our sample to
major merger pairs whose mass ratios are $\leq 2.5$. 
Criterion (4) makes the comparisons between
local pairs in this sample and pairs in high-z samples 
robust.  For high-z galaxies,
it is difficult to distinguish a galaxy pair from a single galaxy if
the separation is $\leq 5 h^{-1}$ kpc, and the probability for chance
pairs is significantly higher if projected separation is much larger
than $20 h^{-1}$ kpc (Kartaltepe et al. 2007).  

A total of 57 pairs were selected according to the criteria.
Component galaxies are classified as either ``S'' or ``E'' using the following 
scheme: First, classical ``eye-ball'' classifications were carried out. 
Independently, two of us (CKX and DD) inspected the
SDSS optical image of every galaxy in the sample and assigned a type to it
according to its morphology. In addtion, we also 
run an automatic classification script. It classifies a galaxy as ``E'' if
$u - r > 2.22$ and $R_{50}/R_{90} < 0.35$, where 
$R_{50}$ and $R_{90}$ are the radii 
containing 50\% and 90\% of the Petrosian flux (Park \& Choi 2005).
Otherwise, the galaxy is classified as ``S'' type.
The final classification is the median of the two eyeball results and the
result of the automatic method.
We excluded 20 pairs in which both components are ``E'' types
(``E+E'' pairs). According to their optical colors and EW(H$\alpha$), 
little star formation is evident, consistent with
IRAS results on E+E pairs (Xu \& Sulentic 1991). 
In order to minimize the contamination due to interlopers, 
single-redshift pairs were also excluded. We would stress 
that, judging from SDSS images, these single-redshift pairs 
do not show any special characteristics compared to the
total sample, therefore dropping them will not introduce 
any significant bias to our study.

The final sample (KPAIR hereafter) for the Spitzer observations 
contains 27 S+S and S+E pairs. Among the 54 component galaxies,
42 are classified as S and 12 as E. They are listed in Table 1.
The mass in the table
(and hearafter) is the so called ``stellar mass'', excluding the
mass of the dark matter and gas.  
It is estimated from the K-band luminosity by assuming
a mass-to-luminosity ratio of 1.32 (solar units), taken from
Cole et al. (2001) for the Salpeter IMF\footnote{
The mass-to-luminosity ratio estimated using the Kennicutt IMF is a factor of
0.55 lower (Cole et al. 2001), and that estimated using the Kroupa IMF
is a factor of 0.59 lower (Kauffmann et al. 2003).}. 
The heliocentric velocities  
are from SDSS-DR3, and the 60$\mu m$ flux densities are from
the IRAS Faint Source Catelog (Moshir et al. 1992). 

\section{Spitzer Observations}
Infrared imaging observations of all pairs but J1315+6207 in Table 1
were carried out in 2005 and 2006 under the Spitzer
Cycle 2 GO Program ``Local Benchmarks for the Evolution 
          of Interacting Galaxies'' (PID \#20187).
These include images in the 
four bands (3.6, 4.5, 5.8 and 8.0 $\mu m$) of the Infrared Array
Camera (IRAC, Fazio et al. 2004) and in three bands (24, 70 and 160 $\mu m$)
of the Multiband Imaging Photometer for Spitzer (MIPS, Rieke et
al. 2004).
KPAIR J1315+6207 (UGC08335a/b) is included in the Spitzer GO-1 survey of
LIRGs (Mazzarella et al. 2009). Its IRAC data were taken from 
Mazzarella et al. (2009), while its MIPS observations were unsuccessful.

\begin{deluxetable}{cclrrrrlrr}
\label{tbl:kpair}
\tabletypesize{\scriptsize}
\setlength{\tabcolsep}{0.05in} 
\tablenum{1}
\tablewidth{0pt}
\tablecaption{KPAIR Galaxy Sample}
\tablehead{
\ch{(1)}   &\ch{(2)}    &\ch{(3)}    &\ch{(4)}&\ch{(5)}&\ch{(6)} &\ch{(7)}
             &\ch{(8)}         &\ch{(9)}  &\ch{(10)}     \\
\ch{Pair ID} & \ch{Galaxy ID} & \ch{RA}  & \ch{Dec} & 
\ch{V$_z$} & \ch{K$_s$} & \ch{log(M)} & \ch{Type} 
& \ch{f$_{60\mu m}$} & \ch{L$_{60\mu m}$} \\
\ch{(KPAIR)}  & \ch{(2MASX)} & \ch{(J2000)} & 
\ch{(J2000)} & \ch{(km s$^{-1}$)} & \ch{(mag)} & 
\ch{(M$_\sun$)}& & \ch{(Jy)} &\ch{(log(L$_\sun$))} \\  
}
\startdata
J0020+0049 & J00202580+0049350 & 00h20m25.8s & +00d49m35s &  5078 & 10.99 & 10.84 & S & 0.62 &  9.68\\
           & J00202748+0050009 & 00h20m27.5s & +00d50m01s &  5480 & 10.50 & 11.04 & E &      &      \\
J0109+0020 & J01093371+0020322 & 01h09m33.7s & +00d20m32s & 13499 & 12.39 & 11.08 & E &      &      \\
           & J01093517+0020132 & 01h09m35.1s & +00d20m13s & 13319 & 12.47 & 11.05 & S &      &      \\
J0118-0013 & J01183417-0013416 & 01h18m34.1s & -00d13m42s & 14219 & 12.05 & 11.27 & S 2 & 3.48 & 11.30\\
           & J01183556-0013594 & 01h18m35.6s & -00d13m59s & 14273 & 12.88 & 10.93 & S &      &      \\
J0211-0039 & J02110638-0039191 & 02h11m06.3s & -00d39m21s &  5920 & 11.42 & 10.77 & S 2 &      &      \\
           & J02110832-0039171 & 02h11m08.3s & -00d39m17s &  6016 & 10.90 & 10.98 & S &      &      \\
J0906+5144 & J09060283+5144411 & 09h06m02.8s & +51d44m41s &  8849 & 11.68 & 11.01 & E &      &      \\
           & J09060498+5144071 & 09h06m05.0s & +51d44m07s &  8852 & 11.95 & 10.90 & S 2 &      &      \\
J0937+0245 & J09374413+0245394 & 09h37m44.1s & +02d45m39s &  6988 & 10.01 & 11.46 & S & 2.00 & 10.43\\
           & J09374506+0244504 & 09h37m45.0s & +02d44m50s &  6790 & 10.45 & 11.29 & E &      &      \\
J0949+0037 & J09494143+0037163 & 09h49m41.4s & +00d37m16s &  1861 & 11.59 &  9.71 & S & 2.27 &  9.36\\
           & J09495263+0037043 & 09h49m52.6s & +00d37m05s &  1918 & 10.98 &  9.95 & S &      &      \\
J1020+4831 & J10205188+4831096 & 10h20m51.9s & +48d31m10s & 15886 & 13.26 & 10.88 & S &      &      \\
           & J10205369+4831246 & 10h20m53.6s & +48d31m24s & 15930 & 12.27 & 11.27 & E R &      &      \\
J1027+0114 & J10272950+0114490 & 10h27m29.5s & +01d14m48s &  6727 & 11.79 & 10.73 & S &      &      \\
           & J10272970+0115170 & 10h27m29.7s & +01d15m16s &  6661 & 10.90 & 11.08 & E &      &      \\
J1043+0645 & J10435053+0645466 & 10h43m50.5s & +06d45m47s &  8262 & 11.96 & 10.83 & S &      &      \\
           & J10435268+0645256 & 10h43m52.7s & +06d45m25s &  8088 & 12.20 & 10.73 & S &      &      \\
J1051+5101 & J10514368+5101195 & 10h51m43.6s & +51d01m20s &  7503 & 10.27 & 11.41 & E & 0.78 & 10.07\\
           & J10514450+5101303 & 10h51m44.5s & +51d01m30s &  7138 & 10.97 & 11.13 & S &      &      \\
J1202+5342 & J12020424+5342317 & 12h02m04.3s & +53d42m32s & 19366 & 12.97 & 11.16 & S &      &      \\
           & J12020537+5342487 & 12h02m05.3s & +53d42m48s & 19156 & 12.43 & 11.37 & E &      &      \\
J1308+0422 & J13082737+0422125 & 13h08m27.4s & +04d22m13s &  7186 & 13.39 & 10.15 & S &      &      \\
           & J13082964+0422045 & 13h08m29.6s & +04d22m05s &  7251 & 12.44 & 10.53 & S &      &      \\
J1332-0301 & J13325525-0301347 & 13h32m55.3s & -03d01m35s & 14297 & 12.95 & 10.90 & S & 0.57 & 10.51\\  
           & J13325655-0301395 & 13h32m56.6s & -03d01m40s & 14000 & 12.19 & 11.21 & S &      &      \\
J1346-0325 & J13462001-0325407 & 13h46m20.0s & -03d25m41s &  6949 & 11.20 & 11.01 & S &      &      \\
           & J13462215-0325057 & 13h46m22.2s & -03d25m06s &  7171 & 11.66 & 10.82 & E 2 &      &      \\
J1400+4251 & J14005782+4251207 & 14h00m57.8s & +42d51m20s &  9689 & 11.89 & 11.01 & S & 2.32 & 10.80\\
           & J14005882+4250427 & 14h00m58.8s & +42d50m42s &  9923 & 12.18 & 10.90 & S &      &      \\
J1425+0313 & J14250552+0313590 & 14h25m05.5s & +03d13m59s & 10693 & 11.98 & 11.06 & E 1 &      &      \\
           & J14250739+0313560 & 14h25m07.4s & +03d13m55s & 10807 & 12.97 & 10.66 & S &      &      \\
J1433+4004 & J14334683+4004512 & 14h33m46.8s & +40d04m52s &  7674 & 10.78 & 11.25 & S & 1.80 & 10.48\\
           & J14334840+4005392 & 14h33m48.4s & +40d05m39s &  7788 & 11.17 & 11.10 & S &      &      \\
J1453+0317 & J14530282+0317451 & 14h53m02.8s & +03d17m46s &  1576 & 10.67 &  9.92 & S & 0.29 &  8.30\\
           & J14530523+0319541 & 14h53m05.2s & +03d19m54s &  1573 & 10.05 & 10.17 & S &      &      \\
J1506+0346 & J15064391+0346364 & 15h06m43.9s & +03d46m36s & 10498 & 11.48 & 11.22 & S & 0.27 &  9.91\\
           & J15064579+0346214 & 15h06m45.8s & +03d46m22s & 10183 & 11.61 & 11.17 & S &      &      \\
J1510+5810 & J15101587+5810425 & 15h10m15.8s & +58d10m43s &  9161 & 11.77 & 11.02 & S & 0.29 &  9.86\\
           & J15101776+5810375 & 15h10m17.8s & +58d10m37s &  9563 & 12.35 & 10.79 & S &      &      \\
J1528+4255 & J15281276+4255474 & 15h28m12.8s & +42d55m48s &  5588 & 10.02 & 11.26 & S & 1.10 &  9.96\\
           & J15281667+4256384 & 15h28m16.6s & +42d56m39s &  5345 & 10.59 & 11.03 & S &      &      \\
J1556+4757 & J15562191+4757172 & 15h56m21.9s & +47d57m17s &  5741 & 12.10 & 10.49 & S & 0.53 &  9.71\\
           & J15562738+4757302 & 15h56m27.4s & +47d57m30s &  5980 & 12.16 & 10.46 & E &      &      \\
J1602+4111 & J16024254+4111499 & 16h02m42.6s & +41d11m50s & 10019 & 11.69 & 11.11 & S & 1.43 & 10.60\\
           & J16024475+4111589 & 16h02m44.7s & +41d11m59s &  9950 & 12.50 & 10.78 & S &      &      \\
J1704+3448 & J17045089+3448530 & 17h04m50.9s & +34d48m53s & 17163 & 13.07 & 11.01 & S & 1.41 & 11.06\\
           & J17045097+3449020 & 17h04m50.9s & +34d49m02s & 16893 & 12.40 & 11.28 & S &      &      \\
J2047+0019 & J20471908+0019150 & 20h47m19.1s & +00d19m15s &  4209 &  9.08 & 11.37 & S & 0.34 &  9.18\\
           & J20472428+0018030 & 20h47m24.3s & +00d18m03s &  3795 &  9.74 & 11.10 & E &      &      \\
J1315+6207 & J13153076+6207447 & 13h15m30.8s & +62d07m45s &  9170 & 11.99 & 10.91 & S & 11.39& 11.47\\
           & J13153506+6207287 & 13h15m35.1s & +62d07m29s &  9176 & 11.54 & 11.09 & S &      &      \\
\hline
\enddata
\tablecomments{
\nid{\bf Descriptions of Columns}:
\begin{description}
\item{(1)} Pair ID. The designations are ``KPAIR J0020+0049'', etc.
\item{(2)} Galaxy ID, taken from 2MASS.
\item{(3)} RA (J2000). 
\item{(4)} Dec (J2000).
\item{(3)} Recession velocity taken from SDSS.
\item{(6)} $\rm K_s$ ($\rm K_{20}$) magnitude taken from 2MASS.
\item{(7)} Logarithm of the stellar mass of the galaxy.
\item{(8)} Morphological type. Galaxies containing known AGNs (via NED) are flagged with 1, 2 or R to indicate type 1, type 2 or radio AGNs.
\item{(9)} The IRAS 60$\mu m$ flux density of the total pair.
\item{(10)} The 60$\mu m$ luminosity ($\rm \nu L_\nu$) of the total pair.
\end{description}
}
\end{deluxetable}

%
\begin{deluxetable}{lccccccccccccc}
\label{tbl:observation2}
\tabletypesize{\footnotesize}
\setlength{\tabcolsep}{0.01in} 
\rotate
\tablenum{2}
\tablewidth{0pt}
\tablecaption{The Spitzer Observations}
\tablehead{
\colhead{}  & \colhead{}  & \colhead{}  & \colhead{} &\multicolumn{2}{c}{\rm IRAC} &\colhead{}
     &\multicolumn{2}{c}{\rm MIPS} &\colhead{} &\multicolumn{4}{c}{\rm On-Source Time} \\
\cline{5-6} \cline{8-9}\cline{11-14} \\
\colhead{\rm Pair ID} & & \colhead{\rm RA} & \colhead{\rm Dec}  & \colhead{\rm AORKEY} &  \colhead{\rm ObsDate} &
    & \colhead{\rm AORKEY} &  \colhead{\rm ObsDate} && \colhead{\rm IRAC} & \colhead{\rm 24$\mu m$} & \colhead{\rm 70$\mu m$} & \colhead{\rm 160$\mu m$} \\
\colhead{\rm (KPAIR)} && \colhead{\rm (J2000)} & \colhead{\rm (J2000)}  & \colhead{\rm} &  \colhead{\rm (yy/mm/dd)} &
    & \colhead{} &  \colhead{\rm (yy/mm/dd)} && \colhead{\rm (sec)} & \colhead{\rm (sec)} & \colhead{\rm (sec)} & \colhead{\rm (sec)} \\
\colhead{(1)} &\colhead{}  & \colhead{(2)}& \colhead{(3)}& \colhead{(4)}
& \colhead{(5)}&  \colhead{} &\colhead{(6)}
& \colhead{(7)}&  \colhead{} & \colhead{(8)}
& \colhead{(9)}& \colhead{(10)} & \colhead{(11)}\\
}
\startdata
    J0020+0049  & & 00h20m26.6s & +00d49m48s & 14276608 & 05/07/22 && 14269440 & 06/07/15 && 144.0 & 165.7 & 125.8 & 83.9\\
    J0109+0020  & & 01h09m34.4s & +00d20m23s & 14276864 & 06/01/01 && 14269696 & 06/07/19 && 144.0 & 165.7 & 125.8 & 83.9\\
    J0118$-$0013  & & 01h18m34.9s & -00d13m51s & 14277120 & 05/08/20 && 14269952 & 06/07/21 && 144.0 & 165.7 & 125.8 & 83.9\\
    J0211$-$0039  & & 02h11m07.3s & -00d39m19s & 14277376 & 05/08/21 && 14270208 & 06/02/14 && 144.0 & 165.7 & 125.8 & 83.9\\
    J0906+5144  & & 09h06m03.8s & +51d44m24s & 14277632 & 05/11/26 && 14270464 & 05/12/04 && 144.0 & 165.7 & 125.8 & 83.9\\
    J0937+0245  & & 09h37m44.6s & +02d45m15s & 14277888 & 05/11/26 && 14270720 & 05/12/03 && 144.0 & 165.7 & 125.8 & 83.9\\
    J0949+0037A & & 09h49m52.6s & +00d37m05s & 14278144 & 05/12/27 && 14270976 & 05/12/03 && 144.0 & 165.7 & 125.8 & 83.9\\
    J0949+0037B & & 09h49m41.4s & +00d37m16s & 14278400 & 05/12/27 && 14271232 & 05/12/03 && 144.0 & 165.7 & 125.8 & 83.9\\
    J1020+4831  & & 10h20m52.8s & +48d31m17s & 14278656 & 05/11/24 && 14271488 & 05/12/01 && 144.0 & 165.7 & 125.8 & 83.9\\
    J1027+0114  & & 10h27m29.5s & +01d15m03s & 14278912 & 05/12/24 && 14271744 & 06/06/07 && 144.0 & 165.7 & 125.8 & 83.9\\
    J1043+0645  & & 10h43m51.6s & +06d45m36s & 14279168 & 06/01/03 && 14272000 & 06/06/07 && 144.0 & 165.7 & 125.8 & 83.9\\
    J1051+5101  & & 10h51m44.2s & +51d01m25s & 14279424 & 05/11/24 && 14272256 & 05/12/05 && 144.0 & 165.7 & 125.8 & 83.9\\
    J1202+5342  & & 12h02m04.8s & +53d42m40s & 14279680 & 05/12/24 && 14272512 & 05/12/05 && 144.0 & 165.7 & 125.8 & 83.9\\
    J1308+0422  & & 13h08m28.6s & +04d22m09s & 14279936 & 05/07/24 && 14272768 & 06/02/21 && 144.0 & 165.7 & 125.8 & 83.9\\
    J1332$-$0301  & & 13h32m55.9s & -03d01m37s & 14280192 & 05/07/24 && 14273024 & 06/03/01 && 144.0 & 165.7 & 125.8 & 83.9\\
    J1346$-$0325  & & 13h46m21.1s & -03d25m23s & 14280448 & 05/07/24 && 14273280 & 06/03/01 && 144.0 & 165.7 & 125.8 & 83.9\\
    J1400+4251  & & 14h00m58.3s & +42d51m02s & 14280704 & 06/02/13 && 14273536 & 06/06/12 && 144.0 & 165.7 & 125.8 & 83.9\\
    J1425+0313  & & 14h25m06.5s & +03d13m58s & 14280960 & 05/07/24 && 14273792 & 06/03/01 && 144.0 & 165.7 & 125.8 & 83.9\\
    J1433+4004  & & 14h33m47.5s & +40d05m15s & 14281216 & 05/07/25 && 14274048 & 06/03/01 && 144.0 & 165.7 & 125.8 & 83.9\\
    J1453+0317A & & 14h53m02.8s & +03d17m46s & 14281472 & 05/07/24 && 14274304 & 06/03/01 && 144.0 & 165.7 & 125.8 & 83.9\\
    J1453+0317B & & 14h53m05.2s & +03d19m54s & 14281728 & 05/07/24 && 14274560 & 06/02/25 && 144.0 & 165.7 & 125.8 & 83.9\\
    J1506+0346  & & 15h06m44.9s & +03d46m29s & 14281984 & 05/07/24 && 14274816 & 06/03/01 && 144.0 & 165.7 & 125.8 & 83.9\\
    J1510+5810  & & 15h10m16.8s & +58d10m40s & 14282240 & 05/07/25 && 14275072 & 06/02/22 && 144.0 & 165.7 & 125.8 & 83.9\\
    J1528+4255  & & 15h28m14.6s & +42d56m13s & 14282496 & 05/07/24 && 14275328 & 06/03/03 && 144.0 & 165.7 & 125.8 & 83.9\\
    J1556+4757  & & 15h56m24.7s & +47d57m24s & 14282752 & 05/07/24 && 14275584 & 06/05/06 && 144.0 & 165.7 & 125.8 & 83.9\\
    J1602+4111  & & 16h02m43.7s & +41d11m55s & 14283008 & 05/07/24 && 14275840 & 05/08/29 && 144.0 & 165.7 & 125.8 & 83.9\\
    J1704+3448  & & 17h04m50.9s & +34d48m58s & 14283776 & 05/07/24 && 14284032 & 05/08/29 && 144.0 & 165.7 & 125.8 & 83.9\\
    J2047+0019A & & 20h47m24.3s & +00d18m03s & 14283264 & 05/10/21 && 14276096 & 05/11/10 && 144.0 & 165.7 & 125.8 & 83.9\\
    J2047+0019B & & 20h47m19.1s & +00d19m15s & 14283520 & 05/10/21 && 14276352 & 05/11/10 && 144.0 & 165.7 & 125.8 & 83.9\\

\hline
\enddata
\end{deluxetable}

\subsection{IRAC}

\subsubsection{Observations}

All but 3 pairs in our sample are smaller than $2'.5$ at optical wavelengths.
Each pair (or each pair component for the 3 large pairs) was observed with
the IRAC instrument in Full Array mode.  High dynamic range settings
and a 12 positions Reuleaux dither pattern were utilized. The dynamic
range setting was designed to incorporate short exposures of 0.4 sec
along with longer exposures of 12 sec, in order to acquire photometry
on potentially saturated nuclei as well as reach the requested
signal-to-noise ratio for the outer galaxy regions. The dither pattern
produces the redundancy needed to eliminate cosmic rays. 
The average surface brightness
sensitivity levels (4$\sigma$) are 0.032, 0.048, 0.160, and 0.316
MJy sr$^{-1}$ in the 4 IRAC bands, respectively.

\subsubsection{Data Reduction}

The Basic Calibrated Data (BCD) 
images were produced in the Spitzer Science Center (SSC) pipeline
and these served as the start of our data reduction process. Custom IDL
tools created by one of us (J.S. Huang of the IRAC instrument team)
were used to apply the additional steps of pointing refinement,
distortion correction, and mosaicking with sigma-clipping for the
rejection of cosmic rays.
The long and short exposures were processed independently for the purpose
of creating mosaics in all four bands. These mosaics were resampled
from the native 1\farcs22 pixel scale to a 0\farcs66 pixel scale.

Aperture photometry was carried out on the images in order to obtain
flux densities in the four IRAC bands. Two different methods 
were applied to different sources:
\begin{description}
\item{(1)} Standard aperture photometry for individual
galaxies. This was performed on well separated
pairs, utilizing both IPAC Skyview and IRAF
APPHOT. Several blank sky regions were chosen from within the IRAC
field containing the galaxy pairs. These regions were used to
calculate the mean background value for subtraction. The standard
deviation of their means  provides estimate of
the uncertainties of background subtraction (Smith et al. 2007). 
Apertures of different shapes (circular, eliptical and polygon) are
chosen according to the situation. 
\item{(2)} Aperture photometry first for the total pair and then
for one of the components. This was performed on pairs
whose two component galaxies are too close to be separated cleanly.
We always chose
the galaxy which has a more regular morphology (point-source-like or
a regular elliptical) for the aperture photometry. This flux is then
subtracted from the flux of the total pair, which yields the flux of the
other component.
\end{description}

In addition to total fluxes of whole galaxies,
circular aperture fluxes on two physical scales corresponding to
$D = 4$ and 10 kpc, respectively, were also
obtianed. These fluxes sample the nucleus and disk emissions.

In the analysis of IRAC extended sources, it has been found that light
is scattered out of the apertures into the array.  The effect is most
severe in the 5.8 and 8.0 $\mu$m channels. The SSC has developed
`best practices' for applying aperture corrections.  These
corrections are dependent upon the IRAC channel and the area within
the aperture (or ``effective circular aperture radii'').  All of the
reported fluxes (Table 3) have this apperture correction applied, which
introduces an uncertainty of $\sim$ 10\% to the flux densities. No
color correction is applied to the IRAC data.

The errors of the IRAC photometry are dominated by two uncertainties:
for bright sources the uncertainty of the aperture correction is
dominant, and for the faint sources the uncertainty of the background
subtraction becomes important. The errors reported in Table
3 were determined as the quadratic sum of the two.

\subsection{MIPS}
\subsubsection{Observations}
The 24$\mu m$ observations used the small-field photometry mode. Both the
70$\mu m$ and the 160$\mu m$ observations used the $3\times 1$ raster map
mode, with a
1/8 array step size in the cross-scan direction. These provided 
images of $\sim 5'$ on one side and no less
than $2'.5$ on the other side,
large enough to include some background regions
for the sky subtraction, as well as to have the central
$2'.5\times 2'.5$ of each map to be fully sampled. 
For the 3 pairs larger than $2'.5$, the same observation configuration
was applied to each component instead of the total pair.
For all fields, the same exposure time and same number of
cycles were applied, which are : ($t_{exp}$ in secs,
$N_{cycle}$) = (10, 1), (10, 1) and (10, 4) at 24, 70 and 160 $\mu m$,
respectively. These corresponded to effective integration times
of 165.7, 125, and 83.9 seconds for the three bands, and yielded
average surface brightness rms noise (4$\sigma$) of
 0.28, 2.0, and 6.0 MJy sr$^{-1}$ in these bands, respectively.

\subsubsection{Data Reduction}
The BCD images were taken directly from the
products of Spitzer Science Center (SSC) pipeline (Version 13).  These
are calibrated images of individual exposures (MIPS Data Handbook). 
The following major artifacts on the BCD images were
corrected in our data reduction process:
\begin{description}
\item{(1)} 'Stripes' in the 70 $\mu m$ images. 
These are caused by flat-fielding variations due either to
the latents of the stim flash or to the long term responsivity variation.
All of our 70 $\mu m$ observations were affacted, some more severely 
than others. Following the recommendations in
the MIPS Data Manual, we performed self-flat-fielding using our
own data.  This was done by calculating for each detector
pixel the 2-$\sigma$ clipped median over all BCD frames 
within the given observation.
Dividing the BCDs by the self-flat-field image removed the artificial stripes. 
\item{(2)} Dark latents in the 
24 $\mu m$ images. These are long time scale (up to
10 hours) artifacts produced by very bright sources. Two of our observations
were affected: AORKEY=14271232 and AORKEY=14270720 (see Table 2). 
In both cases, the 
bright source which causes latents was not in the field of view. 
The same self-flat-fielding method for the correction of 
the artificial stripes in the 70 $\mu m$ images was exploited in
the correction for these artifacts.
\item{(3)} Gradients in the 24 $\mu m$ images. These non-astronomical gradients 
are due to poor flat fielding. If not corrected, they produce artificial 
discontinuities in the mosaicked images. Most of the 24 $\mu m$ BCDs
were affected by this. We fitted a slope to the sky background 
of each individual BCD and subtracted it. This
effectively took out the gradients.
\end{description}
After these corrections, the SSC software MOPEX was used to make the
mosaic image out of the BCD frames for each 24$\mu m$ or 70$\mu m$
MIPS observation.  The 160$\mu m$ maps were taken from the post-BCD
products of the same SSC pipeline. Combined mosaic images were
produced using MOPEX for the 3 large pairs that have MIPS observations
of the two components separately.

Flux densities of individual galaxies in pairs were 
measured from the corresponding images. Given the relatively 
large beams and small separations of pair 
components, it was challenging to carry out these measurements.
Two different methods have been exploited:
\begin{description}
\item{(1)} {\bf Aperture photometry}.  When only one of the two
  components in a pair is detectable or, if both components are bright
  in the given MIPS band, the two components are separable, the flux
  densities were measured using aperture photometry.  Polygon
  apertures were used in order to avoid overlapped apertures, with the
  aperture size changing with the galaxy size.  The aperture
  corrections were estimated using two different methods.  In all
  cases except for the six $160 \mu m$ measurements in which the
  galaxy in question is extended in the $160\mu m$ band, the aperture
  corrections were estimated under the point source assumption,
  exploiting the Point Response Function (PRF) presented in Appendix
  B.  For the 24$\mu m$ and 70$\mu m$ sources that are significantly
  larger than a point source, this is an underestimation of the real
  aperture correction, though the fractional aperture
  correction itself decreases with source size. 
  For the extended 160$\mu m$ sources, the apperture
  corrections were estimated differently.  Because of the relatively
  small size of the $160 \mu m$ images compared to the beam, much of
  the emission of any galaxy that is extended in the 160$\mu m$ falls
  out of the $160 \mu m$ image, and would not be recovered if an
  aperture correction based on the point source assumption were
  applied.  In order to fully recover the missing flux, the following
  method was carried out. First, the 24$\mu m$ image (2 times larger than the
  160$\mu m$ image) was convolved with the PRF of the 160$\mu m$
  band. Then, under the assumption that the $f_{160}/f_{24}$
  ratio is constant across the galaxy in question, the desired
  aperture correction in the 160$\mu m$ band was estimated from the
  ratio between the total flux in the 24$\mu m$ band and that in the
  160$\mu m$ aperture measured on the smoothed 24$\mu m$ image.
  From experiments with individual sources, the following aperture 
  correction errors were assigned: 14\% for weak 
  sources in the 24$\mu m$ band, 5\% for other 24$\mu m$ sources and
  all 70$\mu m$ sources, and 10\% for all 160$\mu m$ sources.
\item{(2)} {\bf PRF fitting}.  This was applied to pairs in which both
  component galaxies are detectable point sources and the two PRFs
  overlap with each other.  
  In these cases, the flux densities of the two components
  were measured by simultaneously fitting two PRFs within a pair,
  positions and brightness of the two PRFs being the free parameters
  (altogether six of them: $x_1$, $y_1$, peak$_1$, $x_2$, $y_2$,
  peak$_2$). The average flux measurement errors are 14\%, 5\% and
  21\% in the 24, 70, and 160$\mu m$ bands, respectively.
\end{description}


MIPS data are calibrated assuming a
nominal spectrum of a blackbody at 10,000K. This is very
different from the dust emission in galaxies. Therefore color corrections,
calculated using the modified blackbody spectra ($\beta = 2$) 
of temperatures of 100K, 20K and 20K, were applied to the 
24$\mu m$ , 70$\mu m $ and 160$\mu m$ band flux densities, respectively.
The corresponding correction factors (multiplicative)
are 0.967, 0.923 and 0.954.

The errors of the MIPS photometry are the quadratic sum of (in the order of
importance): (1) the aperture-correction/model-fitting error (dominant
for all sources except for the faintest ones), (2) the uncertainty of
the background subtraction, and (3) the rms error of the background.
We ignored the photon noise which is never significant.  The
4-$\sigma$ upperlimits for the undetected sources were estimated using
the quadratic 
sum of the uncertainty of the mean background value and the rms
of the background.  The apertures addopted in these estimates
were derived from the K-band Kron radii taken from the 2MASS
database. However, for a given MIPS band, if the Kron radius of a
galaxy is less than a minimum value corresponds to the beam, then the
aperture is derived using the minimum (8$''$, 20$''$, and 40$''$ for
24, 70 and 160$\mu m$, respectively).

\setcounter{figure}{0}
\begin{figure}
\vspace*{-2.0cm}
\plotone{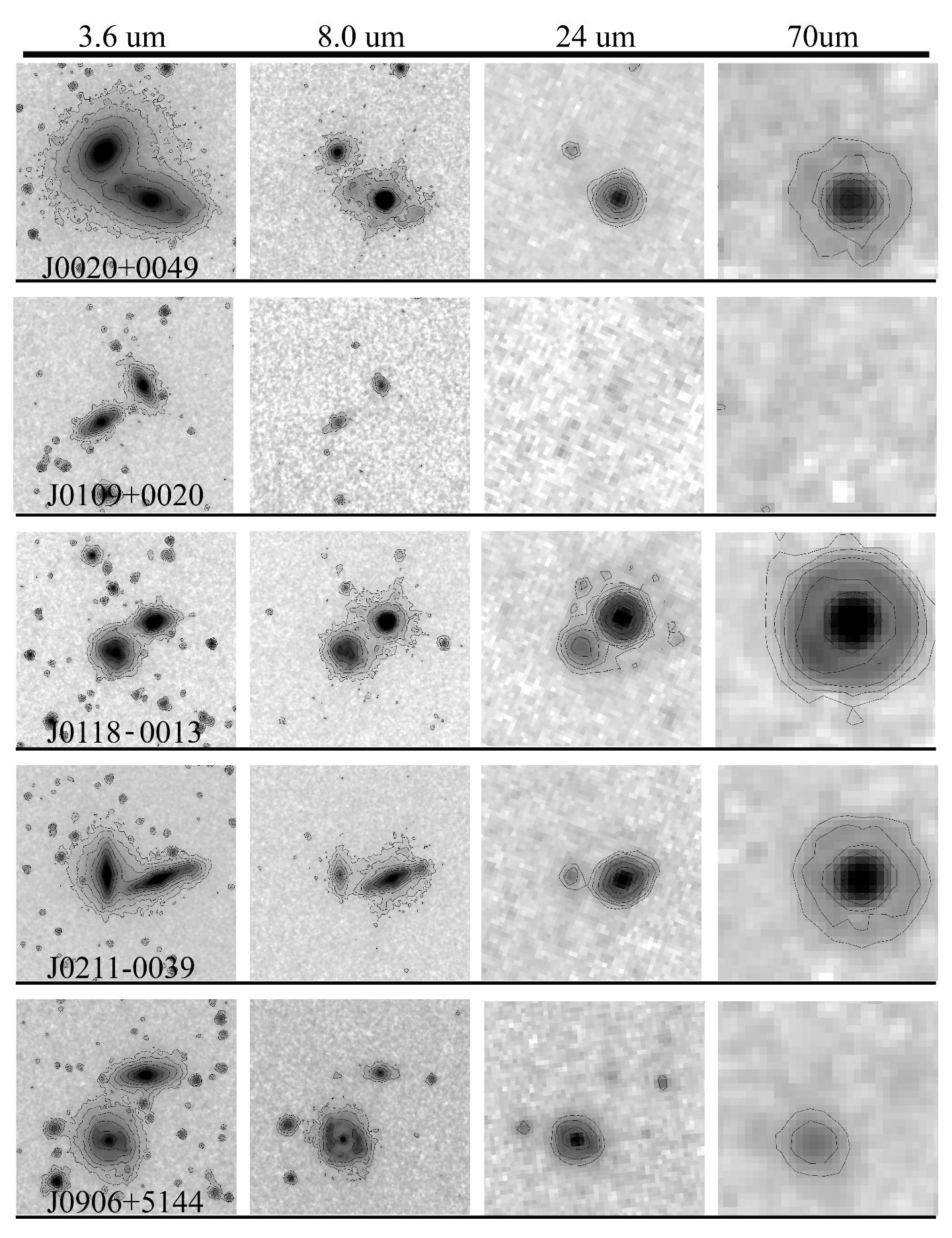}
\vspace*{-1.0cm}
\caption{Spitzer images of galaxy pairs. The contour levels are 
$S_{bg}\; +\; S_0 \times 2^n$ (n=0, 1, 2, ...),
where $S_{bg}$ is the local background level and
$S_0$, the sample median of the 4-$\sigma$ threshold of the corresponding
bands, are 0.032, 0.16, 0.28 and 2.0 MJy sr$^{-1}$ for the 3.6$\mu m$,
8.0$\mu m$, 24$\mu m$, and 70$\mu m$ band, respectively. 
The sizes of all images are $2'\times 2'$ except for 
the three large pairs: J0949+0037, J1453+0317 and J2047+0019, for which a
scale bar of length of 2$'$ is given.}
\end{figure}

\setcounter{figure}{0}
\begin{figure}
\vspace*{-1.0cm}
\plotone{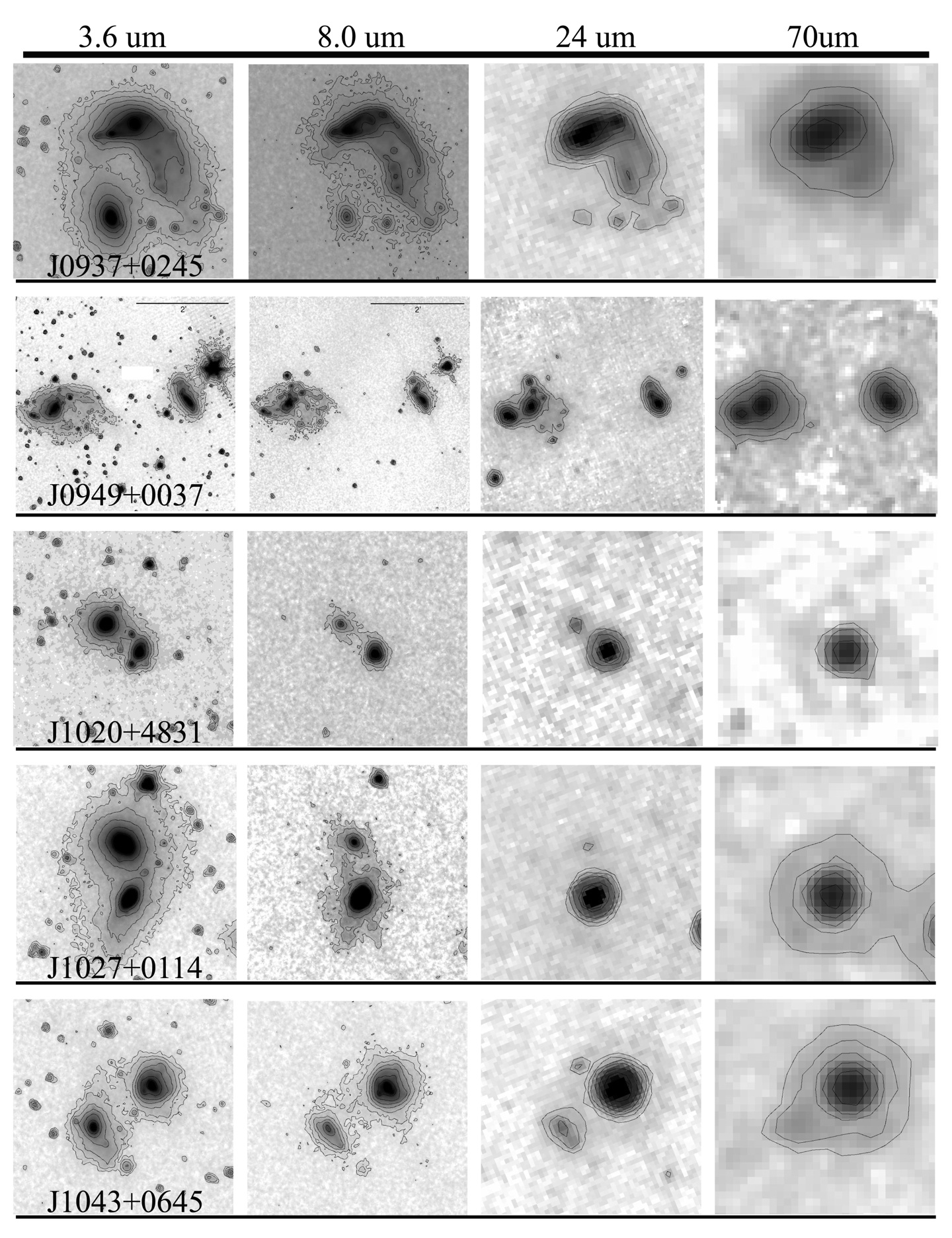}
\caption{Continued.}
\end{figure}

\setcounter{figure}{0}
\begin{figure}
\vspace*{-1.0cm}
\plotone{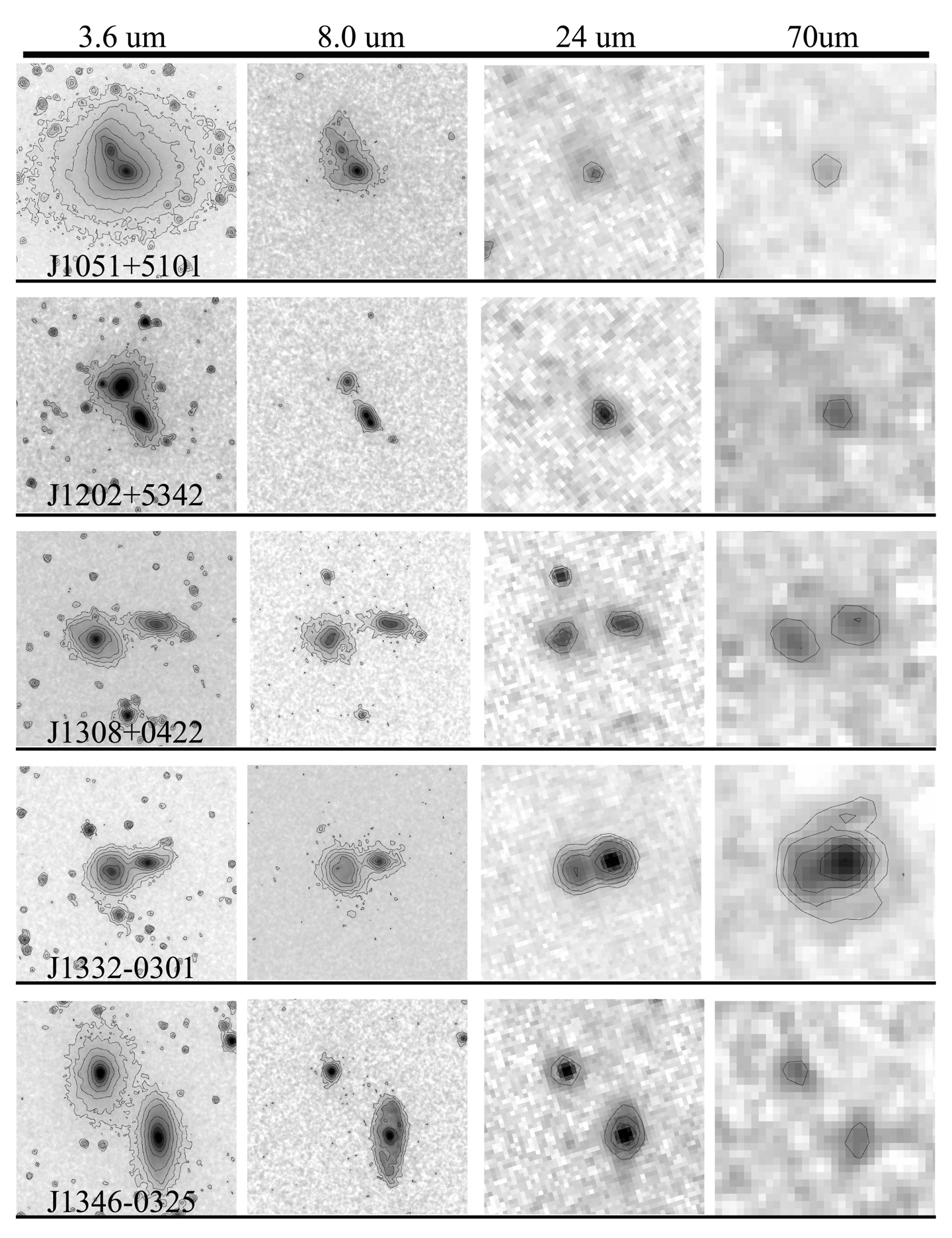}
\caption{Continued.}
\end{figure}

\setcounter{figure}{0}
\begin{figure}
\vspace*{-1.0cm}
\plotone{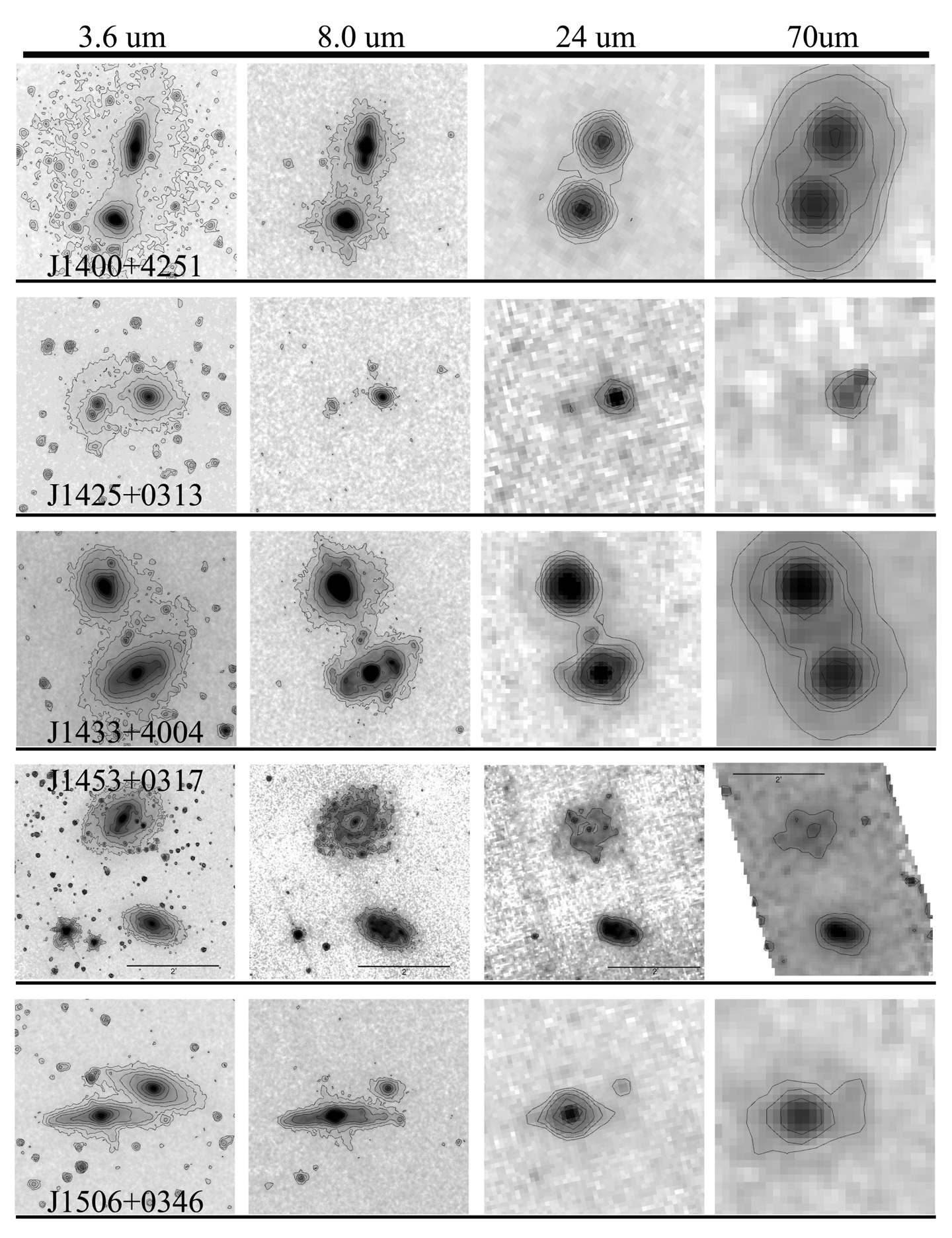}
\caption{Continued.}
\end{figure}

\setcounter{figure}{0}
\begin{figure}
\plotone{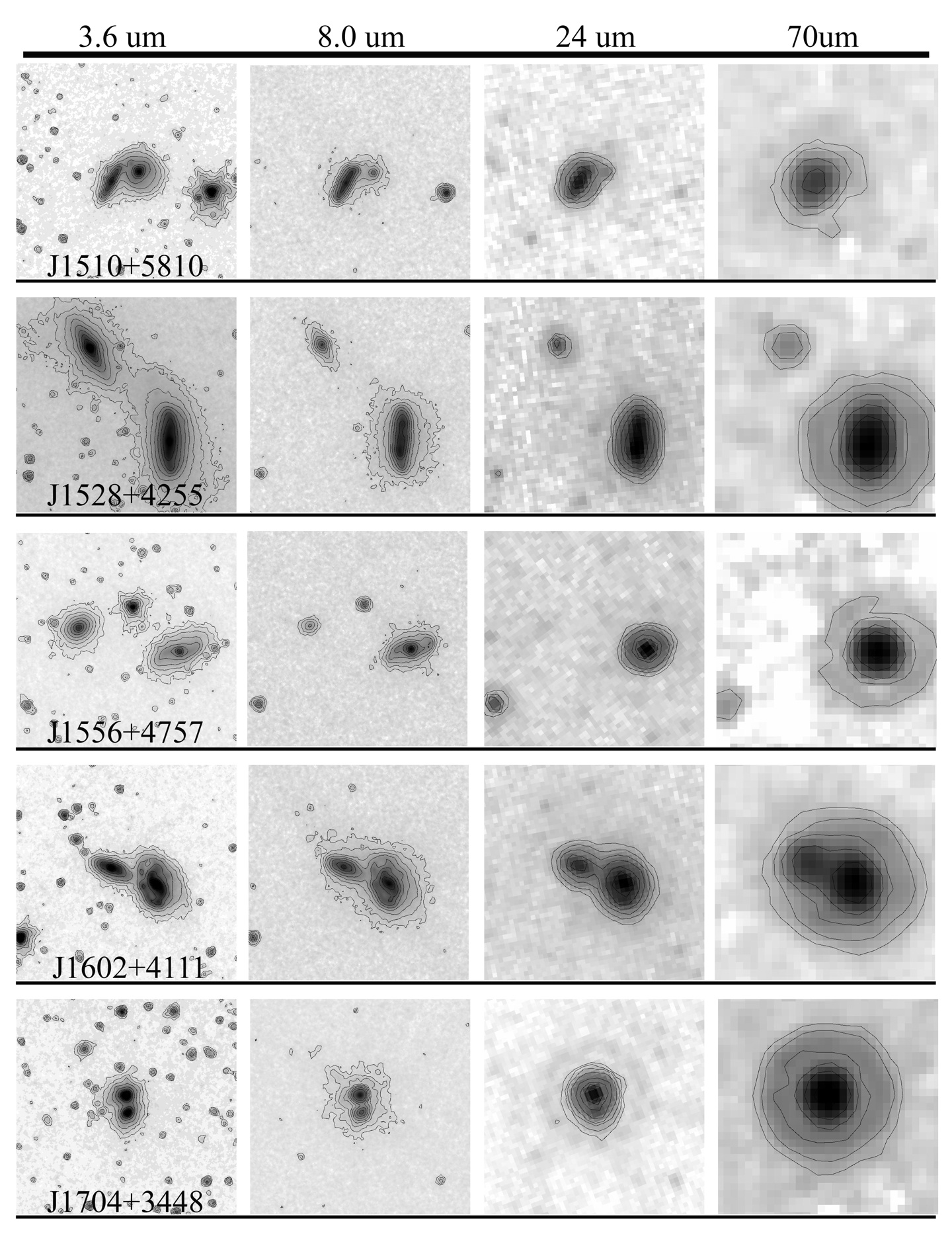}
\caption{Continued.}
\end{figure}
\setcounter{figure}{0}
\begin{figure}
\plotone{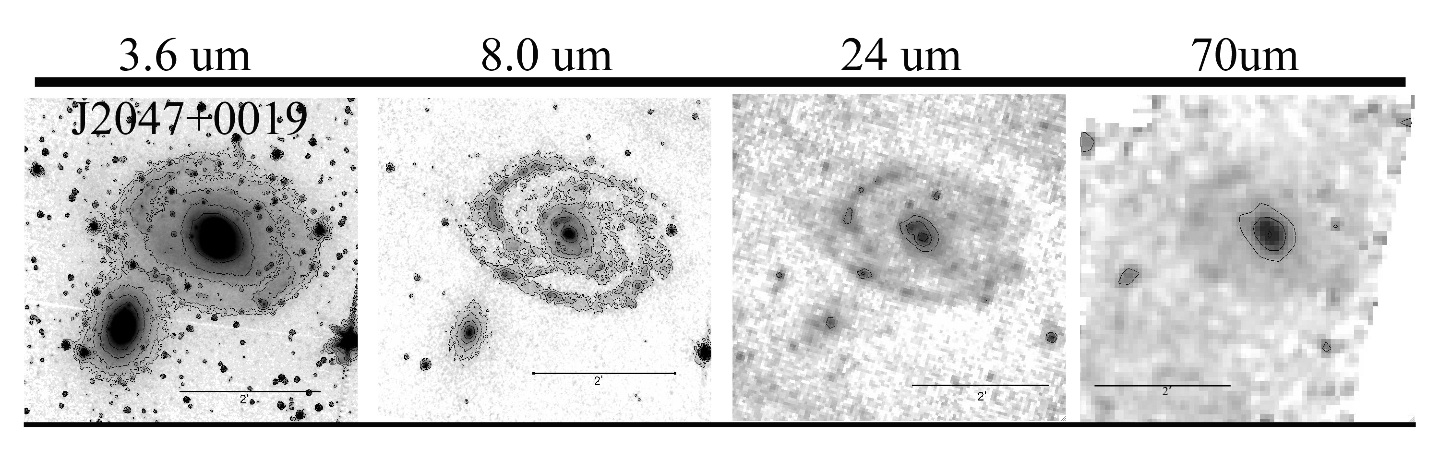}
\caption{Continued.}
\end{figure}
%
%
\setcounter{figure}{1}
\begin{figure}
\plottwo{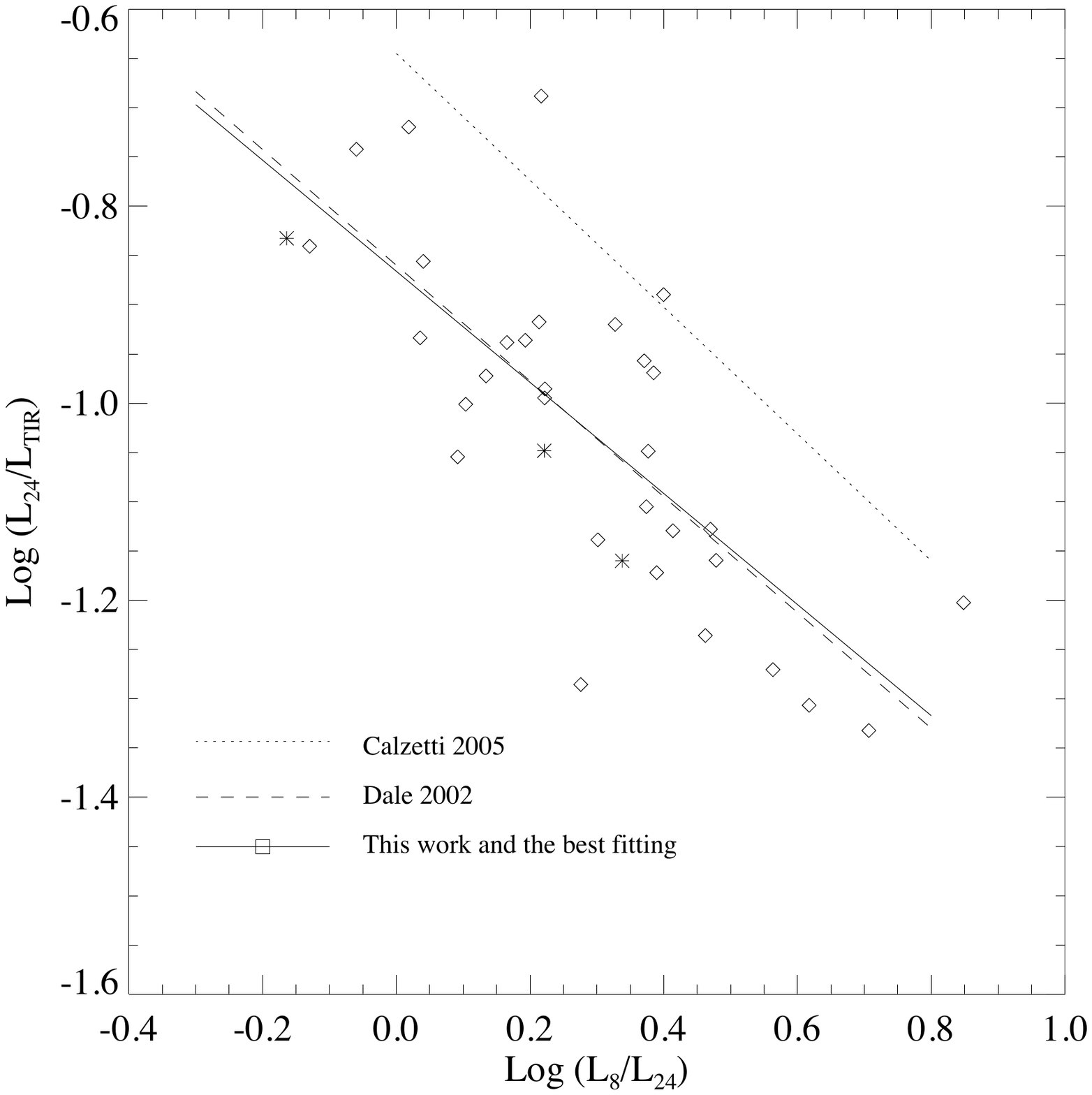}{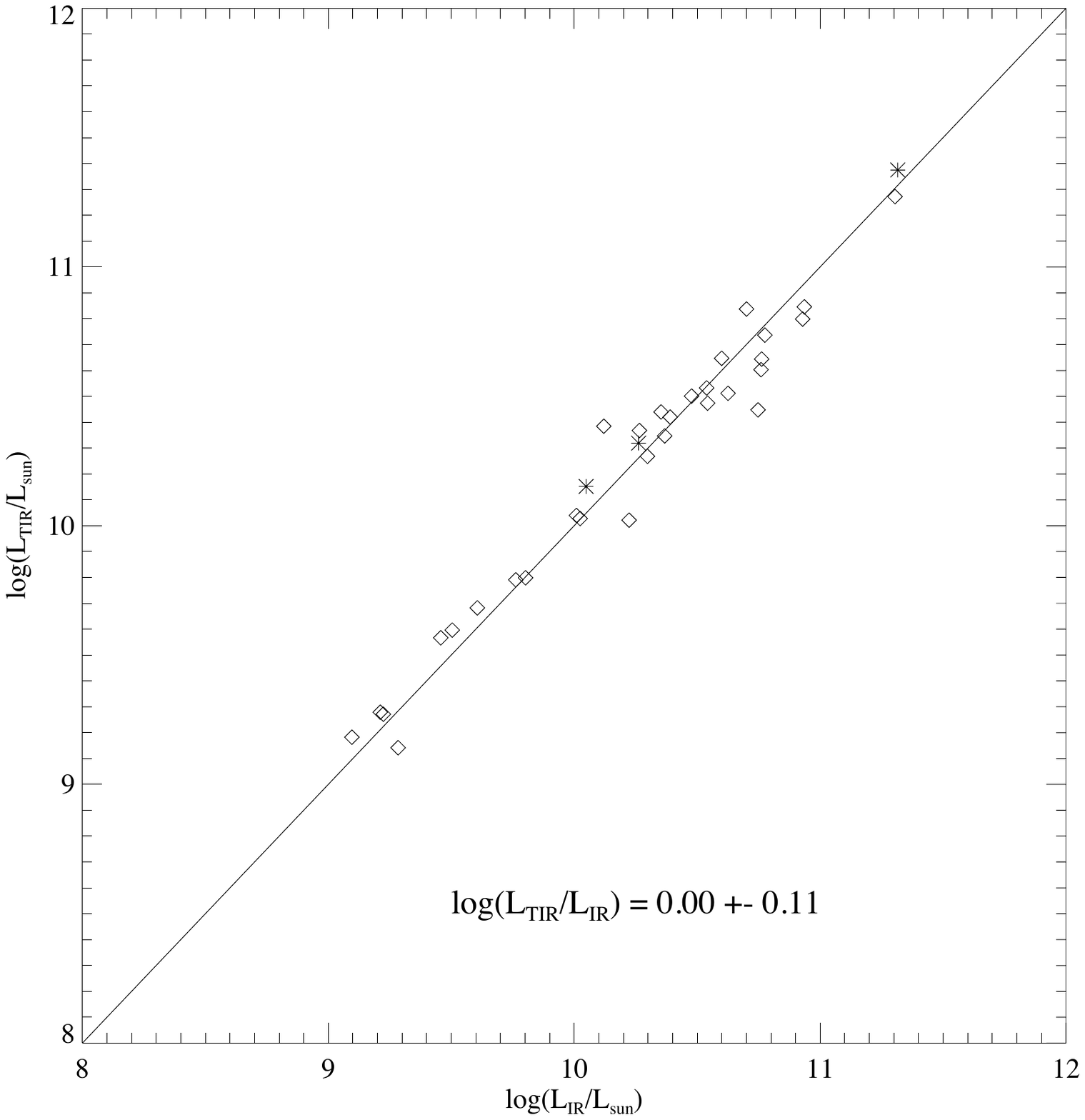}
\caption{{\it Left}:
$\log (L_{24}/L_{TIR})$ vs. $\log (L_8/L_{24})$ plot of spiral
 galaxies (diamonds: without AGNs; eight-point stars: with AGNs) in the KPAIR
sample, detected in all Spitzer bands. 
The solid line is a linear regression of
the data. The dotted line is the relation found by Calzetti et al. (2005)
for star formation regions in M51. The dashed line, also 
taken from Calzetti et al. (2005), is the prediction of the model 
of Dale \& Helou (2002) for galaxies.
{\it Right}: 
$\log (L_{TIR})$ vs. $\log (L_{IR})$ plot of same galaxies, where
$\log (L_{IR}) = \log (L_{24}) + 0.87 + 0.56 \times \log (L_{8}/L_{24})$.
}
\end{figure}

\section{IR Emission of Paired Galaxies --- Images and Catalogs}
Spitzer images at 3.6, 8.0, 24 and 70 $\mu m$ are 
presented in Fig.1 for all pairs in our sample except 
for KPAIR J1315+6207 (UGC08335a/b). 
These are gray scale images overlaid by contours.
The contour levels are $S_{bg}\; +\; S_0 \times 2^n$ (n=0, 1, 2, ...),
where $S_{bg}$ is the local background level and
$S_0 =$ 0.032, 0.16, 0.28 and 2.0 MJy sr$^{-1}$ for the 3.6,
8.0, 24, and 70$\mu m$ band, respectively. The $S_0$ values
are the sample medians of the 4-$\sigma$ threshold of the corresponding
bands. The sizes of all images are $2'\times 2'$ except for 
the three large pairs: J0949+0037, J1453+0317 and J2047+0019, for which a
scale bar of length of 2$'$ is given. We chose not to present the images
in the other 3 Spitzer bands because:
(1) the images of the 4.5$\mu m$ emission
have nearly identical morphology as their 3.6$\mu m$ counterparts;
(2) the 5.8$\mu m$ array is the most noisy among the IRAC detector arrays;
(3) the angular resolution of the 160$\mu m$ band, FWHM=40$''$, 
is so coarse that little information can be gained from the
images in addtion to what is already given by the $\rm f_{160}$ of component
galaxies listed in Table 3.

The flux densities of KPAIR galaxies, in all 7
Spitzer bands, are listed in Table 3. For the 4 IRAC bands, aperture
photometry corresponding to physical diameters of 4 and 10 kpc are
also provided. The methods and error estimates are discussed
previously in Section 3. For non-detections, 4$\sigma$ 
upper-limits are given.  All galaxies in our sample are
detected in IRAC bands. The detection rates are 47/52, 40/52, and
36/52 for the MIPS 24$\mu m$, 70$\mu m$ and 160$\mu m$ bands,
respectively.

Listed are also two estimates of the total IR (3 --- 1000 $\mu m$)
luminosities.  The $\rm L_{TIR}$ is derived using the formula of Dale
et al. (2005):
\begin{equation}
L_{TIR} = 1.559\times L_{24} + 0.7686\times L_{70} + 1.374\times L_{160} 
\end{equation}
where $L_{24}$, $L_{70}$, and $L_{160}$ are the monochromatic luminosities
($\nu L_{\nu}$). Another IR luminosity is defined using
the 24 and 8$\mu m$ flux densities:
\begin{equation}
\log (L_{IR}) = \log (L_{24}) + 0.87 (\pm 0.03) + 0.56 (\pm 0.09) 
\times \log (L_{8}/L_{24})
\end{equation}
where $L_{24}$ and $L_{8}$ are again the monochromatic luminosities
($\nu L_{\nu}$). $L_{IR}$ is an unbiased approximation of
$L_{TIR}$. It was derived from the linear regression of the
$\log (L_{24}/L_{TIR})$ vs. $\log (L_8/L_{24})$ correlation using
data of 34 spiral galaxies in the KPAIR sample that are
detected in all Spitzer bands (left panel of Fig.2), exploiting 
a similar method that was originally developed by Calzetti 
et al. (2005). The result is nearly identical for a linear regression with 
the galaxies containing AGNs (3 of them) excluded.
A comparison between  $\rm log(L_{TIR})$ and $\rm log(L_{IR})$
is plotted in the right panel of Fig.2. Indeed
there is a very tight linear correlation between
them, with a dispersion of only 0.11.
Given the better detection rate of  $\rm L_{IR}$ than that of $\rm L_{TIR}$
and the tight linear correlation between the two, we will use 
 $\rm L_{IR}$ hereafter whenever total IR luminosities are invoked in
calculations.
\begin{deluxetable}{lccccccccccc}
\label{tbl:observation3}
\tabletypesize{\scriptsize}
\setlength{\tabcolsep}{0.05in} 
\rotate
\tablenum{3}
\tablewidth{0pt}
\tablecaption{IR Emission of Paired Galaxies}
\tablehead{
\ch{(1)}   &\ch{(2)}    &\ch{(3)}    &\ch{(4)}&\ch{(5)}&\ch{(6)} &\ch{(7)}
             &\ch{(8)}         &\ch{(9)}           &\ch{(10)} &\ch{(11)}
&\ch{(12)}
\\
\ch{Galaxy ID} &     \ch{Aperture} &
\ch{f$_{3.6\mu m}$} & \ch{f$_{4.5\mu m}$}& 
\ch{f$_{5.8\mu m}$}& \ch{f$_{8\mu m}$}& \ch{f$_{24\mu m}$}
        & \ch{f$_{70\mu m}$}& \ch{f$_{160\mu m}$} & photometry
& \ch{L$_{TIR}$} & \ch{L$_{IR}$} \\
\ch{} &  \ch{(kpc)}   &
 \ch{(mJy)}     & \ch{(mJy)}     & \ch{(mJy)} 
        & \ch{(mJy)}     & \ch{(mJy)}     
        & \ch{(mJy)}     & \ch{(mJy)}   &  & \ch{(L$_\sun$)} & \ch{(L$_\sun$)} 
\\
}
\startdata
J00202580+0049350 &      4 &   6.20$\pm  0.62$ &  4.24$\pm  0.42$ & 10.05$\pm  1.01$ & 28.20$\pm  2.82$ &\\
                  &     10 &  10.66$\pm  1.07$ &  7.08$\pm  0.71$ & 11.85$\pm  1.19$ & 29.82$\pm  2.99$ &\\
                  &  total &  14.34$\pm  1.44$ &  9.26$\pm  0.93$ & 13.10$\pm  1.35$ & 30.84$\pm  3.11$ &   55.53$\pm 2.87$ &  677.27$\pm 73.41$ &1375.07$\pm 152.44$ & AAA &  10.03 &  10.02 \\
J00202748+0050009 &      4 &  10.56$\pm  1.06$ &  6.53$\pm  0.65$ &  4.48$\pm  0.45$ &  2.90$\pm  0.29$ &\\
                  &     10 &  16.25$\pm  1.62$ &  9.93$\pm  0.99$ &  6.65$\pm  0.68$ &  4.46$\pm  0.48$ &\\
                  &  total &  17.39$\pm  1.74$ & 10.61$\pm  1.06$ &  6.97$\pm  0.74$ &  4.78$\pm  0.56$ &    2.68$\pm 0.15$ &    $<$33.87        &          $<$101.60 & AAA &$<$8.96 &   9.01 \\
& & & & & & & & & & & \\
J01093371+0020322 &      4 &   1.93$\pm  0.19$ &  1.24$\pm  0.12$ &  0.74$\pm  0.07$ &  0.45$\pm  0.05$ &\\
                  &     10 &   3.34$\pm  0.33$ &  2.15$\pm  0.21$ &  1.32$\pm  0.13$ &  0.88$\pm  0.09$ &\\
                  &  total &   3.97$\pm  0.40$ &  2.60$\pm  0.26$ &  1.46$\pm  0.23$ &  0.98$\pm  0.21$ &     $<$0.47       &           $<$10.57 &          $<$64.39  & AAA &$<$9.58 &$<$9.03 \\
J01093517+0020132 &      4 &   1.27$\pm  0.13$ &  0.83$\pm  0.08$ &  0.52$\pm  0.05$ &  0.30$\pm  0.03$ &\\
                  &     10 &   2.65$\pm  0.26$ &  1.78$\pm  0.18$ &  1.17$\pm  0.12$ &  0.70$\pm  0.07$ &\\
                  &  total &   3.52$\pm  0.35$ &  2.36$\pm  0.24$ &  1.48$\pm  0.21$ &  0.86$\pm  0.17$ &     $<$1.03       &           $<$11.40 &          $<$91.13  & AAA & $<$9.58 &$<$8.99 \\
& & & & & & & & & & & \\
J01183417-0013416 &      4 &   4.58$\pm  0.46$ &  3.99$\pm  0.40$ &  8.44$\pm  0.84$ & 33.48$\pm  3.35$ &\\
                  &     10 &   6.69$\pm  0.67$ &  5.87$\pm  0.59$ & 13.59$\pm  1.36$ & 56.31$\pm  5.63$ &\\
                  &  total &   7.46$\pm  0.75$ &  6.41$\pm  0.64$ & 14.17$\pm  1.48$ & 57.06$\pm  5.71$ & 249.73$\pm 34.96$ &2817.82$\pm 305.30$ &2749.61$\pm 293.58$ & PPP &  11.41 &  11.36 \\
J01183556-0013594 &      4 &   0.51$\pm  0.05$ &  0.31$\pm  0.03$ &  0.43$\pm  0.05$ &  1.92$\pm  0.19$ &\\
                  &     10 &   1.86$\pm  0.19$ &  1.25$\pm  0.13$ &  2.18$\pm  0.23$ &  9.21$\pm  0.92$ &\\
                  &  total &   4.05$\pm  0.41$ &  2.83$\pm  0.29$ &  4.97$\pm  0.87$ & 20.20$\pm  2.03$ &   24.99$\pm 3.50$ & 305.32$\pm  33.19$ &  549.92$\pm 73.94$ & PPP & 10.55  &  10.67 \\
& & & & & & & & & & & \\
J02110638-0039191 &      4 &   6.74$\pm  0.67$ &  4.95$\pm  0.50$ & 11.35$\pm  1.14$ & 30.82$\pm  3.08$ &\\
                  &     10 &  10.22$\pm  1.02$ &  7.32$\pm  0.73$ & 15.72$\pm  1.61$ & 42.62$\pm  4.26$ &\\
                  &  total &  11.62$\pm  1.16$ &  7.75$\pm  0.78$ & 16.17$\pm  1.74$ & 41.71$\pm  4.17$ &   75.18$\pm 3.89$ &1034.44$\pm 112.08$ &2184.97$\pm 237.30$ & AAA & 10.33  &  10.28 \\
J02110832-0039171 &      4 &   8.28$\pm  0.83$ &  5.42$\pm  0.54$ &  3.82$\pm  0.39$ &  2.83$\pm  0.28$ &\\
                  &     10 &  12.81$\pm  1.28$ &  8.27$\pm  0.83$ &  5.90$\pm  0.69$ &  4.73$\pm  0.48$ &\\
                  &  total &  12.87$\pm  1.29$ &  8.29$\pm  0.83$ &  5.72$\pm  0.71$ &  4.32$\pm  0.45$ &    3.25$\pm 0.17$ &          $<$13.77  &          $<$102.84 & AAA &$<$9.36 &   9.13 \\
& & & & & & & & & & & \\
J09060283+5144411 &      4 &   4.25$\pm  0.42$ &  2.78$\pm  0.28$ &  1.69$\pm  0.17$ &  1.07$\pm  0.11$ &\\
                  &     10 &   6.42$\pm  0.64$ &  4.31$\pm  0.43$ &  2.68$\pm  0.31$ &  1.78$\pm  0.18$ &\\
                  &  total &   7.97$\pm  0.80$ &  5.63$\pm  0.57$ &  3.61$\pm  1.18$ &  2.21$\pm  0.36$ &    0.69$\pm 0.10$ &           $<$6.89  &           $<$71.07 & WAA &$<$9.07 &   9.03 \\
J09060498+5144071 &      4 &   1.80$\pm  0.18$ &  1.21$\pm  0.12$ &  1.05$\pm  0.11$ &  1.60$\pm  0.16$ &\\
                  &     10 &   4.26$\pm  0.43$ &  2.88$\pm  0.29$ &  3.07$\pm  0.35$ &  5.91$\pm  0.59$ &\\
                  &  total &   7.04$\pm  0.71$ &  4.80$\pm  0.49$ &  5.83$\pm  0.94$ & 13.10$\pm  1.32$ &   18.06$\pm 0.93$ & 191.21$\pm  20.76$ &  885.31$\pm 93.44$ & AAA & 10.17  &  10.08 \\
& & & & & & & & & & & \\
J09374413+0245394 &      4 &   8.53$\pm  0.85$ &  5.70$\pm  0.57$ &  7.41$\pm  0.74$ & 20.82$\pm  2.08$ &\\
                  &     10 &  21.66$\pm  2.17$ & 15.26$\pm  1.53$ & 26.77$\pm  2.71$ & 75.67$\pm  7.57$ &\\
                  &  total &  21.66$\pm  2.17$ & 15.26$\pm  1.53$ & 26.77$\pm  2.71$ & 75.67$\pm  7.57$ &  183.80$\pm 9.50$ &2133.90$\pm 231.20$ &5989.62$\pm 629.23$ & AAA & 10.85  &  10.73 \\
J09374506+0244504 &      4 &  11.42$\pm  1.14$ &  7.03$\pm  0.70$ &  4.60$\pm  0.46$ &  2.83$\pm  0.28$ &\\
                  &     10 &  17.38$\pm  1.74$ & 10.74$\pm  1.07$ &  7.03$\pm  0.82$ &  4.57$\pm  0.48$ &\\
                  &  total &  21.03$\pm  2.10$ & 12.90$\pm  1.30$ &  8.79$\pm  1.72$ &  6.81$\pm  0.84$ &    2.74$\pm 0.15$ &           $<$16.45 &          $<$207.44 & AAA &$<$9.31 &   9.34 \\  
& & & & & & & & & & & \\
J09494143+0037163 &      4 &   9.95$\pm  1.00$ &  6.73$\pm  0.67$ & 12.11$\pm  1.25$ & 24.63$\pm  2.49$ &\\
                  &     10 &  17.61$\pm  1.79$ & 10.99$\pm  1.10$ & 21.11$\pm  2.82$ & 35.16$\pm  4.13$ &\\
                  &  total &  13.29$\pm  1.34$ &  8.74$\pm  0.88$ & 17.63$\pm  1.98$ & 31.51$\pm  3.33$ &   87.09$\pm 4.50$ &1254.62$\pm 135.96$ &1364.04$\pm 154.12$ & AAA &  9.27  &   9.23 \\
J09495263+0037043 &      4 &  10.84$\pm  1.09$ &  7.67$\pm  0.77$ & 14.83$\pm  1.51$ & 33.65$\pm  3.38$ &\\                                                                               
                  &     10 &  23.41$\pm  2.36$ & 17.94$\pm  1.80$ & 31.27$\pm  3.64$ & 71.17$\pm  7.44$ &\\                                                                               
                  &  total &  27.34$\pm  2.99$ & 20.07$\pm  2.07$ & 35.00$\pm  7.65$ & 81.32$\pm 11.33$ &  192.14$\pm 9.94$ &3028.93$\pm 328.16$ &4019.44$\pm 421.57$ & AAA &  9.68  &   9.62 \\
& & & & & & & & & & & \\
J10205188+4831096 &      4 &   0.68$\pm  0.07$ &  0.52$\pm  0.05$ &  0.61$\pm  0.06$ &  2.82$\pm  0.28$ &\\
                  &     10 &   1.50$\pm  0.15$ &  1.12$\pm  0.11$ &  1.46$\pm  0.15$ &  5.80$\pm  0.58$ &\\
                  &  total &   2.15$\pm  0.22$ &  1.58$\pm  0.16$ &  1.73$\pm  0.27$ &  7.20$\pm  0.73$ &   14.76$\pm 2.07$ &  182.84$\pm 19.85$ &  276.01$\pm 34.89$ & PAA &  10.39 &  10.42 \\
J10205369+4831246 &      4 &   1.40$\pm  0.14$ &  1.00$\pm  0.10$ &  0.63$\pm  0.06$ &  0.45$\pm  0.04$ &\\
                  &     10 &   2.85$\pm  0.29$ &  2.03$\pm  0.20$ &  1.36$\pm  0.14$ &  1.14$\pm  0.12$ &\\
                  &  total &   4.25$\pm  0.43$ &  2.91$\pm  0.29$ &  1.92$\pm  0.27$ &  2.04$\pm  0.23$ &    1.90$\pm 0.27$ &           $<$7.56 &          $<$72.13   & PAA &$<$9.78 &   9.72 \\
& & & & & & & & & & & \\
J10272950+0114490 &      4 &   5.81$\pm  0.58$ &  4.38$\pm  0.44$ & 12.01$\pm  1.20$ & 36.28$\pm  3.63$ &\\
                  &     10 &   8.27$\pm  0.83$ &  5.91$\pm  0.59$ & 14.29$\pm  1.45$ & 42.18$\pm  4.22$ &\\
                  &  total &   8.27$\pm  0.83$ &  5.91$\pm  0.59$ & 14.29$\pm  1.45$ & 42.18$\pm  4.22$ &   53.17$\pm 2.75$ &  692.21$\pm 75.02$ &1598.47$\pm 173.13$ & AAA &  10.28 &  10.32 \\
J10272970+0115170 &      4 &   6.80$\pm  0.68$ &  4.49$\pm  0.45$ &  2.90$\pm  0.29$ &  1.96$\pm  0.20$ &\\
                  &     10 &  11.84$\pm  1.18$ &  7.81$\pm  0.78$ &  5.28$\pm  0.59$ &  3.93$\pm  0.40$ &\\
                  &  total &  16.04$\pm  1.61$ & 10.77$\pm  1.08$ &  8.09$\pm  1.36$ &  6.54$\pm  0.69$ &    1.75$\pm 0.25$ &           $<$22.95 &          $<$62.12  & WAA &$<$9.10 &   9.22 \\
& & & & & & & & & & & \\
J10435053+0645466 &      4 &   3.48$\pm  0.35$ &  2.72$\pm  0.27$ &  7.53$\pm  0.75$ & 23.48$\pm  2.35$ &\\
                  &     10 &   7.29$\pm  0.73$ &  5.50$\pm  0.55$ & 15.40$\pm  1.56$ & 50.30$\pm  5.03$ &\\
                  &  total &   8.75$\pm  0.88$ &  6.55$\pm  0.66$ & 17.58$\pm  1.91$ & 57.33$\pm  5.74$ &  164.80$\pm 8.52$ &1005.01$\pm 108.90$ &1661.42$\pm 175.97$ & APP &  10.62 &  10.79 \\
J10435268+0645256 &      4 &   2.01$\pm  0.20$ &  1.39$\pm  0.14$ &  1.62$\pm  0.17$ &  3.59$\pm  0.36$ &\\
                  &     10 &   3.86$\pm  0.39$ &  2.67$\pm  0.27$ &  3.33$\pm  0.42$ &  7.83$\pm  0.78$ &\\
                  &  total &   5.55$\pm  0.56$ &  3.83$\pm  0.39$ &  4.30$\pm  1.20$ &  9.93$\pm  1.02$ &   10.10$\pm 0.52$ &   94.53$\pm 10.40$ &  460.59$\pm 67.12$ & APP &  9.81 &   9.83 \\
& & & & & & & & & & & \\
J10514368+5101195 &      4 &  10.13$\pm  1.01$ &  6.38$\pm  0.64$ &  4.30$\pm  0.43$ &  3.25$\pm  0.33$ &\\
                  &     10 &  18.89$\pm  1.89$ & 12.01$\pm  1.20$ &  7.87$\pm  0.83$ &  5.94$\pm  0.60$ &\\
                  &  total &  18.89$\pm  1.89$ & 12.01$\pm  1.20$ &  7.87$\pm  0.83$ &  5.94$\pm  0.60$ &    3.28$\pm 0.14$ &    55.62$\pm 6.17$ &  307.21$\pm 56.37$ & AWW &  9.51 &   9.40 \\
J10514450+5101303 &      4 &   4.15$\pm  0.41$ &  2.66$\pm  0.27$ &  1.75$\pm  0.18$ &  1.14$\pm  0.11$ &\\
                  &     10 &  10.84$\pm  1.08$ &  7.02$\pm  0.70$ &  4.90$\pm  0.56$ &  3.41$\pm  0.35$ &\\
                  &  total &  10.84$\pm  1.08$ &  7.02$\pm  0.70$ &  4.90$\pm  0.56$ &  3.41$\pm  0.35$ &           $<$2.49 &            $<$17.44 &          $<$105.45& AAA &$<$9.39&$<$9.05 \\
& & & & & & & & & & & \\
J12020424+5342317 &      4 &   0.64$\pm  0.06$ &  0.45$\pm  0.04$ &  0.26$\pm  0.03$ &  0.42$\pm  0.04$ &\\
                  &     10 &   1.70$\pm  0.17$ &  1.21$\pm  0.12$ &  0.88$\pm  0.09$ &  1.91$\pm  0.19$ &\\
                  &  total &   2.51$\pm  0.25$ &  1.78$\pm  0.18$ &  1.30$\pm  0.19$ &  3.12$\pm  0.31$ &    4.96$\pm 0.29$ &   48.56$\pm  5.45$ &  364.87$\pm 52.62$ & AAA & 10.44 &  10.18 \\
J12020537+5342487 &      4 &   0.87$\pm  0.09$ &  0.69$\pm  0.07$ &  0.31$\pm  0.03$ &  0.26$\pm  0.03$ &\\
                  &     10 &   2.43$\pm  0.24$ &  1.70$\pm  0.17$ &  1.00$\pm  0.10$ &  0.66$\pm  0.07$ &\\
                  &  total &   3.91$\pm  0.39$ &  2.84$\pm  0.28$ &  1.73$\pm  0.27$ &  1.27$\pm  0.13$ &           $<$0.27 &           $<$6.88 &          $<$32.91   & AAA&$<$10.11&$<$9.72 \\
& & & & & & & & & & & \\
J13082737+0422125 &      4 &   1.16$\pm  0.12$ &  0.78$\pm  0.08$ &  1.16$\pm  0.12$ &  2.87$\pm  0.29$ &\\
                  &     10 &   2.07$\pm  0.21$ &  1.36$\pm  0.14$ &  2.07$\pm  0.27$ &  5.31$\pm  0.53$ &\\
                  &  total &   2.35$\pm  0.24$ &  1.48$\pm  0.16$ &  1.99$\pm  0.67$ &  5.71$\pm  0.61$ &    5.91$\pm 0.31$ &    87.80$\pm 9.67$ &  336.59$\pm 49.16$ & AAP &  9.58 &   9.48 \\
J13082964+0422045 &      4 &   2.49$\pm  0.25$ &  1.59$\pm  0.16$ &  1.60$\pm  0.16$ &  2.85$\pm  0.28$ &\\
                  &     10 &   4.04$\pm  0.40$ &  2.58$\pm  0.26$ &  2.92$\pm  0.34$ &  5.96$\pm  0.60$ &\\
                  &  total &   4.68$\pm  0.47$ &  2.98$\pm  0.30$ &  3.38$\pm  0.64$ &  7.43$\pm  0.77$ &    5.38$\pm 0.28$ &    87.55$\pm 9.64$ & 376.16$\pm  63.60$ & AAP &  9.61 &   9.53 \\
& & & & & & & & & & & \\
J13325525-0301347 &      4 &   1.38$\pm  0.14$ &  1.04$\pm  0.10$ &  1.57$\pm  0.16$ &  5.60$\pm  0.56$ &\\
                  &     10 &   2.60$\pm  0.26$ &  1.91$\pm  0.19$ &  2.89$\pm  0.29$ & 10.49$\pm  1.05$ &\\
                  &  total &   3.25$\pm  0.33$ &  2.34$\pm  0.23$ &  3.30$\pm  0.35$ & 11.50$\pm  1.15$ &   46.50$\pm 6.51$ &  497.48$\pm 53.92$ &  575.53$\pm 67.59$ & PPP & 10.68 &  10.64 \\
J13325655-0301395 &      4 &   0.88$\pm  0.09$ &  0.60$\pm  0.06$ &  0.46$\pm  0.05$ &  1.23$\pm  0.12$ &\\
                  &     10 &   2.82$\pm  0.28$ &  1.89$\pm  0.19$ &  2.23$\pm  0.22$ &  7.44$\pm  0.74$ &\\
                  &  total &   5.29$\pm  0.53$ &  3.62$\pm  0.36$ &  4.50$\pm  0.48$ & 15.98$\pm  1.60$ &   15.94$\pm 2.23$ &  176.01$\pm 19.19$ & 770.98$\pm  85.21$ & PPP & 10.53 &  10.52 \\
& & & & & & & & & & & \\
J13462001-0325407 &      4 &   5.26$\pm  0.53$ &  3.65$\pm  0.36$ &  2.93$\pm  0.30$ &  3.11$\pm  0.31$ &\\
                  &     10 &   9.04$\pm  0.90$ &  6.08$\pm  0.61$ &  5.38$\pm  0.59$ &  6.97$\pm  0.70$ &\\
                  &  total &  11.83$\pm  1.19$ &  7.84$\pm  0.79$ &  7.64$\pm  1.59$ & 11.03$\pm  1.12$ &   13.99$\pm 0.72$ &   51.49$\pm  5.85$ &  692.82$\pm 74.59$ & AAA &  9.80 &   9.79 \\
J13462215-0325057 &      4 &   5.42$\pm  0.54$ &  3.60$\pm  0.36$ &  2.59$\pm  0.26$ &  2.25$\pm  0.22$ &\\
                  &     10 &   7.36$\pm  0.74$ &  4.86$\pm  0.49$ &  3.31$\pm  0.40$ &  2.75$\pm  0.28$ &\\
                  &  total &   9.06$\pm  0.91$ &  5.96$\pm  0.60$ &  3.23$\pm  1.09$ &  3.28$\pm  0.36$ &    6.15$\pm 0.32$ &    41.13$\pm 4.83$ &           $<$77.92 & AAA &$<$9.13&   9.34 \\
& & & & & & & & & & & \\
J14005782+4251207 &      4 &   3.80$\pm  0.38$ &  3.41$\pm  0.34$ &  8.21$\pm  0.82$ & 26.57$\pm  2.66$ &\\
                  &     10 &   6.88$\pm  0.69$ &  5.37$\pm  0.54$ & 11.80$\pm  1.19$ & 40.61$\pm  4.06$ &\\
                  &  total &  10.07$\pm  1.02$ &  7.17$\pm  0.72$ & 14.42$\pm  2.01$ & 49.30$\pm  4.93$ &   94.87$\pm 13.28$ &1218.70$\pm 132.05$ &1697.41$\pm 184.78$& PPP & 10.76 &  10.81 \\
J14005882+4250427 &      4 &   5.36$\pm  0.54$ &  4.22$\pm  0.42$ & 12.09$\pm  1.21$ & 39.95$\pm  3.99$ &\\
                  &     10 &   7.41$\pm  0.74$ &  5.80$\pm  0.58$ & 15.95$\pm  1.60$ & 54.19$\pm  5.42$ &\\
                  &  total &   8.86$\pm  0.89$ &  6.35$\pm  0.64$ & 16.23$\pm  1.91$ & 55.36$\pm  5.54$ &  190.59$\pm 26.68$ &1470.53$\pm 159.33$ &1786.75$\pm 204.99$& PPP & 10.87 &  10.97 \\
& & & & & & & & & & & \\
J14250552+0313590 &      4 &   3.63$\pm  0.36$ &  2.69$\pm  0.27$ &  2.16$\pm  0.22$ &  2.38$\pm  0.24$ &\\
                  &     10 &   5.68$\pm  0.57$ &  3.99$\pm  0.40$ &  3.18$\pm  0.33$ &  3.48$\pm  0.35$ &\\
                  &  total &   6.72$\pm  0.67$ &  4.67$\pm  0.47$ &  3.55$\pm  0.43$ &  3.74$\pm  0.39$ &    9.66$\pm 0.50$ & 111.32$\pm  12.18$ & 526.61$\pm  60.98$ & AAA & 10.11 &   9.83 \\
J14250739+0313560 &      4 &   1.18$\pm  0.12$ &  0.77$\pm  0.08$ &  0.48$\pm  0.05$ &  0.31$\pm  0.03$ &\\
                  &     10 &   1.63$\pm  0.16$ &  1.05$\pm  0.10$ &  0.66$\pm  0.10$ &  0.52$\pm  0.06$ &\\
                  &  total &   1.63$\pm  0.16$ &  1.05$\pm  0.10$ &  0.66$\pm  0.10$ &  0.52$\pm  0.06$ &    1.39$\pm 0.19$ &           $<$9.15 &          $<$104.79  & WAA &$<$9.55&   8.98 \\
& & & & & & & & & & & \\
J14334683+4004512 &      4 &   6.67$\pm  0.67$ &  4.75$\pm  0.48$ &  8.79$\pm  0.88$ & 23.43$\pm  2.34$ &\\
                  &     10 &  11.98$\pm  1.20$ &  8.12$\pm  0.81$ & 12.27$\pm  1.24$ & 30.52$\pm  3.05$ &\\
                  &  total &  19.27$\pm  1.93$ & 12.93$\pm  1.30$ & 19.74$\pm  2.32$ & 47.21$\pm  4.73$ &   86.65$\pm 4.48$ &  840.66$\pm 91.10$ &1747.03$\pm 185.16$ & PPP & 10.49 &  10.57 \\
J14334840+4005392 &      4 &  10.27$\pm  1.03$ &  7.11$\pm  0.71$ & 19.60$\pm  1.96$ & 60.41$\pm  6.04$ &\\
                  &     10 &  14.36$\pm  1.44$ &  9.95$\pm  0.99$ & 27.36$\pm  2.74$ & 85.42$\pm  8.54$ &\\
                  &  total &  16.50$\pm  1.65$ & 11.43$\pm  1.15$ & 29.35$\pm  3.06$ & 91.71$\pm  9.17$ &  117.17$\pm 6.06$ &1337.09$\pm 144.88$ &2553.35$\pm 269.55$ & PPP & 10.66 &  10.79 \\
& & & & & & & & & & & \\
J14530282+0317451 &      4 &  17.75$\pm  1.78$ & 13.12$\pm  1.31$ & 18.41$\pm  1.94$ & 42.39$\pm  4.24$ &\\
                  &     10 &  26.86$\pm  2.69$ & 26.41$\pm  2.64$ & 22.96$\pm  4.43$ & 57.36$\pm  5.74$ &\\
                  &  total &  26.86$\pm  2.69$ & 26.41$\pm  2.64$ & 22.96$\pm  4.43$ & 57.36$\pm  5.74$ &   47.05$\pm 2.43$ &  680.55$\pm 73.76$ &3028.63$\pm 320.69$ & AAA &   9.18 &   9.11 \\
J14530523+0319541 &      4 &  25.20$\pm  2.52$ & 16.15$\pm  1.62$ & 13.88$\pm  1.52$ & 23.56$\pm  2.38$ &\\
                  &     10 &  45.99$\pm  4.61$ & 31.66$\pm  3.19$ & 31.69$\pm  4.94$ & 90.85$\pm  9.08$ &\\
                  &  total &  49.18$\pm  4.94$ & 35.19$\pm  3.59$ & 33.88$\pm  7.17$ &117.68$\pm 11.77$ &   50.07$\pm 2.59$ &  830.04$\pm 89.96$ &2432.15$\pm 259.60$ & AAA &   9.14 &   9.29 \\
& & & & & & & & & & & \\
J15064391+0346364 &      4 &   4.37$\pm  0.44$ &  2.73$\pm  0.27$ &  1.84$\pm  0.18$ &  1.30$\pm  0.13$ &\\
                  &     10 &   7.20$\pm  0.72$ &  4.59$\pm  0.46$ &  3.18$\pm  0.33$ &  2.37$\pm  0.24$ &\\
                  &  total &   9.04$\pm  0.90$ &  5.67$\pm  0.57$ &  3.81$\pm  0.52$ &  2.92$\pm  0.32$ &    2.22$\pm 0.31$ &   36.79$\pm  4.35$ &          $<$141.61 & WAA &$<$9.82 &   9.45 \\
J15064579+0346214 &      4 &   4.45$\pm  0.45$ &  4.34$\pm  0.43$ &  6.10$\pm  0.61$ & 11.95$\pm  1.19$ &\\
                  &     10 &   7.22$\pm  0.72$ &  6.25$\pm  0.62$ &  8.89$\pm  0.89$ & 20.05$\pm  2.00$ &\\
                  &  total &   9.61$\pm  0.96$ &  7.72$\pm  0.77$ & 10.95$\pm  1.19$ & 26.50$\pm  2.66$ &   47.67$\pm 2.47$ &  340.83$\pm 36.98$ &1440.24$\pm 156.70$ & AAA &  10.55 &  10.57 \\
& & & & & & & & & & & \\
J15101587+5810425 &      4 &   3.47$\pm  0.35$ &  2.22$\pm  0.22$ &  1.51$\pm  0.15$ &  1.22$\pm  0.12$ &\\
                  &     10 &   5.54$\pm  0.55$ &  3.61$\pm  0.36$ &  2.69$\pm  0.29$ &  2.83$\pm  0.28$ &\\
                  &  total &   6.28$\pm  0.63$ &  4.07$\pm  0.41$ &  3.16$\pm  0.35$ &  4.13$\pm  0.41$ &    2.86$\pm 0.40$ &    31.00$\pm 3.62$ &          $<$62.28  & PPA &$<$9.37 &   9.50 \\
J15101776+5810375 &      4 &   2.46$\pm  0.25$ &  1.78$\pm  0.18$ &  2.87$\pm  0.29$ &  8.28$\pm  0.83$ &\\
                  &     10 &   4.83$\pm  0.48$ &  3.41$\pm  0.34$ &  6.28$\pm  0.64$ & 18.22$\pm  1.82$ &\\
                  &  total &   5.01$\pm  0.50$ &  3.50$\pm  0.35$ &  6.45$\pm  0.65$ & 18.62$\pm  1.86$ &   27.91$\pm 3.91$ &  384.68$\pm 41.70$ &1160.61$\pm 124.37$ & PPA &  10.39 &  10.30 \\
& & & & & & & & & & & \\
J15281276+4255474 &      4 &  15.92$\pm  1.59$ & 10.57$\pm  1.06$ & 14.35$\pm  1.44$ & 33.82$\pm  3.38$ &\\
                  &     10 &  31.74$\pm  3.17$ & 20.64$\pm  2.06$ & 32.36$\pm  3.26$ & 74.22$\pm  7.42$ &\\
                  &  total &  33.68$\pm  3.37$ & 24.95$\pm  2.50$ & 37.27$\pm  4.63$ & 81.05$\pm  8.16$ &   93.80$\pm 4.85$ &1524.48$\pm 165.19$ &3472.94$\pm 365.14$ & AAA & 10.43 &  10.41 \\
J15281667+4256384 &      4 &  14.02$\pm  1.40$ &  8.93$\pm  0.89$ &  6.26$\pm  0.63$ &  5.04$\pm  0.50$ &\\
                  &     10 &  18.60$\pm  1.86$ & 11.71$\pm  1.17$ &  7.95$\pm  0.89$ &  6.17$\pm  0.63$ &\\
                  &  total &  24.34$\pm  2.44$ & 13.20$\pm  1.33$ &  7.98$\pm  1.89$ &  7.21$\pm  0.92$ &    4.25$\pm 0.22$ &  123.94$\pm 13.59$ &  252.58$\pm 32.84$ & AAA &  9.29 &   9.23 \\
& & & & & & & & & & & \\
J15562191+4757172 &      4 &   2.94$\pm  0.29$ &  2.11$\pm  0.21$ &  5.31$\pm  0.53$ & 15.16$\pm  1.52$ &\\
                  &     10 &   5.61$\pm  0.56$ &  3.84$\pm  0.39$ &  7.95$\pm  0.85$ & 21.14$\pm  2.11$ &\\
                  &  total &   6.48$\pm  0.65$ &  4.40$\pm  0.45$ &  8.85$\pm  1.22$ & 22.15$\pm  2.22$ &   48.79$\pm 2.44$ & 538.73$\pm  58.40$ &1163.05$\pm 125.91$ & AAA & 10.06 &  10.03 \\
J15562738+4757302 &      4 &   3.63$\pm  0.36$ &  2.26$\pm  0.23$ &  1.45$\pm  0.15$ &  0.90$\pm  0.09$ &\\
                  &     10 &   4.93$\pm  0.49$ &  3.06$\pm  0.31$ &  1.79$\pm  0.36$ &  1.24$\pm  0.13$ &\\
                  &  total &   5.07$\pm  0.51$ &  3.12$\pm  0.32$ &  1.76$\pm  0.64$ &  1.33$\pm  0.16$ &           $<$0.35 &           $<$7.34 &          $<$30.31   & AAA &$<$9.05&$<$8.66 \\
& & & & & & & & & & & \\
J16024254+4111499 &      4 &   2.89$\pm  0.29$ &  2.15$\pm  0.22$ &  6.78$\pm  0.68$ & 22.67$\pm  2.27$ &\\
                  &     10 &   7.02$\pm  0.70$ &  5.04$\pm  0.50$ & 20.88$\pm  2.11$ & 53.41$\pm  5.34$ &\\
                  &  total &  11.04$\pm  1.11$ &  7.88$\pm  0.79$ & 22.89$\pm  2.43$ & 77.56$\pm  7.76$ &  109.52$\pm 15.33$ &1110.58$\pm 120.33$ &2171.84$\pm 237.03$& PPP & 10.82 &  10.96 \\
J16024475+4111589 &      4 &   1.72$\pm  0.17$ &  1.17$\pm  0.12$ &  3.29$\pm  0.33$ &  9.50$\pm  0.95$ &\\
                  &     10 &   3.64$\pm  0.36$ &  2.58$\pm  0.26$ &  6.74$\pm  0.69$ & 21.92$\pm  2.19$ &\\
                  &  total &   3.64$\pm  0.36$ &  2.58$\pm  0.26$ &  6.74$\pm  0.69$ & 21.92$\pm  2.19$ &   26.83$\pm 3.76$ &  460.22$\pm 49.90$ &1143.07$\pm 146.32$ & PPP & 10.46 &  10.39 \\
& & & & & & & & & & & \\
J17045089+3448530 &      4 &   1.11$\pm  0.11$ &  0.84$\pm  0.08$ &  1.06$\pm  0.11$ &  4.62$\pm  0.46$ &\\
                  &     10 &   2.47$\pm  0.25$ &  1.80$\pm  0.18$ &  2.63$\pm  0.27$ & 13.13$\pm  1.31$ &\\
                  &  total &   2.82$\pm  0.28$ &  2.07$\pm  0.21$ &  2.94$\pm  0.31$ & 15.88$\pm  1.59$ &   28.93$\pm 4.05$ &    82.28$\pm 9.07$ &  360.76$\pm 48.50$ & PPP & 10.49 &  10.80 \\
J17045097+3449020 &      4 &   2.56$\pm  0.26$ &  2.25$\pm  0.23$ &  4.56$\pm  0.46$ & 23.63$\pm  2.36$ &\\
                  &     10 &   5.07$\pm  0.51$ &  3.99$\pm  0.40$ &  8.83$\pm  0.88$ & 49.86$\pm  4.99$ &\\
                  &  total &   5.56$\pm  0.56$ &  4.25$\pm  0.43$ &  8.71$\pm  0.88$ & 47.99$\pm  4.80$ &  131.22$\pm 18.37$ &1518.03$\pm 164.47$ &1636.24$\pm 175.34$& PPP & 11.31 &  11.36 \\
& & & & & & & & & & & \\
J20471908+0019150 &      4 &  33.53$\pm  3.35$ & 18.87$\pm  1.89$ & 14.31$\pm  1.44$ & 10.86$\pm  1.09$ &\\
                  &     10 &  61.20$\pm  6.12$ & 34.60$\pm  3.46$ & 27.27$\pm  2.82$ & 25.13$\pm  2.55$ &\\
                  &  total & 132.43$\pm 13.43$ & 75.75$\pm  7.63$ & 66.41$\pm 16.73$ &101.26$\pm 13.41$ &  121.03$\pm 6.27$ &  462.86$\pm 50.18$ &3048.90$\pm 326.39$ & AAA & 10.02 &  10.24 \\
J20472428+0018030 &      4 &  21.05$\pm  2.10$ & 13.11$\pm  1.31$ &  8.92$\pm  0.90$ &  5.76$\pm  0.58$ &\\
                  &     10 &  36.97$\pm  3.70$ & 23.40$\pm  2.34$ & 15.72$\pm  1.74$ & 10.50$\pm  1.13$ &\\
                  &  total &  50.90$\pm  5.12$ & 31.05$\pm  3.11$ & 22.91$\pm  4.75$ & 16.23$\pm  2.88$ &   11.93$\pm 0.62$ &           $<$36.64 &          $<$117.51 & AAA &$<$9.10&   9.32 \\
& & & & & & & & & & & \\
J13153076+6207447&
                         4 &   6.00$\pm  0.60$ &  4.00$\pm  0.40$ & 15.00$\pm  1.50$ & 46.00$\pm  4.60$ &\\
                  &     10 &   8.00$\pm  0.80$ &  6.00$\pm  0.60$ & 18.00$\pm  1.80$ & 54.00$\pm  5.40$ &\\
                  &  total &   8.00$\pm  0.80$ &  6.00$\pm  0.60$ & 18.00$\pm  1.80$ & 54.00$\pm  5.40$ &                   &                    &                    &     & 11.01 &  11.01 \\
J13153506+6207287 &
                         4 &  16.00$\pm  1.60$ & 15.00$\pm  1.50$ & 59.00$\pm  5.90$ &179.00$\pm 17.90$ &\\
                  &     10 &  17.00$\pm  1.70$ & 16.00$\pm  1.60$ & 62.00$\pm  6.20$ &187.00$\pm 18.70$ &\\
                  &  total &  17.00$\pm  1.70$ & 16.00$\pm  1.60$ & 62.00$\pm  6.20$ &187.00$\pm 18.70$ &                   &                    &                    &     & 11.65 &  11.65 \\
\hline
\enddata
\tablecomments{
\nid{\bf Descriptions of Columns}:
\begin{description}
\item{(1)} Galaxy ID.
\item{(2)} Diameter of IRAC aperture.
\item{(3)} Flux density in the IRAC 3.6$\mu m$ band.
\item{(4)} Flux density in the IRAC 4.5$\mu m$ band.
\item{(5)} Flux density in the IRAC 5.8$\mu m$ band.
\item{(6)} Flux density in the IRAC 8.0$\mu m$ band.
\item{(7)} Flux density in the MIPS 24$\mu m$ band.
\item{(8)} Flux density in the MIPS 70$\mu m$ band.
\item{(9)} Flux density in the MIPS 160$\mu m$ band.
\item{(10)} Photometry methods for measurements of MIPS flux
  densities. ``A'' stands for aperture photometry, ``W'' for weak
  source, and ``P'' for PRF fitting. The three letters correspond to
  the 24, 70 and 160$\mu m$ bands, respectively. For example, ``AAP''
  means that flux densities in both the 24 and 70$\mu m$ bands were
  measured by aperture photometry while the 160$\mu m$ flux density
  was measured by PRF fitting.
\item{(11)} Logarithm of the $\rm L_{TIR}$, calculated from
$\rm L_{24\mu m}$, $\rm L_{70\mu m}$ and $\rm L_{160\mu m}$ using the formula of
Dale \& Helou (2002).
\item{(12)} Logarithm of the $\rm L_{IR}$, an unbiased estimator of the
 $\rm L_{TIR}$ calculated from $\rm L_{24\mu m}$ and $\rm L_{8\mu m}$.
\end{description}
}
\end{deluxetable}

The images and photometry show very diversified IR emission 
properties among KPAIR galaxies. We verify that, except for those with
AGN's, paired E galaxies have very low dust emission 
($\rm \log(L_{IR}/L_\sun) \sim 9$--10),
indicating little star formation occurring in them. Among the
paired spiral galaxies, the majority have rather moderate IR luminosity
($\sim 10^{10} L_\sun$). There are six known AGN's (Table 1), 
three in S galaxies and three in E galaxies. There are four LIRGs 
(one of them being an AGN), but no ULIRGs, in the sample.
Some detailed notes on individual pairs can be found in Appendix A.

\section{Star Formation Enhancement}
\subsection{Control Sample of Single Late-type Galaxies}
In this section, star formation rates of 39 non-AGN spiral galaxies in the
KPAIR sample are compared with those of single galaxies in a control sample. 
In order to have a clean comparison, we chose to select a
control sample that contains also 39 spiral galaxies, each matching a spiral
galaxy in the pair sample of the same mass. 

The selection of the control sample
was confined to two Spitzer data archives: (1) The SWIRE survey
of Lockman field and ELAIS-N1 field (covering $\sim 20$ deg$^2$, 
Lonsdale et al. 2004), which provides an IRAC 
3.6$\mu m$ band selected sample of field galaxies (Surace et al. 2005). 
The restriction to the Lockman and ELAIS-N1 fields 
is because these regions in the SWIRE survey have good SDSS
spectroscopic coverages (Abazajian et al. 2005);
. (2) The SINGS survey of nearby galaxies 
(Kennicutt et al. 2004), which observed a heterogeneously selected sample
of 75 well-known galaxies, including normal late and 
early types, AGNs, and starbursts.

The criteria for a galaxy to be considered in the selection of
the control sample are:
\begin{description}
\item{(1)} has published Spitzer IRAC and MIPS data; 
\item{(2)} has spectroscopic redshift, and $z \leq 0.1$; 
\item{(3)} $K_{s}$, taken from the ${\rm K_{20}}$ of 2MASS, $\leq 13.5$ mag ;
\item{(4)} is a late-type galaxy outside any interacting system;
\item{(5)} does not have a known AGN.
\end{description}

Paired spiral galaxies with $\log (M_{star}/M_\sun) \geq 10.3$ were matched with
single galaxies in the two SWIRE fields. 
Their Spitzer flux densities were taken from the SWIRE Data Release 2 
(Surace et al. 2005). For both IRAC 8$\mu m$ and MIPS 24$\mu m$ bands, the
Kron fluxes (and the associated errors) are adopted. For $\rm f_{24}$, an 
additional 15\% aperture correction is applied (Shupe et al. 2008).
SDSS images were inspected and galaxies showing
signs of interaction or being in pairs were excluded. These single galaxies
were separated into two morphological 
types, ``S'' or ``E'', using the same scheme as for the classification of
paired galaxies  (see Section 2). 
Then, each non-AGN spiral 
galaxy in the KPAIR sample is matched by one of the 88 single 
spiral galaxies so selected, according to the following requirements:
(i) The single galaxy should have
$\log (M_{star})$ within $\pm 0.1$ of that of the paired galaxy.
(ii) Among all single spiral galaxies fulfilling the
requirement (i), the chosen one should
have the minimum redshift difference from that of the paired galaxy. 
It should be pointed out that, 
despite the requirement (ii), there is still a significant
difference between redshift distributions of the paired galaxies
and of single galaxies in the control sample: the medians are
$z = 0.031$ and $z = 0.046$, respectively. This is because of the pair 
selection criterion (1) (see Section 2) that requires the primaries are
brighter $K_{s} = 12.5$ mag, while the control sample is selected from
galaxies brighter than $K_{s} = 13.5$ mag. We argue that the redshift difference
will not introduce any bias (in particular the Malmquist bias)
into the comparisons between the two samples
that are mass-matched, because all mass-normalized properties (e.g. the
light-to-mass ratio, SFR/M, etc) shall not depend on redshift in these
local samples\footnote{
This insensitivity of mass-normalized properties
to any selection effect is the major reason for us to choose the one-to-one
mass-matching method of control sample selection.}, 
for which the cosmic evolutionary effects are negligible.
Indeed, the requirement (ii) is arguably disposable,
and any galaxy fulfilling the requirement (i) could have been included in
the control sample.
In order to assess the uncertainties due to the particular choice of the
selected control sample, a Monte Carlo analysis (100 repeats) was carried out.
In each of the 100 realizations, an alternative control sample was selected
by relaxing the requirement (ii) and choosing arbitrarily the match of any
KPAIR galaxy among galaxies
fulfilling the requirement (i). Then the same 
statistics of the star formation properties that were calculated using
the official control sample (see the following sections) 
were repeated using the alternative control sample. We confirm that
for any of these statistics,
the mean of the 100 realizations is consistent with the result derived
from the official control sample within 1-$\sigma$.


\begin{deluxetable}{lccccccclcccccc}
\label{tbl:control}
\tabletypesize{\tiny}
\setlength{\tabcolsep}{0.05in} 
\rotate
\tablenum{4}
\tablewidth{0pt}
\tablecaption{Galaxies in the Control Sample and Their KPAIR Matches
(Non-AGN Spirals)}
\tablehead{
\ch{(1)}   &\ch{(2)}    &\ch{(3)}    &\ch{(4)}&\ch{(5)}&\ch{(6)} &\ch{(7)}
             &\ch{(8)}         &\ch{(9)}           &\ch{(10)} &\ch{(11)}
 &\ch{(12)}           &\ch{(13)} &\ch{(14)} & \ch{(15)}
\\
\ch{Galaxy ID} &\ch{   Coordinates  }&\ch{  z } &\ch{K$_s$}  
&\ch{f$_{8\mu m}$}    & \ch{f$_{24\mu m}$} 
&\ch{log(M)} &\ch{log(sSFR)}
 &\ch{KPAIR Galaxy ID}& \ch{z} &\ch{log(M)}
 &\ch{log(sSFR)} & \ch{$\epsilon$} & \ch{CAT} & \ch{SEP}
\\
           &\ch{(J2000)}&  &\ch{(mag)}&\ch{ (mJy) }       &\ch{ (mJy) }    
&\ch{($\rm M_\sun$)}  &\ch{($\rm yr^{-1}$)}  
&                    &          &\ch{($\rm M_\sun$)}  &\ch{($\rm yr^{-1}$)} & & &
\\
}
\startdata
    LCK-287434 &  10h58m58.0s+58d08m01s&  0.0320&  12.17&    7.94$\pm$  0.02&    6.76$\pm$  0.05&  10.88& -10.81 &   J00202580+0049350& 0.0176&  10.84& -10.57&  0.24& SE2       & 0.50\\
    LCK-178064 &  10h46m48.3s+56d34m04s&  0.0450&  12.24&    3.21$\pm$  0.02&    1.93$\pm$  0.04&  11.15& -11.23 &   J01093517+0020132& 0.0447&  11.05& $<-11.58$& $<-0.35$& SE2 & 1.12\\
    LCK-320371 &  10h38m32.4s+57d24m01s&  0.0471&  12.79&    4.86$\pm$  0.02&    3.67$\pm$  0.05&  10.96& -10.78 &   J01183556-0013594& 0.0475&  10.93& -10.03&  0.75& SS2       & 1.06\\
    LCK-523686 &  10h42m24.5s+58d27m31s&  0.0452&  12.75&    1.70$\pm$  0.02&    0.65$\pm$  0.04&  10.94& -11.38 &   J02110832-0039171& 0.0199&  10.98& -11.62& -0.24& SS1       & 0.56\\
    LCK-415950 &  10h52m11.7s+58d26m17s&  0.0317&  10.93&   28.18$\pm$  0.02&  100.16$\pm$  0.07&  11.37& -10.49 &   J09374413+0245394& 0.0230&  11.46& -10.51& -0.02& SE1       & 0.68\\
    LCK-086596 &  10h56m29.0s+56d54m38s&  0.0470&  13.05&    4.72$\pm$  0.03&    1.04$\pm$  0.03&  10.86& -10.93 &   J10205188+4831096& 0.0531&  10.88& -10.23&  0.70& SE2       & 0.88\\
    EN1-158103 &  16h07m36.6s+53d57m31s&  0.0298&  12.48&   19.97$\pm$  0.03&   24.32$\pm$  0.04&  10.70& -10.22 &   J10272950+0114490& 0.0223&  10.73& -10.17&  0.05& SE2       & 0.65\\
    EN1-360222 &  16h00m59.3s+54d43m52s&  0.0429&  13.15&   13.96$\pm$  0.02&   16.53$\pm$  0.04&  10.74& -10.10 &   J10435053+0645466& 0.0273&  10.83&  -9.81&  0.29& SS1       & 1.27\\
    EN1-010947 &  16h11m 9.8s+53d09m47s&  0.0367&  12.99&    0.79$\pm$  0.01&   $<0.30$         &  10.67& $<-11.63$ &   J10435268+0645256& 0.0273&  10.73& -10.67& $>0.96$& SS2  & 1.27\\
    LCK-162208 &  10h44m38.2s+56d22m11s&  0.0240&  10.96&   39.40$\pm$  0.05&   86.93$\pm$  0.05&  11.12& -10.43 &   J10514450+5101303& 0.0244&  11.13& $<-11.69$& $<-1.26$& SE2 & 0.15\\
    EN1-018834 &  16h10m47.6s+53d25m21s&  0.0631&  13.17&    1.92$\pm$  0.01&    2.03$\pm$  0.03&  11.06& -10.96 &   J12020424+5342317& 0.0642&  11.16& -10.75&  0.21& SE2       & 0.87\\
    LCK-233199 &  10h50m52.4s+57d35m07s&  0.0269&  12.49&    7.63$\pm$  0.02&    8.64$\pm$  0.04&  10.60& -10.65 &   J13082964+0422045& 0.0241&  10.53& -10.77& -0.12& SS1       & 1.29\\
    LCK-019297 &  10h47m04.5s+56d20m25s&  0.0469&  12.79&   25.24$\pm$  0.01&   48.28$\pm$  0.05&  10.96&  -9.89 &   J13325525-0301347& 0.0472&  10.90& -10.03& -0.13& SS2       & 0.79\\
    LCK-703238 &  10h45m00.5s+59d44m11s&  0.0444&  12.07&    4.48$\pm$  0.03&    2.00$\pm$  0.05&  11.20& -11.21 &   J13325655-0301395& 0.0472&  11.21& -10.46&  0.75& SS1       & 0.79\\
    LCK-050667 &  10h49m55.9s+56d49m50s&  0.0457&  12.84&    1.92$\pm$  0.01&    1.79$\pm$  0.05&  10.92& -11.13 &   J13462001-0325407& 0.0236&  11.01& -10.99&  0.14& SE1       & 1.28\\
    LCK-027930 &  10h48m52.7s+56d20m10s&  0.0458&  12.49&   13.53$\pm$  0.02&   12.10$\pm$  0.04&  11.06& -10.43 &   J14005782+4251207& 0.0327&  11.01&  -9.98&  0.45& SS1       & 1.37\\
    LCK-071868 &  10h54m09.3s+56d49m15s&  0.0466&  12.82&   22.30$\pm$  0.02&   24.18$\pm$  0.05&  10.94& -10.04 &   J14005882+4250427& 0.0327&  10.90&  -9.70&  0.34& SS2       & 1.37\\
    EN1-516050 &  15h58m23.9s+54d55m52s&  0.0381&  13.10&    8.76$\pm$  0.02&    7.92$\pm$  0.06&  10.66& -10.38 &   J14250739+0313560& 0.0359&  10.66& -11.45& -1.08& SE2       & 1.31\\
    LCK-641925 &  10h36m25.7s+58d33m22s&  0.0272&  11.05&   $<1.89$         &   $<5.00$         &  11.19& $<-11.67$ &   J14334683+4004512& 0.0258&  11.25& -10.46&$>1.22$& SS1   & 1.22\\
    LCK-400414 &  10h49m18.4s+58d20m43s&  0.0281&  11.31&   10.97$\pm$  0.03&    6.04$\pm$  0.05&  11.12& -11.11 &   J14334840+4005392& 0.0258&  11.10& -10.08&  1.03& SS2       & 1.22\\
    LCK-534543 &  10h44m45.1s+58d27m17s&  0.0314&  11.48&   30.42$\pm$  0.03&   28.42$\pm$  0.06&  11.14& -10.49 &   J15064391+0346364& 0.0345&  11.22& -11.54& -1.06& SS1       & 1.10\\
    LCK-136060 &  11h00m54.4s+57d46m32s&  0.0483&  12.56&    2.26$\pm$  0.01&    1.50$\pm$  0.04&  11.08& -11.24 &   J15064579+0346214& 0.0345&  11.17& -10.37&  0.87& SS2       & 1.10\\
    LCK-172179 &  10h45m53.9s+56d30m23s&  0.0461&  12.61&    1.72$\pm$  0.01&    3.11$\pm$  0.05&  11.02& -11.14 &   J15101587+5810425& 0.0312&  11.02& -11.30& -0.15& SS1       & 0.53\\
    LCK-564807 &  10h45m55.5s+59d09m16s&  0.0446&  13.19&    6.85$\pm$  0.02&    5.91$\pm$  0.04&  10.76& -10.46 &   J15101776+5810375& 0.0312&  10.79& -10.27&  0.19& SS2       & 0.53\\
    LCK-621286 &  10h33m19.8s+58d04m33s&  0.0454&  11.83&    2.09$\pm$  0.01&    0.79$\pm$  0.04&  11.32& -11.67 &   J15281276+4255474& 0.0182&  11.26& -10.62&  1.05& SS1       & 1.32\\
    LCK-038716 &  10h48m57.6s+56d37m12s&  0.0469&  12.68&    4.59$\pm$  0.02&    4.16$\pm$  0.06&  11.00& -10.82 &   J15281667+4256384& 0.0182&  11.03& -11.57& -0.75& SS2       & 1.32\\
    LCK-582705 &  10h48m04.4s+59d20m41s&  0.0286&  12.91&    1.25$\pm$  0.02&    1.09$\pm$  0.04&  10.49& -11.32 &   J15562191+4757172& 0.0195&  10.49& -10.23&  1.09& SE1       & 1.32\\
    LCK-329416 &  10h40m26.2s+57d26m23s&  0.0472&  12.53&    7.23$\pm$  0.02&    6.08$\pm$  0.05&  11.07& -10.70 &   J16024254+4111499& 0.0333&  11.11&  -9.92&  0.78& SS1       & 0.64\\
    LCK-040350 &  10h49m10.4s+56d38m10s&  0.0460&  13.40&    1.11$\pm$  0.01&    4.00$\pm$  0.04&  10.70& -10.89 &   J16024475+4111589& 0.0333&  10.78& -10.17&  0.72& SS2       & 0.64\\
    EN1-346329 &  16h02m45.0s+54d32m01s&  0.0636&  13.42&    4.67$\pm$  0.02&    4.12$\pm$  0.04&  10.97& -10.51 &   J17045089+3448530& 0.0568&  11.01&  -9.98&  0.52& SS2       & 0.63\\
    LCK-182514 &  10h46m21.0s+56d45m55s&  0.0673&  12.92&    6.60$\pm$  0.02&    5.14$\pm$  0.05&  11.21& -10.57 &   J17045097+3449020& 0.0568&  11.28&  -9.69&  0.88& SS1       & 0.63\\
    LCK-515902 &  10h41m39.9s+58d19m02s&  0.0723&  12.69&    2.30$\pm$  0.01&    2.81$\pm$  0.05&  11.37& -11.04 &   J20471908+0019150& 0.0133&  11.37& -10.90&  0.14& SE1       & 0.99\\
    LCK-347435 &  10h43m10.5s+57d38m51s&  0.0468&  13.02&    5.72$\pm$  0.01&    6.79$\pm$  0.05&  10.87& -10.54 &   J13153076+6207447& 0.0306&  10.91&  -9.67&  0.87& SS2       & 1.34\\
    LCK-048281 &  10h50m10.8s+56d43m37s&  0.0481&  12.62&    2.57$\pm$  0.02&    1.33$\pm$  0.05&  11.05& -11.20 &   J13153506+6207287& 0.0306&  11.09&  -9.21&  1.99& SS1       & 1.34\\
      NGC0024 &  00h09m56.7s-24d57m44s&  0.0019&   9.22&  168.6$\pm$  3.8&  125.2$\pm$ 4.6&   9.63& -10.72 &   J09494143+0037163& 0.0063&   9.71& -10.25&  0.47& SS2             & 2.04\\
      NGC2403 &  07h36m51.4s+65d36m09s&  0.0004&   6.45& 5138.8$\pm$622.9& 5830.4$\pm$53.2&   9.99& -10.25 &   J09495263+0037043& 0.0063&   9.95& -10.11&  0.14& SS1             & 2.04\\
      NGC0925 &  02h27m16.9s+33d34m45s&  0.0018&   8.59&  709.3$\pm$ 79.8&  827.4$\pm$26.4&  10.06& -10.27 &   J13082737+0422125& 0.0241&  10.15& -10.44& -0.17& SS2             & 1.29\\
      NGC3049 &  09h54m49.6s+09d16m18s&  0.0050&  10.40&  170.2$\pm$  3.0&  434.2$\pm$ 1.6&   9.91& -10.01 &   J14530282+0317451& 0.0052&   9.92& -10.59& -0.58& SS2             & 1.42\\
      NGC3184 &  10h18m17.0s+41d25m28s&  0.0020&   7.62& 1733.2$\pm$ 88.9& 1437.8$\pm$44.1&  10.31& -10.33 &   J14530523+0319541& 0.0052&  10.17& -10.65& -0.31& SS1             & 1.42\\
\hline
\enddata
\tablecomments{
\nid{\bf Descriptions of Columns}:
\begin{description}
\item{(1)} Galaxy ID in the control sample.
\item{(2)} RA (J2000) and Dec (J2000).
\item{(3)} Redshift taken from SDSS.
\item{(4)} $\rm K_s$ ($\rm K_{20}$) magnitude taken from 2MASS.
\item{(5)} IRAC 8$\mu m$ flux density (mJy) taken from DR2 of SWIRE survey.
\item{(6)} MIPS 24$\mu m$ flux density (mJy) taken from DR2 of SWIRE survey.
\item{(7)} Logarithm of the mass (in $\rm M_\sun$).
\item{(8)} Logarithm of the specific SFR ($\log (SFR/M)$).
\item{(9)} ID of the matched KPAIR galaxy.
\item{(10)} Redshift of the KPAIR galaxy.
\item{(11)} Logarithm of the mass (in $\rm M_\sun$) of the KPAIR galaxy.
\item{(12)} Logarithm of the specific SFR (log(SFR/M)) of the KPAIR galaxy.
\item{(13)} SFR enhancement, $\rm \epsilon = \log ((SFR/M)_{KPAIR-S})
  - \log ((SFR/M)_{control})$, of the KPAIR galaxy.
\item{(14)} Category of the KPAIR galaxy. For example, ``SE1''
  indicates that the galaxy is the primary of an S+E pair, and ``SS2''
  denotes a secondary in an S+S pair.
\item{(15)} Scale free separation parameter: SEP$ = s/(r_1 + r_2)$,
  where $s$ is the pair separation,$r_1$ the K-band Kron radius of the
  primary, and $r_2$ the K-band Kron radius of the secondary.
\end{description}
}
\end{deluxetable}

There are no single spiral galaxies among the selected SWIRE sources that have
$\log (M/M_\sun) < 10.3$, while 5 KPAIR S galaxies have mass less than
this limit.  These paired spirals were matched by non-interacting normal spiral
galaxies in the SINGS sample. There are only 6 such SINGS galaxies in
this mass range. The same requirements that were applied
to the SWIRE galaxies were applied here, except for NGC~3418 whose mass
differs from that of its match (J14530523+0319541) by 0.14 dex (the
closest match), slightly exceeding the limit of 0.10 dex. The 8 and
24$\mu m$ data of SINGS galaxies were taken from Smith et
al. (2007). All galaxies in the control sample, together with their matches in
the KPAIR sample, are listed in Table 4.

\subsection{Star Formation Enhancement in Paired Non-AGN Spirals}
The SFR of a galaxy can be estimated from the
IR luminosity using the formula of Kennicutt (1998)
\begin{equation}
\rm SFR\; (M_\sun yr^{-1})= 4.510\times 10^{-44}\times L_{IR}\;\; (ergs\; s^{-1}).
\end{equation}
Note that this estimate is contaminated by the dust emission powered
by the radiation of old stars. Also, it does not include the UV radiation
of young stars that is not absorbed by dust. 
For an average normal spiral galaxy, both biases are at the
$\sim 30\%$ level (Buat \& Xu 1996). Under the assumption that they
affect the KPAIR sample and the control sample in the same way, 
this estimator is exploited in the comparison between the SFR of the
two samples. However, when comparing our results with other works
using different star formation indicators (e.g. the UV continuum
and the optical emission lines), the biases should be taken into
account.

There are two $L_{IR}$ upper limits in the sample of KPAIR spirals,
and two in the control sample.  In what follows, in any of the statistical analyses 
involving $L_{IR}$,  all upper-limits were replaced by numbers of half of the values 
and treated as detections. We confirm that there is
no significant difference whether these upper limits are or are not included 
in the analyses.

In Fig.3, a histogram of $\rm log(L_{IR})$ ($\rm log(SFR)$) of non-AGN spirals
in the KPAIR sample (KPAIR-S) is compared to that of the single spirals in the
control sample (CONTROL). There is a striking difference between the 
two distributions: while the distribution of CONTROL has a single prominent peak
at $\rm log(L_{IR}) = 9.75$ (SFR$\rm  \sim 1 M_\sun yr^{-1}$), the distribution
of KPAIR-S has a second peak at much enhanced $\rm L_{IR}$ level of 
$\rm log(L_{IR}) = 10.65$, corresponding to SFR$\rm  \sim 8 M_\sun yr^{-1}$.
There is a significant excess of KPAIR S galaxies in the high $\rm L_{IR}$
end. Indeed, while 3 non-AGN spiral galaxies in the KPAIR sample are in the LIRG category ($\rm log(L_{IR}) > 11$), none of the galaxies in the control sample 
is a LIRG. The Kolmogorov-Smirnov test (K-S test) of the SFR distributions 
yields a rather low probability of 3.9\%, or equivalently
a rejection at 96.1\% confidence level, for the null hypotheses 
that the two samples are drawn from the same population 
The mean $\rm log(L_{IR})$ for the KPAIR-S sample is
$\rm log(L_{IR}) = 10.13 \pm 0.12$, corresponding to a mean
$\rm log(SFR) = 0.36 \pm 0.12$. For the CONTROL sample, 
the means are $\rm log(L_{IR}) = 9.84 \pm 0.08$ and
$\rm log(SFR) = 0.07 \pm 0.08$. The Student's t-test
yields a score of 2.32, corresponding to a probability of 2.6\% 
for the null hypotheses that the means of the two samples are equal, 
consistent with the result of the K-S test.

\begin{figure}
\plotone{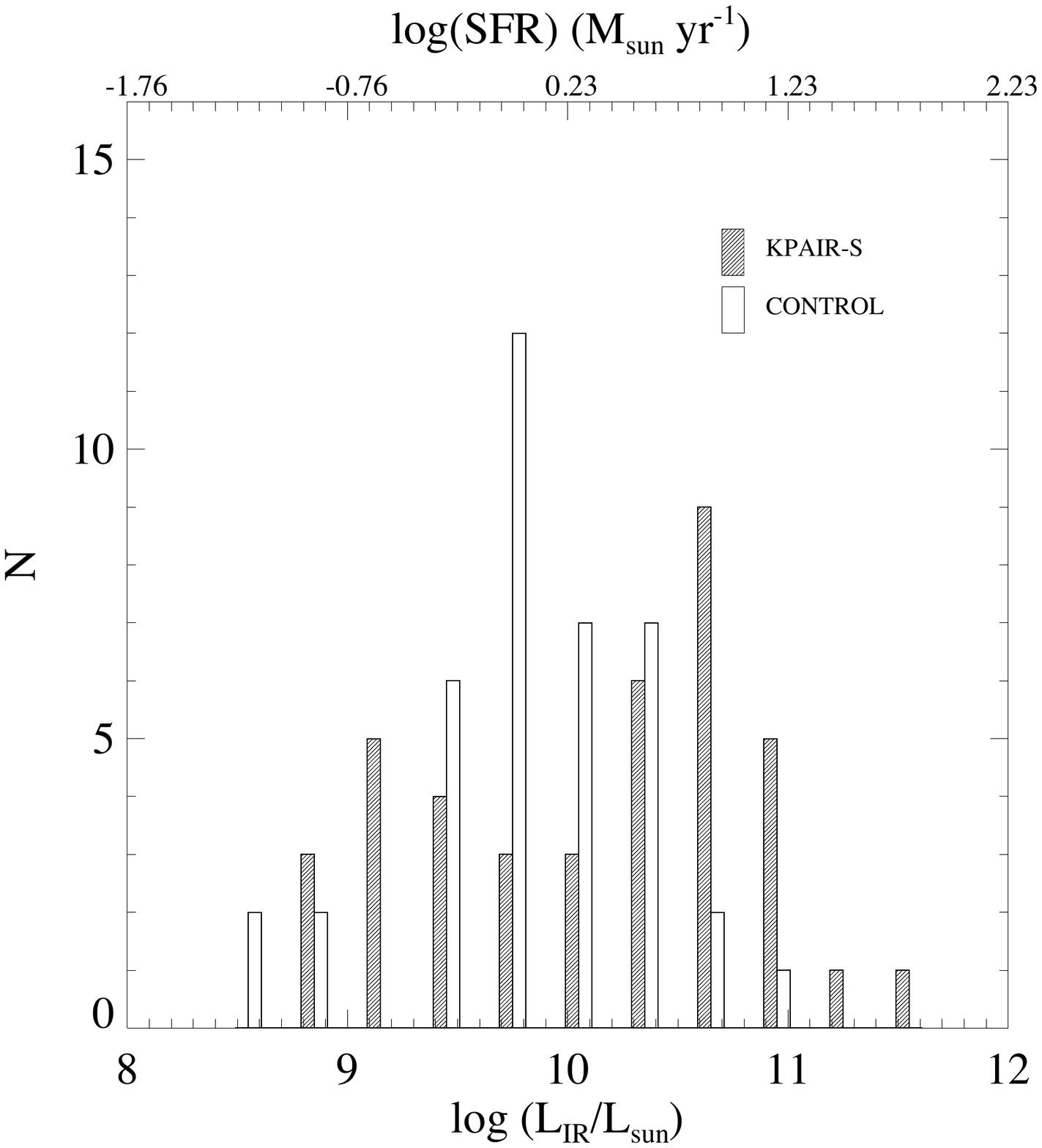}
\caption{Histograms of $\log (L_{IR})$ distributions of the
non-AGN spirals in the KPAIR sample (``KPAIR-S'') and of the control sample.
Corresponding $\log (SFR)$ values are marked on the top.}
\end{figure}

Fig.4 is a comparison of histograms of sSFR, i.e. SFR per 
mass (SFR/M), of the same two samples. It shows a similar shift of the 
distribution of KPAIR S galaxies toward the higher SFR/M bins compared
to the control sample, although the difference is slightly
less prominent than that shown in Fig.3. 
Indeed, eight KPAIR S galaxies have SFR/M$\rm > 10^{-10} yr^{-1}$ while
 only one galaxy in the control sample has such high SFR/M.
The K-S test
of the SFR/M distribution finds a low probability for the null hypotheses
of 3.9\%.
The mean values of $\rm log (SFR/M)$ are $-10.50\pm 0.10$ and
$-10.78\pm 0.08$ for the non-AGN spirals in the KPAIR sample and in
the control sample, respectively.
The score of the Student's t-test of the means is 2.21, and 
the probability for the null hypotheses is only 3.3\%.

\begin{figure}
\plotone{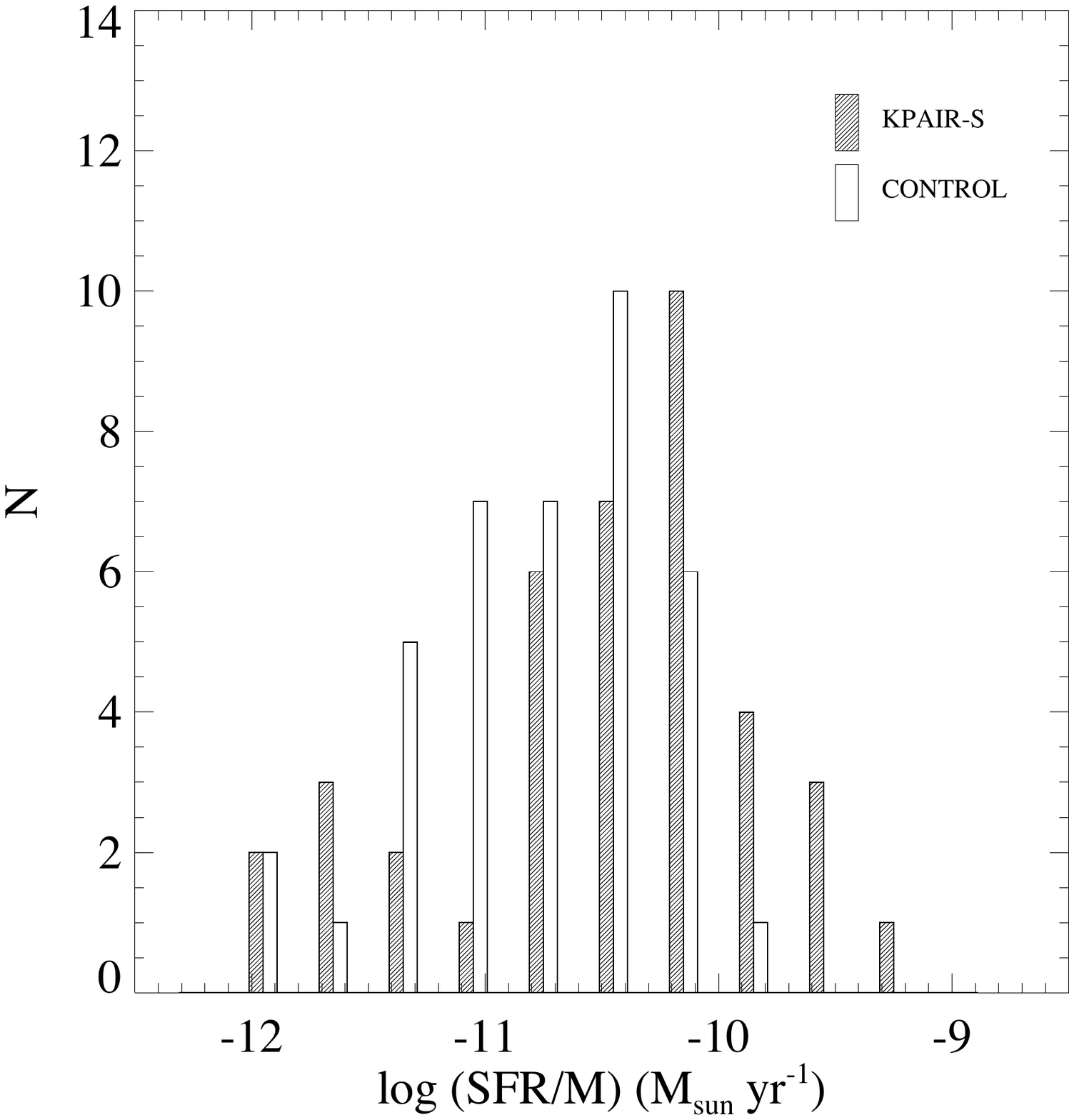}
\caption{Histograms of $\log (SFR/M))$ distributions of 
the non-AGN spirals in the KPAIR sample (``KPAIR-S'') 
and of the control sample.}
\end{figure}

\subsection{Mass Dependence of SFR/M Enhancement in Paired Spirals}
\begin{deluxetable}{ccccccc}
\label{tbl:SFM}
\tabletypesize{\normalsize}
\setlength{\tabcolsep}{0.05in} 
\rotate
\tablenum{5}
\tablewidth{0pt}
\tablecaption{Mean Specific Star Formation Rate (SFR/M) in Mass Bins}
\tablehead{
 \colhead{mass bin ($M_\sun$)} &
 &  \multicolumn{5}{c}{mean log(SFR/M) ($\rm yr^{-1}$)} \\
\cline{1-1}\cline{3-7} \\
 & & \colhead{KPAIR-S} & \colhead{CONTROL} &
 & \colhead{S in S+S} & \colhead{S in S+E}
}
\startdata
 $9.7 < log(M) \leq 10.2$ & & -10.41$\pm$0.10 (5)  &  -10.32$\pm$0.11  (5)  &&-10.41$\pm$0.10  (5) &                      \\
$10.4 < log(M) \leq 10.8$ & & -10.53$\pm$0.18 (7)  &  -10.84$\pm$0.23  (7)  &&-10.47$\pm$0.15  (4) & -10.62$\pm$0.42  (3) \\
$10.8 < log(M) \leq 11.2$ & & -10.48$\pm$0.18 (20) &  -10.77$\pm$0.10 (20)  &&-10.23$\pm$0.20 (14) & -11.07$\pm$0.29  (6) \\
$11.2 < log(M) \leq 11.6$ & & -10.60$\pm$0.21 (7)  &  -11.06$\pm$0.22  (7)  &&-10.55$\pm$0.30  (5) & -10.70$\pm$0.19  (2) \\
\hline
\enddata
\end{deluxetable}
\oneskip

In order to study mass dependence of
the SFR and its enhancement in galaxy pairs, we binned
the spirals both in the KPAIR sample and in the control sample into
four mass bins, and calculated the means of
$\log (SFR/M)$ for individual bins.  The results are listed
in Table 5 and plotted in Fig.5.
Galaxies in the control sample show a clear trend of decreasing
sSFR with increasing mass, as has already been well documented 
in the literature (Kauffmann et al. 2004; Brinchmann et al. 2004;
Schiminovich et al. 2007; Zheng et al. 2007).
On the other hand, the sSFR of spirals in pairs 
is nearly constant with mass. In the lowest mass bin of 
$\rm 9.7 < \log (M/M_\sun) < 10.2$, there is no enhancement
of the sSFR of the paired galaxies compared to that
of the control sample. And at the high mass end, in
the bin of $\rm 11.3 < \log (M/M_\sun) < 11.6$,
the mean sSFR of the paired galaxies is about
3 times higher than that of the control sample. In between,
there is a weak enhancement in the two intermediate
mass bins.
%
\begin{figure}
\plotone{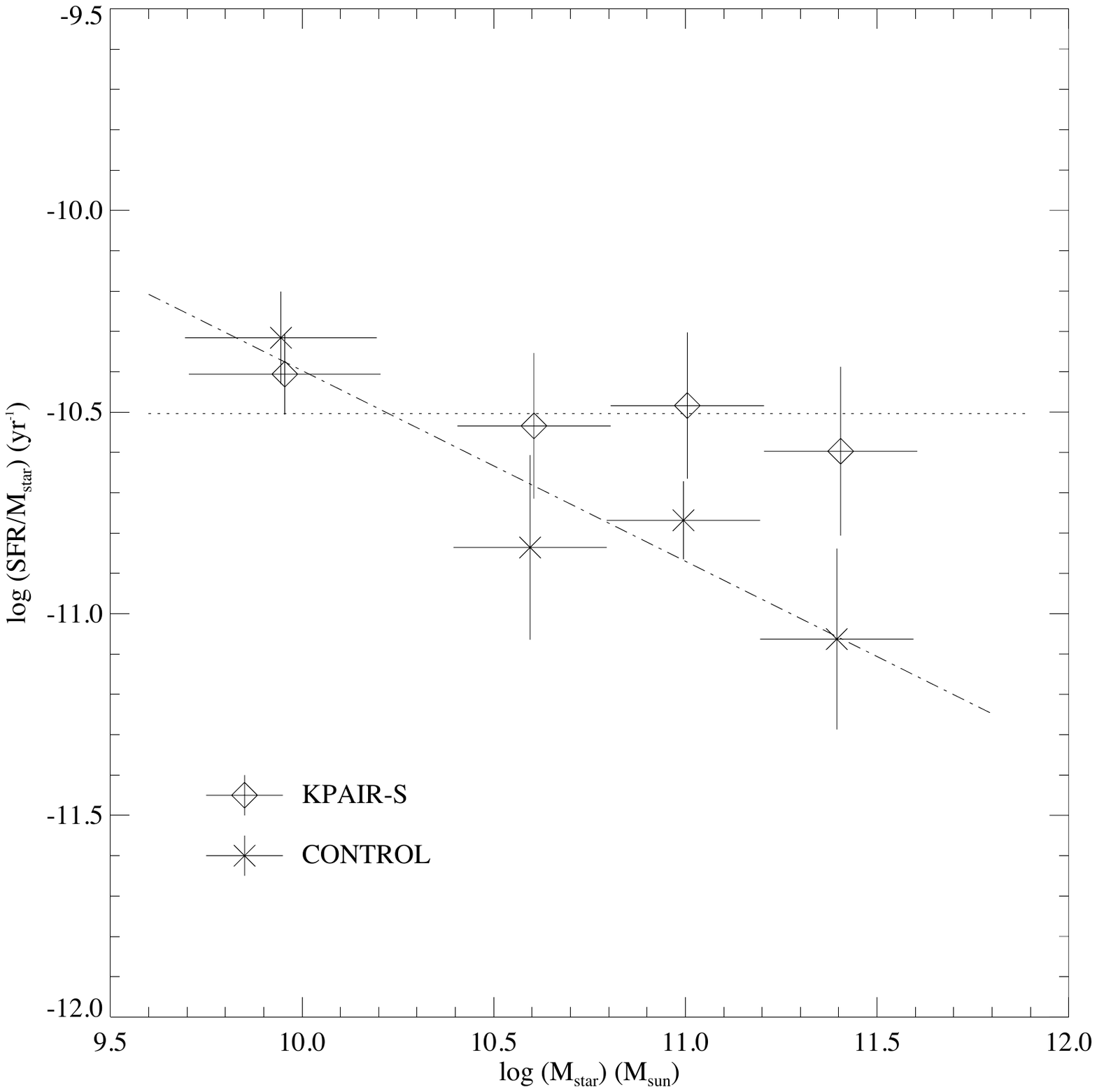}
\caption{Comparison of means of
log(SFR/M) of paired and single spirals in mass bins.
The dotted line marks the mean $SFR/M$ of non-AGN spirals
in KPAIR, and the dotted-dashed line marks the linear regression of 
those in the control sample.
}
\end{figure}

\subsection{Comparison of Non-AGN Spirals in S+S and in S+E Pairs}
Sulentic (1989) and Xu \& Sulentic (1991) found significant FIR 
enhancement in S+E pairs in their IRAS studies, under the
assumption that the ellipticals
in these pairs are FIR quiet. However, the ISO observations of
Domingue et al. (2003) demonstrated 
that this assumption is invalid. In this section, we address again the
question whether spiral galaxies in S+E pairs have similar levels
of SFR enhancement as those in S+S pairs.

In Fig.6, the non-AGN spirals in the KPAIR sample is decomposed 
into two subsamples,
one for galaxies in S+S pairs (28 galaxies) and the other for galaxies
in S+E pairs (11 galaxies), and the $\rm log (SFR/M)$ distribution of
each of them is compared to that of the control sample in one of the two
panels.  Because here the samples being compared do not have the
same numbers of sources, the distributions are normalized (i.e. in
fractions). It shows that the distribution of the S galaxies in S+E
pairs is {\it not} significantly different from that of the
control sample: the K-S test finds a 91\% probability
that the two samples are drawn from the same population.
Therefore, all the enhancement found in Fig.4 is due to spirals 
in the S+S pairs. For the comparison between
the spirals in S+S sample and the control sample, the K-S test
yields a probability of only 2\% for the null hypotheses.
The average values of $\rm log (SFR/M)$ are $-10.36\pm 0.11$ and
$-10.88\pm 0.19$ for spirals in S+S sample and in S+E sample, 
respectively.
\begin{figure}
\plottwo{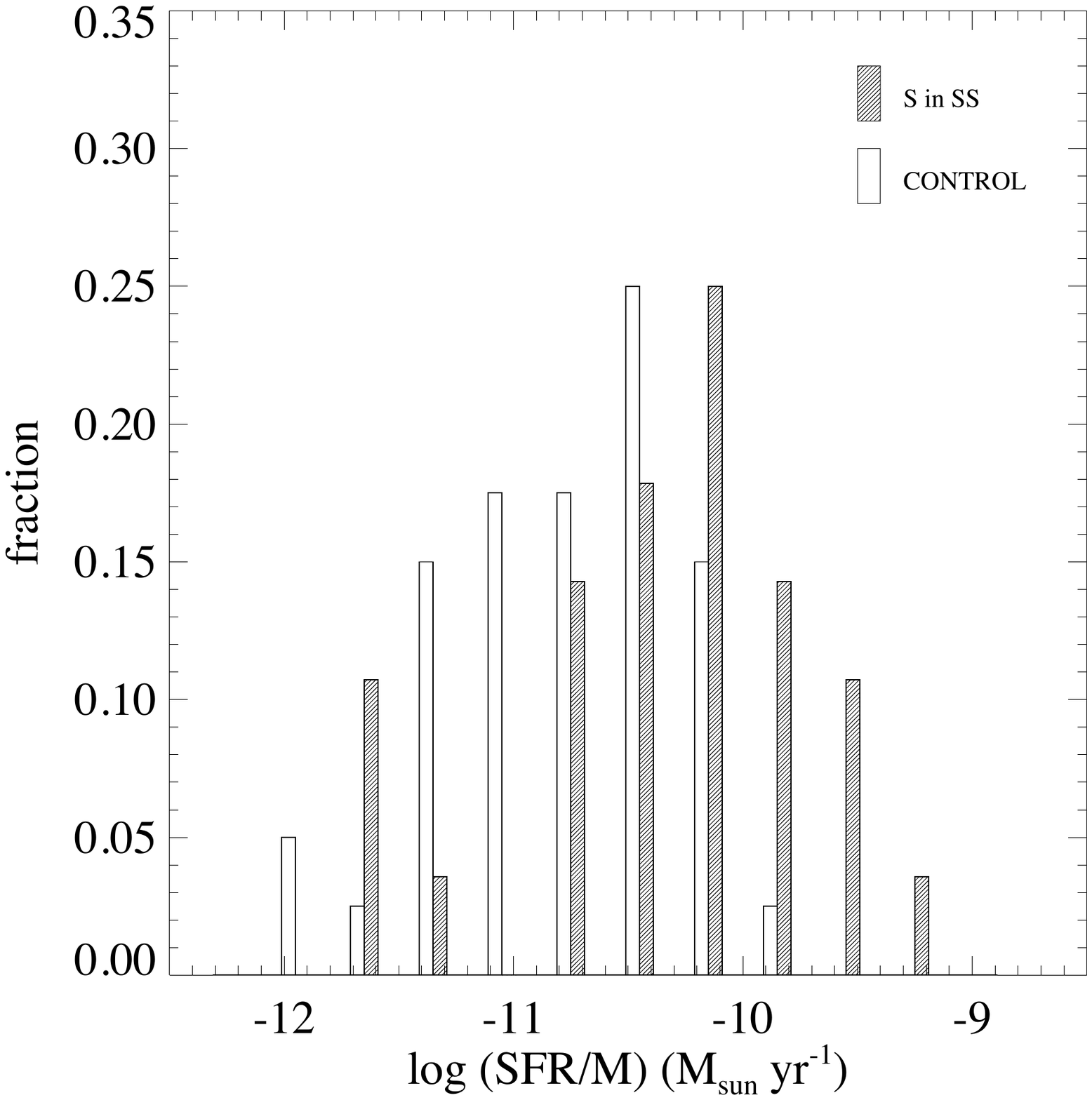}{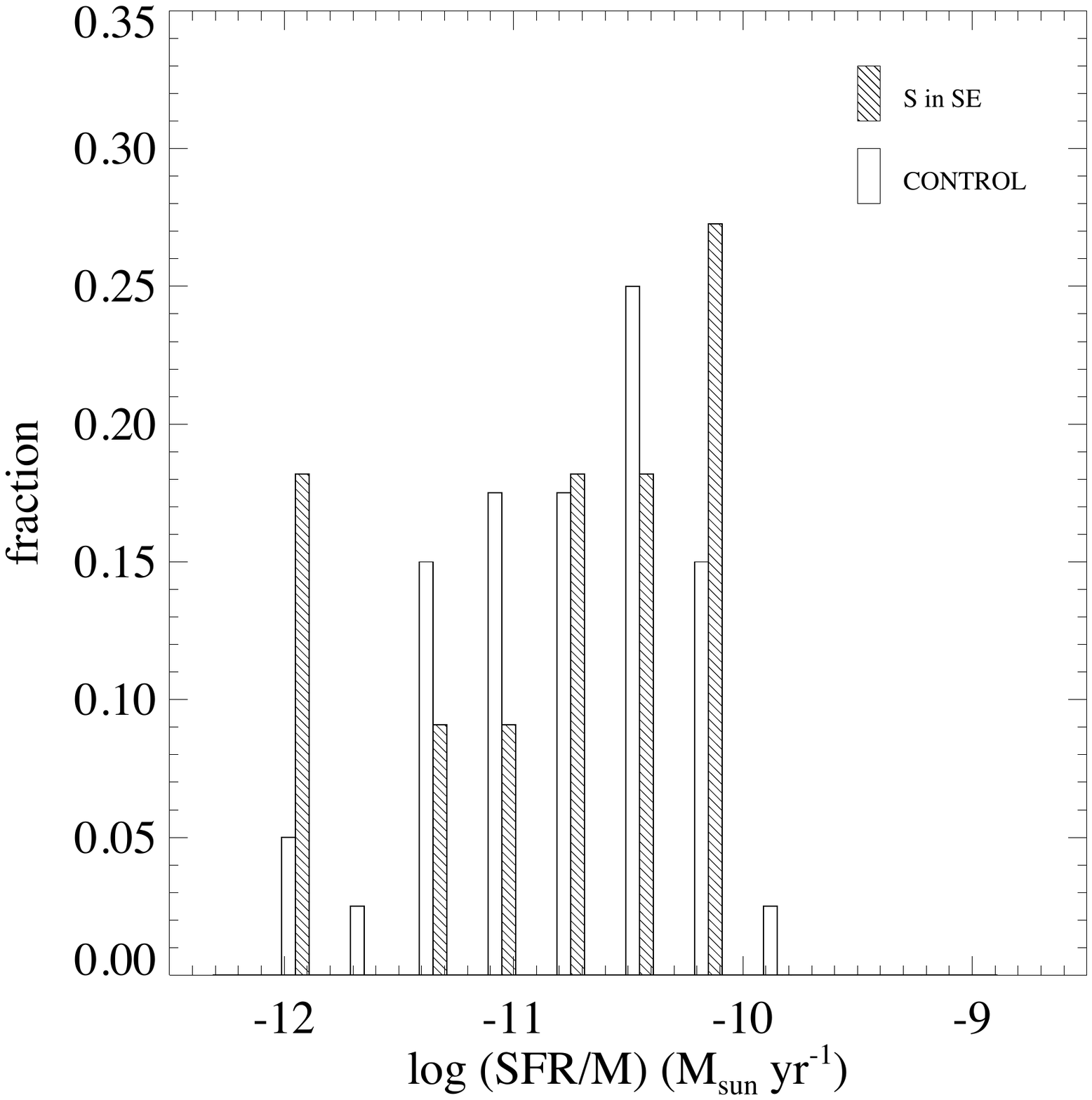}
\caption{{\it Left}: Histograms of $\log (SFR/M))$ distributions of
  the non-AGN spirals in S+S pairs (28 galaxies) and of the
  control sample.  {\it Right}: Histograms of $\log (SFR/M))$
  distributions of the non-AGN spirals in S+E pairs (11
  galaxies) and of the control sample.  }
\end{figure}

In Fig.7, the mass dependence of SFR/M is plotted for
non-AGN spirals in S+S pairs and in S+E pairs, separately. 
No significant enhancement
is found for spirals in S+E pairs in any mass bin.
For those in S+S pairs, they have a similar trend as the
total sample (Fig.5), showing a rather constant
sSFR that is slightly above the mean of the total sample of paired 
spirals.
%
\begin{figure}
\plotone{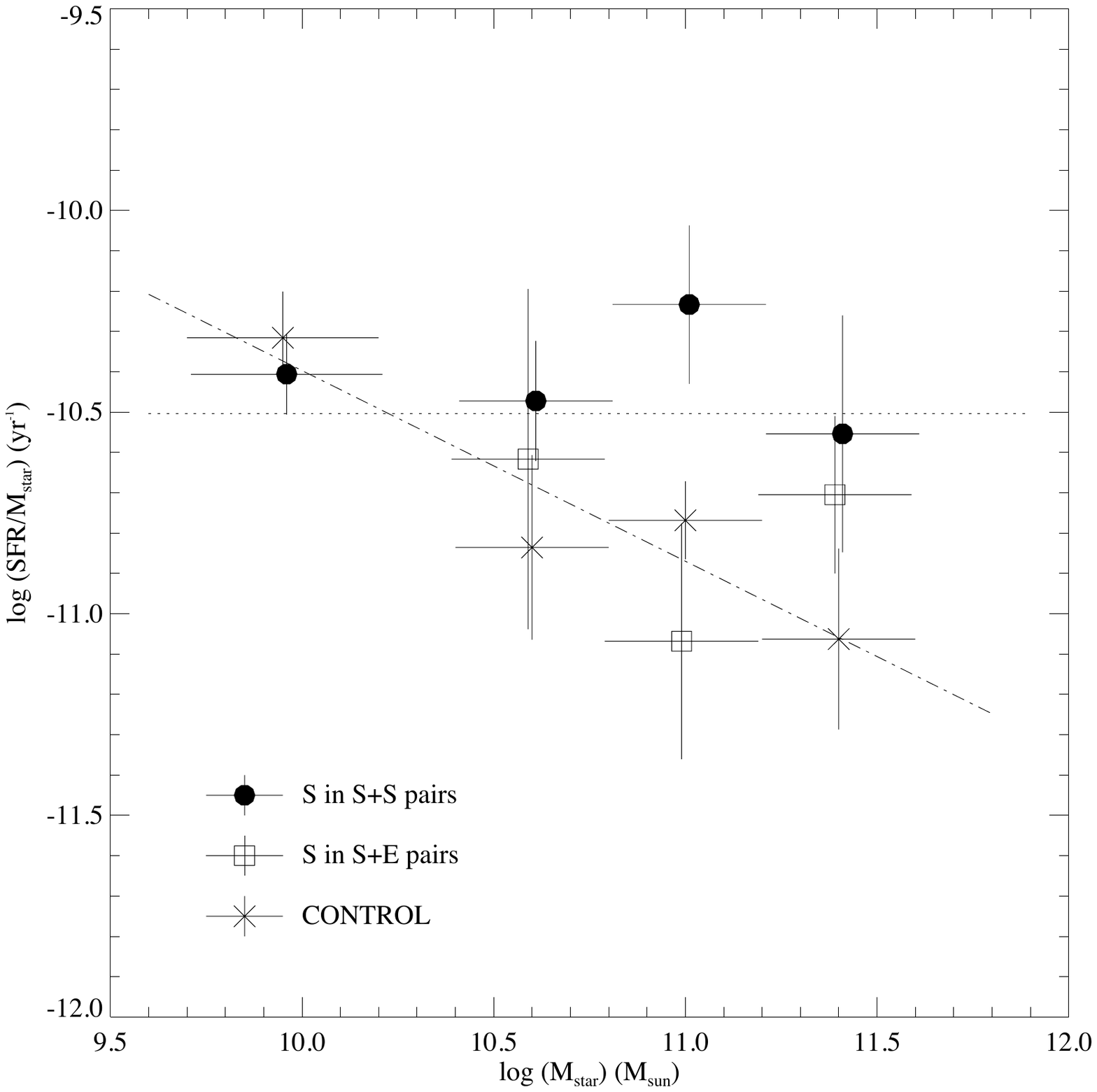}
\caption{Comparison of means of
log(SFR/M) of spirals in S+S pairs, in
S+E pairs, and in the control sample.
The dotted line marks the mean log(SFR/M) of all non-AGN spirals
in KPAIR, and the dotted-dashed line marks the linear regression of 
those in the control sample.
}
\end{figure}

Fig.8 is a $\log (SFR/M))$ versus $\log (M))$ plot for individual
non-AGN spirals in S+S and S+E pairs, compared to their counterparts in
the control sample. All spirals in S+S pairs 
are detected. The upper limits for spirals in S+E pairs
(two) and in the control sample (two) are shown by 
upside-down triangles and downward arrows, respectively.
The two solid lines delineate the regions occupied by the LIRGs and ULIRGs.
All the four LIRGs are in S+S pairs, while there are no ULIRGs in any of the 
samples. The dotted-dashed line marks the result of Brinchmann et al. (2004)
for SDSS galaxies. It shows a very similar trend in
the sSFR vs. mass relation as that revealed by the data of 
single spirals in our control sample, although the SFR
in Brinchmann et al. (2004) was
derived using the optical emission lines data 
while it was derived using the IR luminosity in this work.
The dashed line marks $SFR/M = 1/t_{Hubble}$, with 
the Hubble time $\rm t_{Hubble}=13$ Gyr.
Galaxies above this line have enhanced SFRs compared
to a constant SFR over the Hubble time.
There is indeed an excess of galaxies in S+S 
pairs above this line, while none of the spirals in S+E pairs has enhanced
SFR.

%
\begin{figure}
\plotone{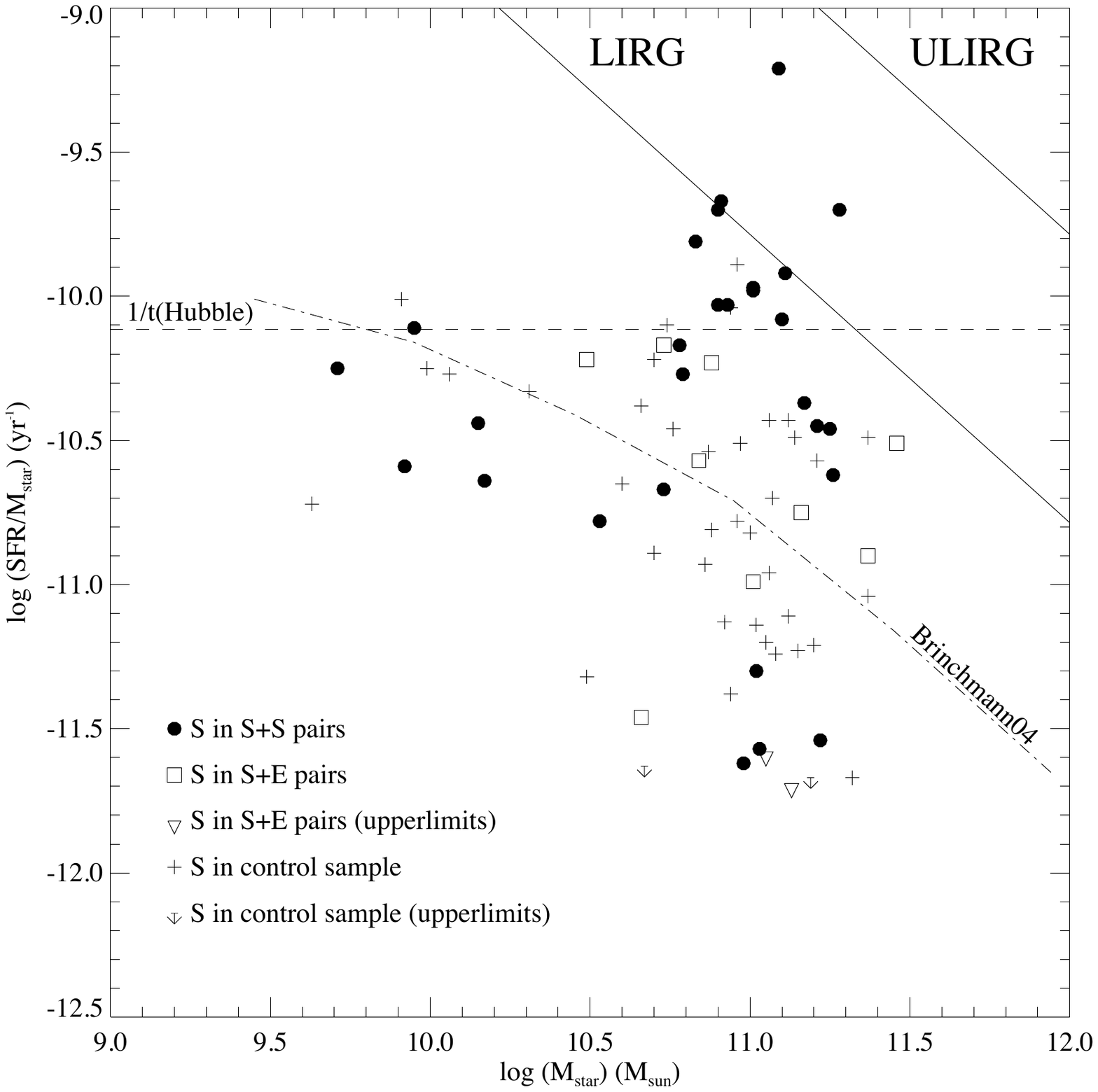}
\caption{
The sSFR (SFR/M) versus mass 
plot for the non-AGN spirals in pairs and those in the control sample.
The dashed line marks $SFR/M = 1/t_{Hubble}$, with 
the Hubble time $\rm t_{Hubble}=13$ Gyr.
The dotted-dashed line is the result of Brinchmann et al. (2004)
for SDSS galaxies, corrected for the IMF
and the Hubble constant differences
(the Kroupa IMF and ${\rm H}_0= 70\; ({\rm km~sec}^{-1} {\rm Mpc}^{-1})$ 
in Brinchmann et al. 2004; the Salpeter IMF and 
${\rm H}_0= 75\; ({\rm km~sec}^{-1} {\rm Mpc}^{-1})$ in this work).
}
\end{figure}

We define a star formation enhancement indicator, $\epsilon$, 
for each non-AGN spiral in KPAIR:
\begin{equation}
\rm \epsilon = \log( (SFR/M)_{KPAIR-S}) -  \log((SFR/M)_{control})
\end{equation}
where $(SFR/M)_{KPAIR-S}$ and $(SFR/M)_{control})$ are
the sSFR of the paired
galaxy and that of its match in the control sample, respectively.

Fig.9 is a plot of means of $\epsilon$ in the $\rm \log (M)$ bins. 
For the S+S sub-sample, there is a clear mass
dependence of $\epsilon$, which can
be expressed by its linear regression:
\begin{equation}
\rm <\epsilon>_{S+S} = 0.03(\pm 0.14) + 0.47(\pm 0.15) \times 
\log \left( {M\over 10^{10} \times M_\sun} \right). \;\; 
\end{equation}
This relation is confined to the mass range covered by our samples:
$10.0 \lsim \log(M/M_\sun) \lsim 11.5$. 
For the S+E subsample, the mean $\epsilon$ is consistent with
0 in all mass bins.
%
\begin{figure}
\plotone{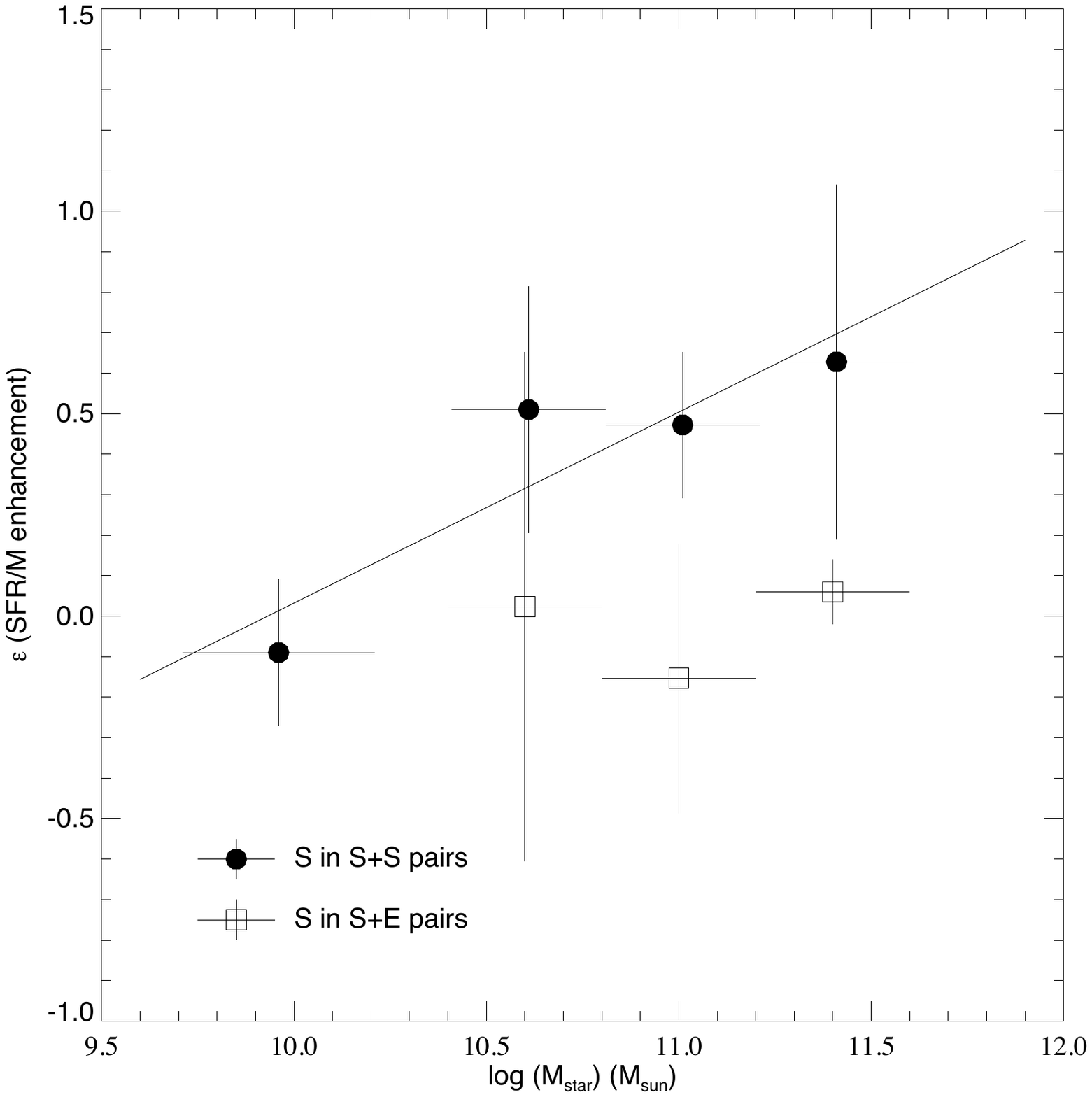}
\caption{
Plot of means of SFR/M enhancement, 
$\rm \epsilon = \log( (SFR/M)_{KPAIR-S}) -  \log((SFR/M)_{control})$,
in the $\rm \log (M)$ bins. The solid line is the linear
regression of the $\log (M)$ dependence of $\epsilon$
of spirals in S+S pairs:
$\rm <\epsilon>_{S+S} = 0.03 + 0.47 \times 
\log [M/( 10^{10} \times M_\sun)]$.
}
\end{figure}

There is a known dependence of sSFR on local environment in the
sense that galaxies in higher local density environments tend to have
lower sSFR (Kauffmann et al. 2004). Could the difference in the SFR/M
of spirals in S+S pairs and in S+E pairs be due to different mean
local densities of the two types of pairs? If S+E pairs are
preferentially found in the denser environment, then the lower SFR/M
of the spirals in these pairs compared to those in S+S pairs is just
another consequence of the SFR-environment relation. We made the
following test for this hypothesis. Fig.10 shows the selection
function of the parent sample of KPAIR sample (Section 2). In the
plot, dots are the 59312 DR3/2MASS galaxies in the parent sample, and
eight-point stars are the non-AGN spirals in the KPAIR sample. The
highest redshift of paired spirals is at $z \sim 0.06$, corresponding
to a limiting mass of $\rm M_{limit} = 10^{10.9}\; M_\sun$. Adopting
this limiting mass, around each non-AGN spirals in the KPAIR sample we
counted neighbors within the parent sample using the following 
criteria: (1) M $\rm \geq M_{limit}$; (2) distance $\rm \leq 2$ Mpc; (3) 
$\rm \delta V_{z} \leq 1000$ km s$^{-1}$. In Fig.11, means of
the neighbor counts ($\rm N_{neighbor}$)
around spirals in S+S pairs and in S+E pairs in
the individual mass bins are compared with each other. According to
Fig.11, there is no evidence for
the spirals in S+E pairs having systematically higher local density than
those in S+S pairs. Therefore the difference 
in the SFR/M of spirals in the two different types of pairs
is not due to the local-density dependence.
Interestingly, in Fig.11, there is a clear indication for the
density/mass correlation, consistent with the literature (Kauffmann et al. 
2004).
%
\begin{figure}
\plotone{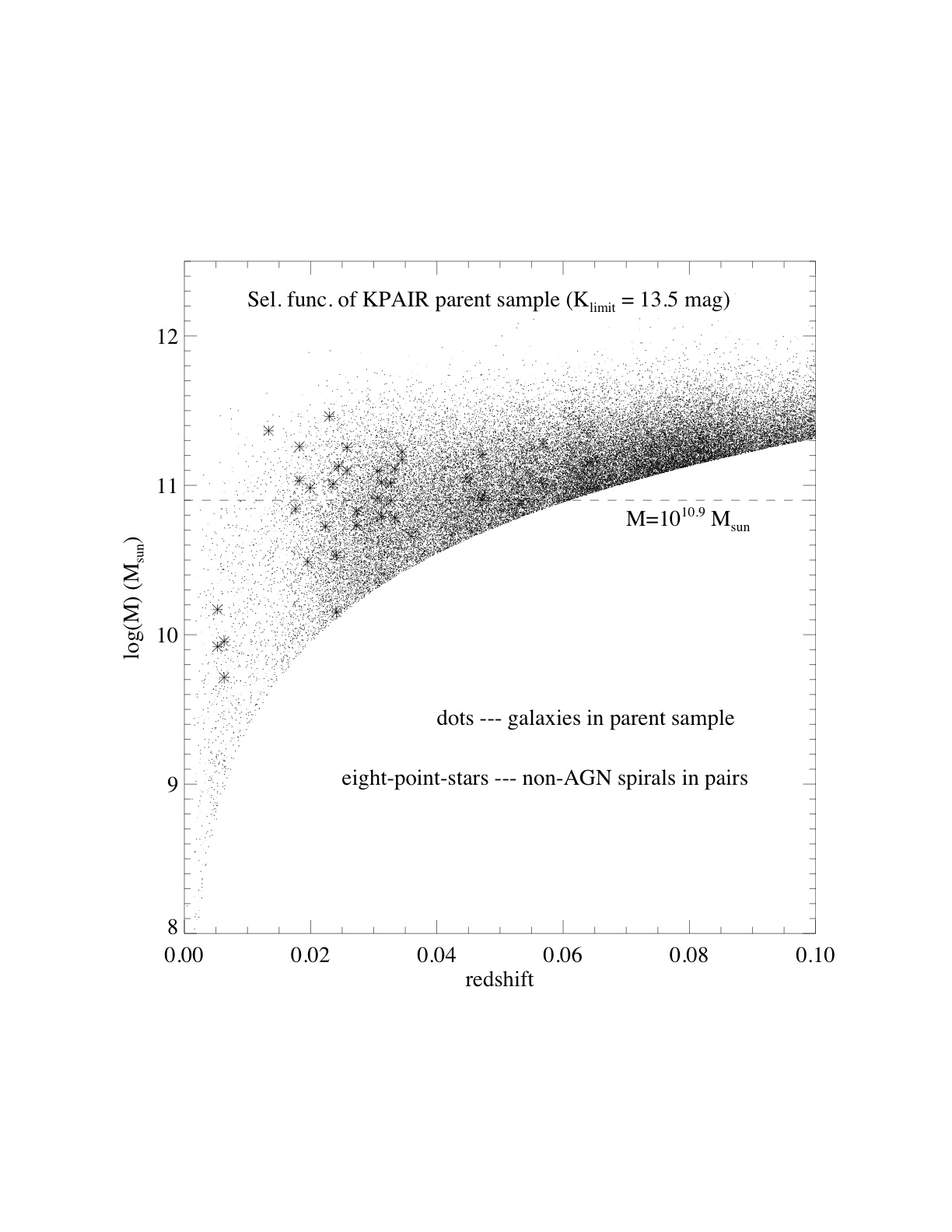}
\caption{
Selection function of the parent sample of the KPAIR sample.
{\it Symbols}: dots are galaxies in the parent sample,
eight-point stars are the non-AGN spirals in the KPAIR sample.
The dashed line marks the limiting mass ($\rm 10^{10.9}\; M_\sun$) for
the neighbor selection.
}
\end{figure}
 %
\begin{figure}
\plotone{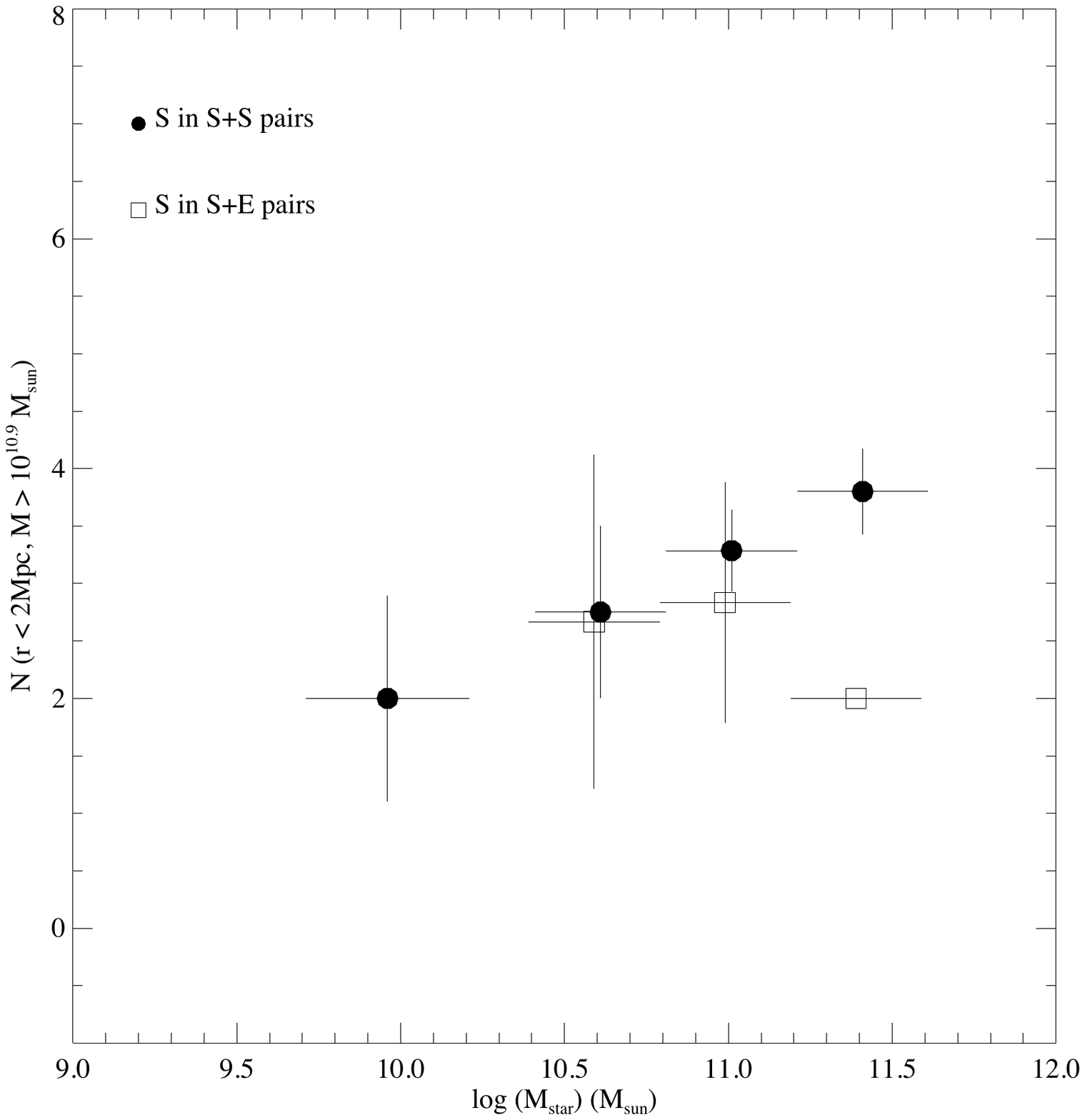}
\caption{
Mean neighbor counts in individual mass bins. Filled circles: results of
non-AGN spirals in S+S pairs. Open squares: results of non-AGN spirals
in S+E pairs. Horizontal error bars show the bin width of the mass bins.
Vertical error bars present the standard deviations of the means.
}
\end{figure}

\subsection{Comparison of Primaries and Secondaries in KPAIR}
Previous studies (Ellison et al. 2008; Woods \& Geller 2007) have found
that the secondaries in minor-merger pairs (mass ratio $> 3$) have
higher SFR enhancement than the primaries. We checked whether this
is also true for spirals in KPAIR, which includes 
only major-merger pairs. The answer is negative. As shown in Fig.12,
there is no significant difference between the mean SFR/M of primaries
and that of secondaries in any mass bins studied in this work. The K-S
test finds a 62\% probability for the null hypothesis that SFR/M distributions
of the primaries and of the secondaries are drawn from the same population.

\begin{figure}
\plotone{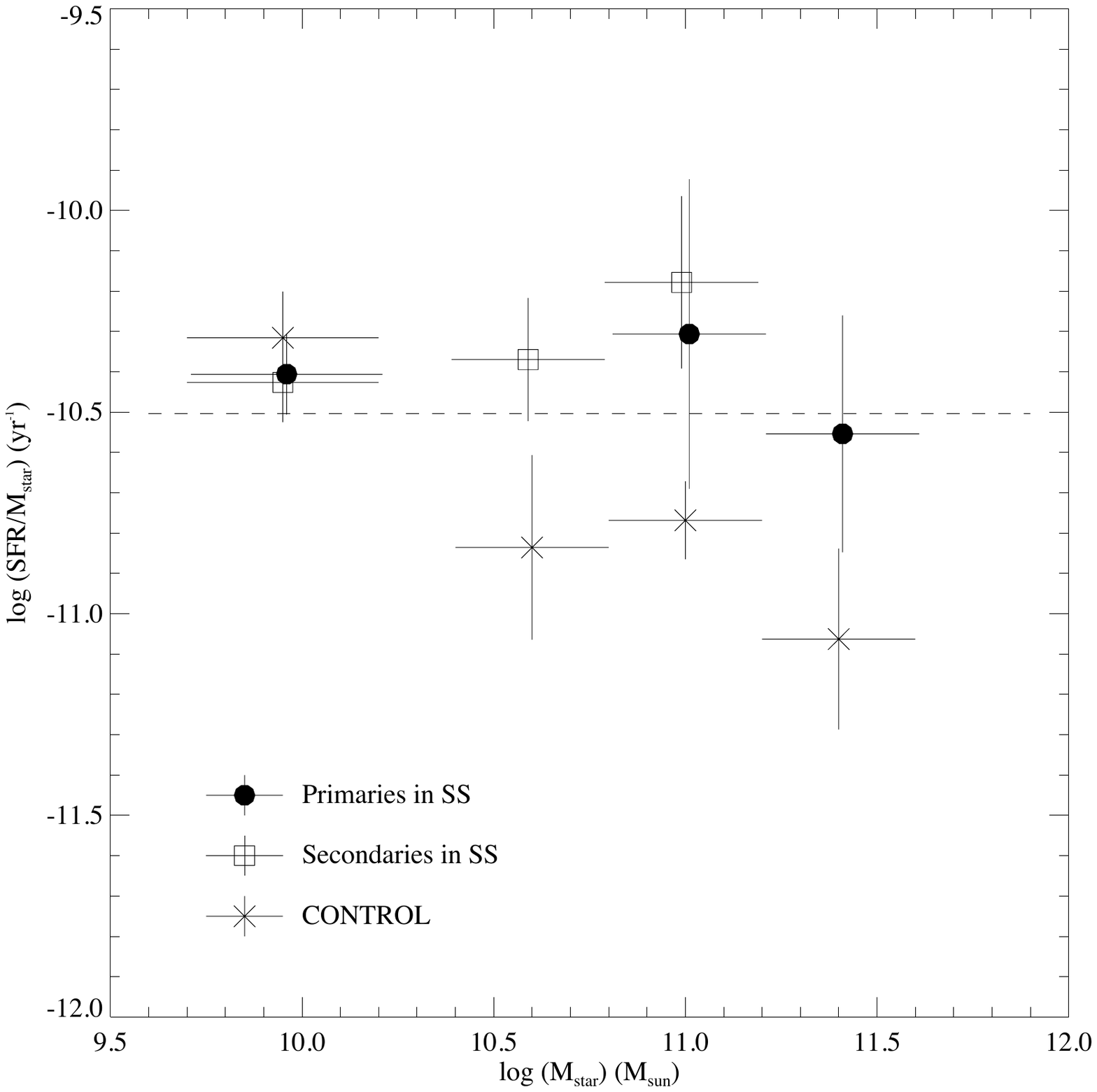}
\caption{Average
sSFR's of primaries and secondaries in KPAIR
pairs. The mass bins are the same as in Fig.5.
The dotted line marks the mean $SFR/M$ of the 
KPAIR-S sample.}
\end{figure}

\subsection{Enhancement in One or Two Components?}

In an early IR study of interacting pairs, Joseph et al. (1984) found that
among all the 22 pairs for which they obtained $\rm K-L$ colors for both 
components, only one component of each pair exhibited  $\rm K-L$ excess.
They interpreted the result as evidence for single-component star formation
enhancement in interacting pairs. Contrarily, in an IRAS study of 
isolated pairs, Xu \& Sulentic (1991) argued that they saw indications of 
IR enhancement in both components in the close interacting (CLO) S+S pairs,
though their results were not conclusive because most of their
CLO pairs are unresolved by IRAS. With the much improved angular resolution of
Spitzer which resolved all pairs in the KPAIR sample, 
we can address directly the question whether 
the SFR is enhanced in only one component or in both components 
in close interacting S+S pairs.

Given our result that only more massive galaxies in S+S pairs 
have SFR/M enhancement (Fig.7 and Fig.9), we picked the 10 pairs
(out of 15 in total) whose two components are both non-AGN and more
massive than $\rm 10^{10.7} M_\sun$. Fig.13 is a SFR/M plot of individual
pairs, each pair in a separate column, showing the log(SFR/M) values 
of both components as well as of their counterparts in the control sample. The
pairs are sorted according to the log(SFR/M) that is the higher one
among the two. 

\begin{figure}
\plotone{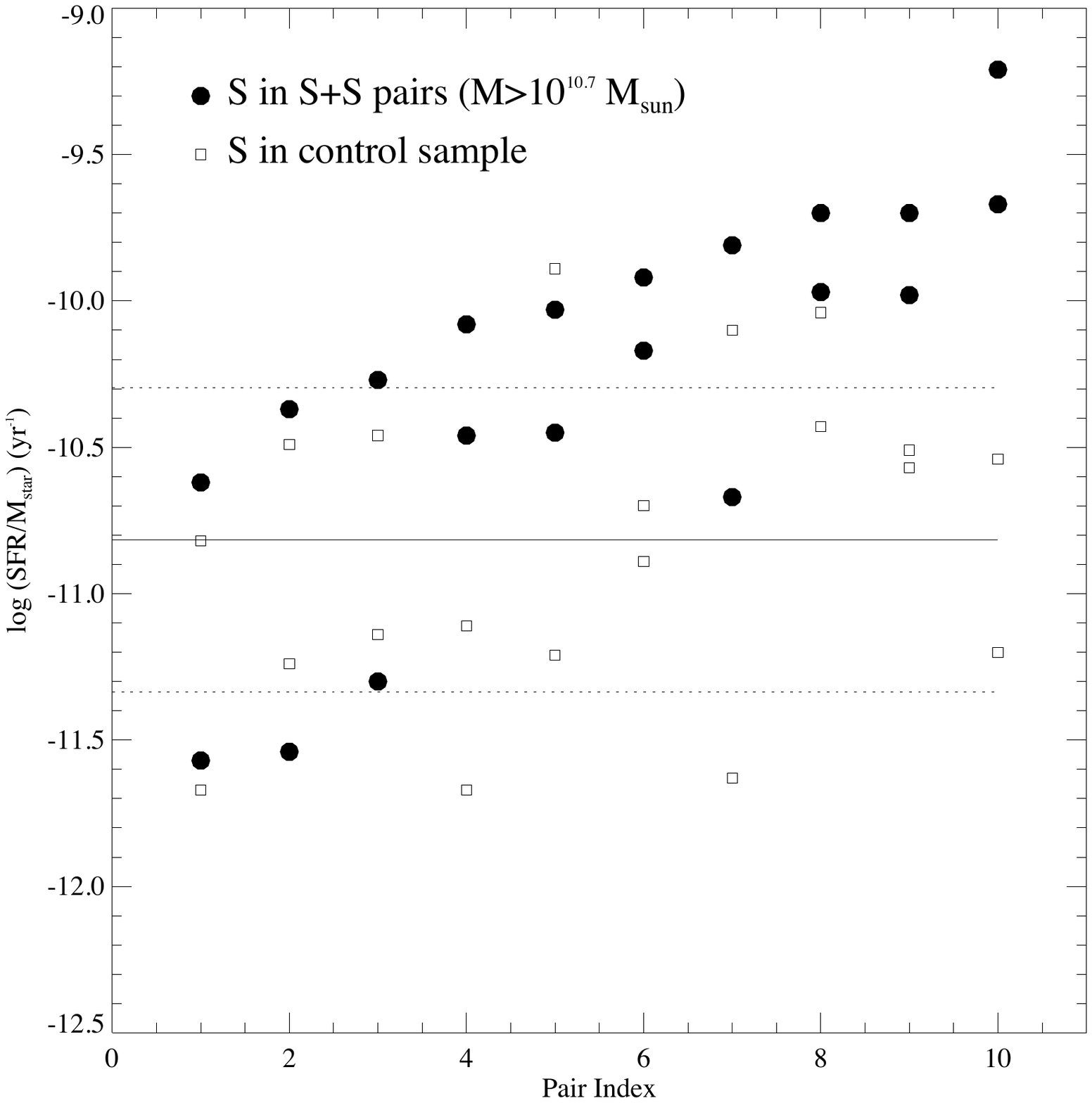}
\caption{Log(SFR/M) plot of 10 S+S pairs with $\rm M \geq 10^{10.7}M_\sun$.
Filled circles represent paired galaxies, and open squares represent
their counterparts in the control sample. Pairs are arranged according to
the log(SFR/M) value that is the higher one among the two. The solid line
and the dotted lines mark the mean log(SFR/M) and the $\pm 1\sigma$ boundaries,
respectively, with the mean and the $\sigma$ (rms) 
derived using the 20 control galaxies included in this plot.
}
\end{figure}

The results in Fig.13 can be summarized as the following: 
\begin{description}
\item{(1)} Massive galaxies ($\rm M \geq 10^{10.7} M_\sun$)
in close major-merger S+S pairs have very diversified 
star formation activity levels, from very quiescent (similar to
``red and dead'' galaxies) to strong starburst (e.g. in LIRGs).
\item{(2)} In individual pairs, the SFR/M values of the two components
show a certain level of concordance (``Holmberg effect''):
When one component has a strong star formation
enhancement, the other is usually enhanced as well 
(with only one exception: Pair \#7 = J1043+0645). On the other hand, if one
component is a ``red and dead'' galaxy, the other one usually
shows no sign of star formation enhancement, either\footnote{
This is in agreement with the lack of SFR/M enhancement 
for spirals in S+E pairs.}.
\end{description}

Fig.14 confirms the ``Holmberg effect'' between the SFR/M of the
two pair components. While the values of 
$\rm log(SFR/M)$ of the two components are highly correlated,
no correlation is found between $\rm log(SFR/M)$ values of their
matches in the control sample, as is expected. 

\begin{figure}
\plotone{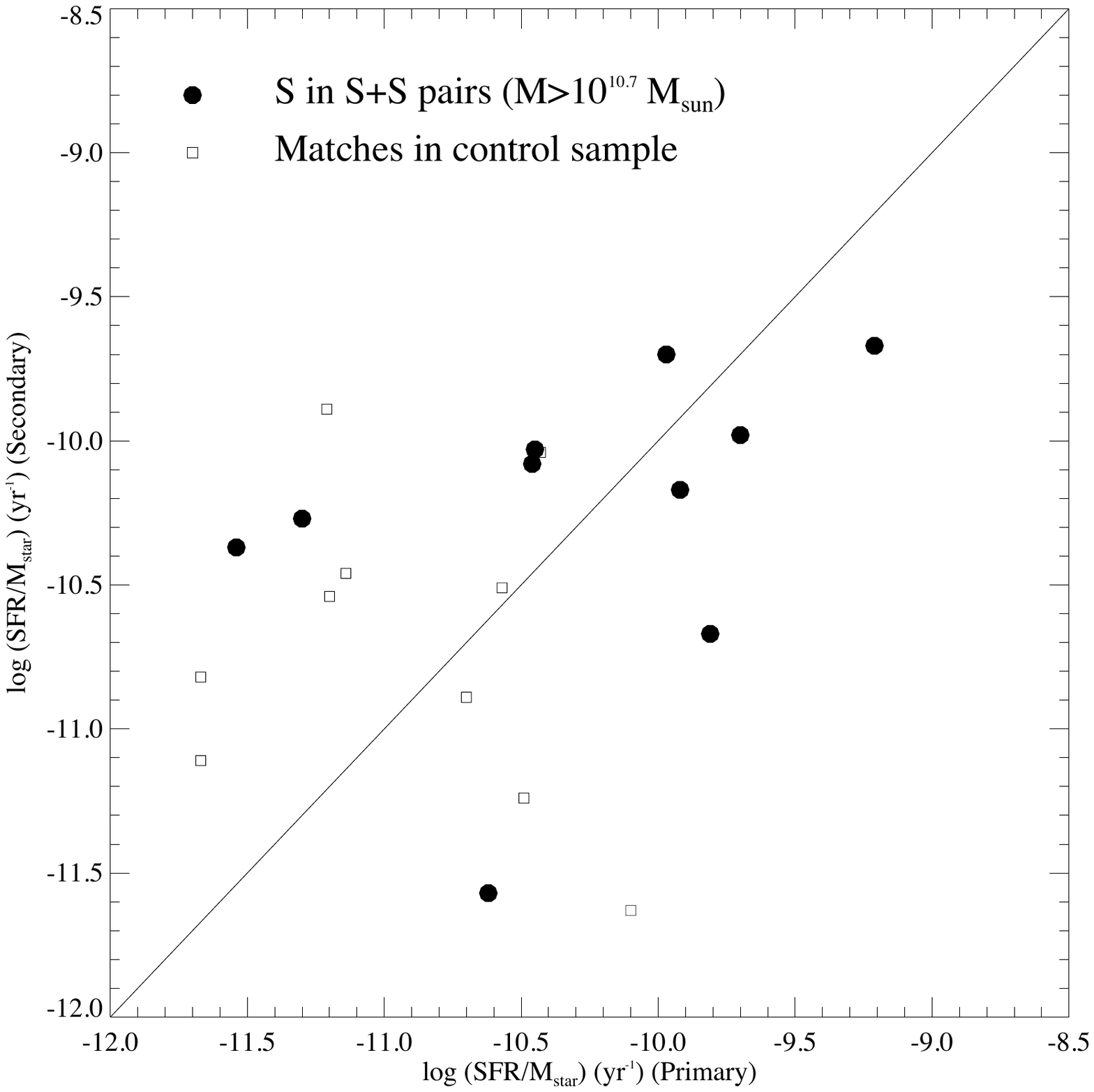}
\caption{Correlation plot of Log(SFR/M) of two components in the 
S+S pairs in Fig.13.
Filled circles represent the pairs, and open squares represent the
their counterparts in the control sample.
}
\end{figure}

The result (2) is apparently in contradiction with that of Joseph et
al. (1984). This might be explained by the differences between our
study and that of Joseph et al. (1984).  Several pairs in the
sample of Joseph et al. (1984) belong to the category of minor-mergers
(e.g. Arp~283 and Arp~294), for which indeed only one component (the
secondary) is usually enhanced (Woods \& Geller 2007).  Furthermore,
the study of Joseph et al. (1984) was confined to the nuclei of the
interacting galaxies, and the $\rm K-L$ excess is only sensitive to
very hot dust emission from compact starbursts.  It is possible that
these compact nuclear starbursts have shorter time scales, and
therefore significantly lower chance to occur simultaneously
in both components, compared to the star formation enhancement over entire
galaxy bodies as probed by our Spitzer observations.

Fig.15 is a plot of log(SFR/M) versus neighbor counts ($\rm
N_{neighbor}$, as defined in Section 5.4) for the same paired galaxies 
plotted in Fig.13.  There is no discernible dependence of log(SFR/M)
on $\rm N_{neighbor}$ in the plot. Therefore, it is something 
other than the local density that determines whether the two galaxies
in an S+S pair should or should not have enhanced star formation activity.

\begin{figure}
\plotone{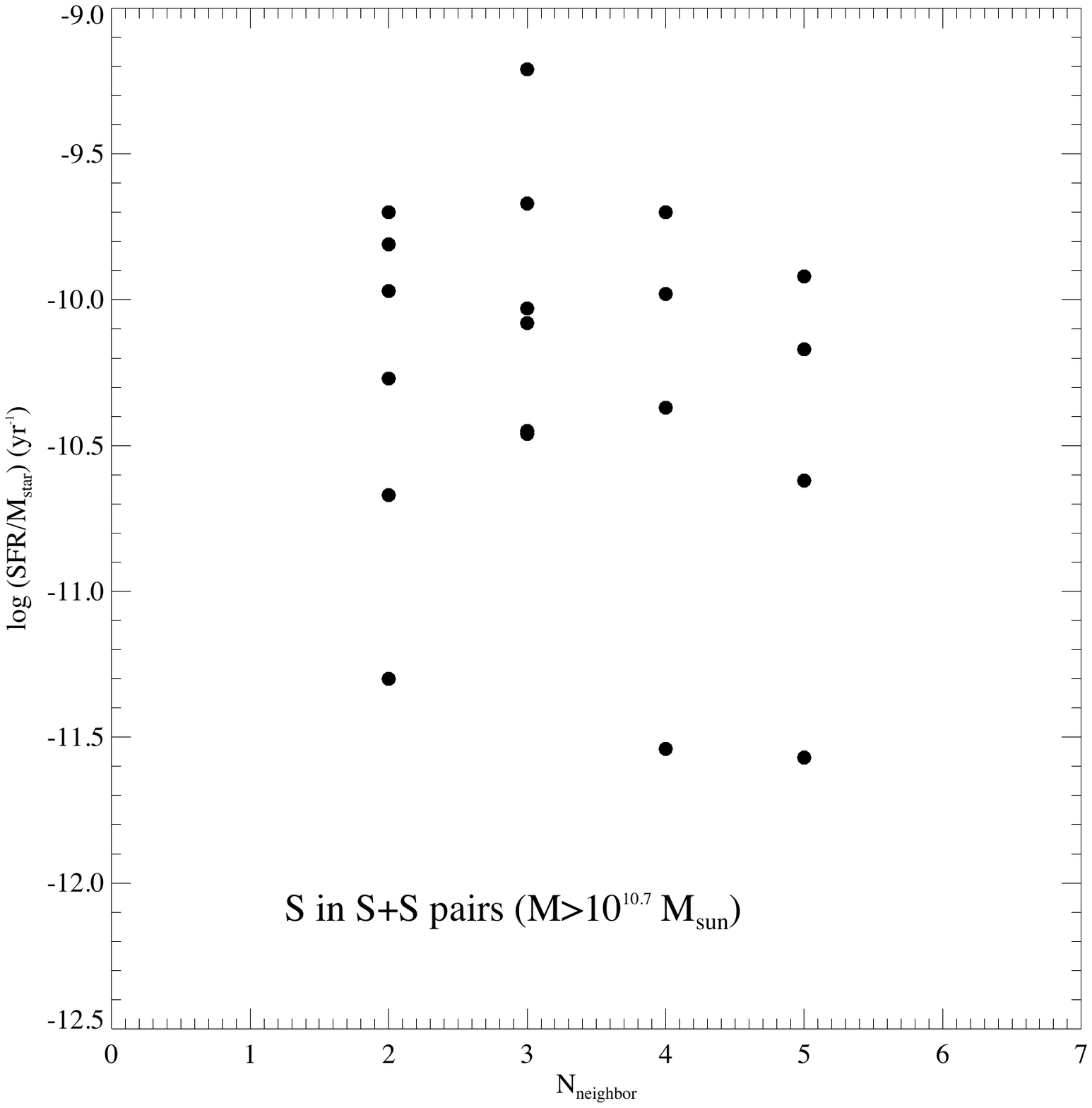}
\caption{Plot of log(SFR/M) versus neighbor counts for the same paired galaxies
plotted in Fig.13.}
\end{figure}

\subsection{Separation and SFR/M Enhancement}

Previous studies (Xu \& Sulentic 1991;
Barton et al. 2000; Lambas et al. 2003; Nikolic et al. 2004; Alonso et
al. 2004; Woods et al. 2006; Barton et al. 2007; Ellison et al. 2008)
found that the star formation enhancement in pairs of separation
$\rm \lsim 20\; h^{-1}$ kpc is much stronger than those of larger
separations. Using our sample, we can address the question whether 
there is still dependence of the SFR on separation for pairs within
this separation limit. We define a normalized separation parameter:
\begin{equation}
\rm SEP = {s \over r_1 + r_2}
\end{equation}
where $\rm s$ is the projected separation, and $\rm r_1$ and $\rm r_2$
are the K-band Kron radii (taken from 2MASS) of the primary and the
secondary, respectively, in the same units as those of $\rm s$ (kpc or
arcsec). In the ideal case of a pair of two round galaxies, the two
component galaxies overlap with each other when $\rm SEP < 1$.

Fig.16 is a histogram of $\rm SEP$ distribution of non-AGN spirals
in KPAIR. The median is at $\rm SEP \sim 1$ (mean $\rm SEP = 1.05\pm
0.40$). Fig.17 is a plot of mean log(SFR/M) v.s. log(M) of spirals
in S+S pairs, separated into two subsamples of $\rm SEP < 1$ and $\rm
SEP \geq 1$. For both subsamples, the log(SFR/M) v.s. log(M) relation
scatters around the mean of the total sample without any obvious
trend, and no significant difference between the two subsamples is
detected.  This seems to suggest that the separation
is not an important parameter any more once the two galaxies
are close enough. There might be several
conflicting factors affecting the star formation activity vs. SEP relation 
in close
major-merger pairs. On the one hand, colliding pairs with $\rm SEP <
1$ may undergo collisionally triggered starbursts in the regions where
the two galaxies overlap, as in the case of the Antennae Galaxies (Xu
et al. 2001). On the other hand, Gao \& Solomon (1999) found a
correlation between molecular gas content and the pair separation,
suggesting a progressive gas depletion due to prolonged star formation
activity. It should also be noticed that $\rm SEP$ is derived from the
projection of the real, 3-dimensional separation. Any dependence on
the true separation can be significantly disturbed by the projection
effect.

\begin{figure}
\plotone{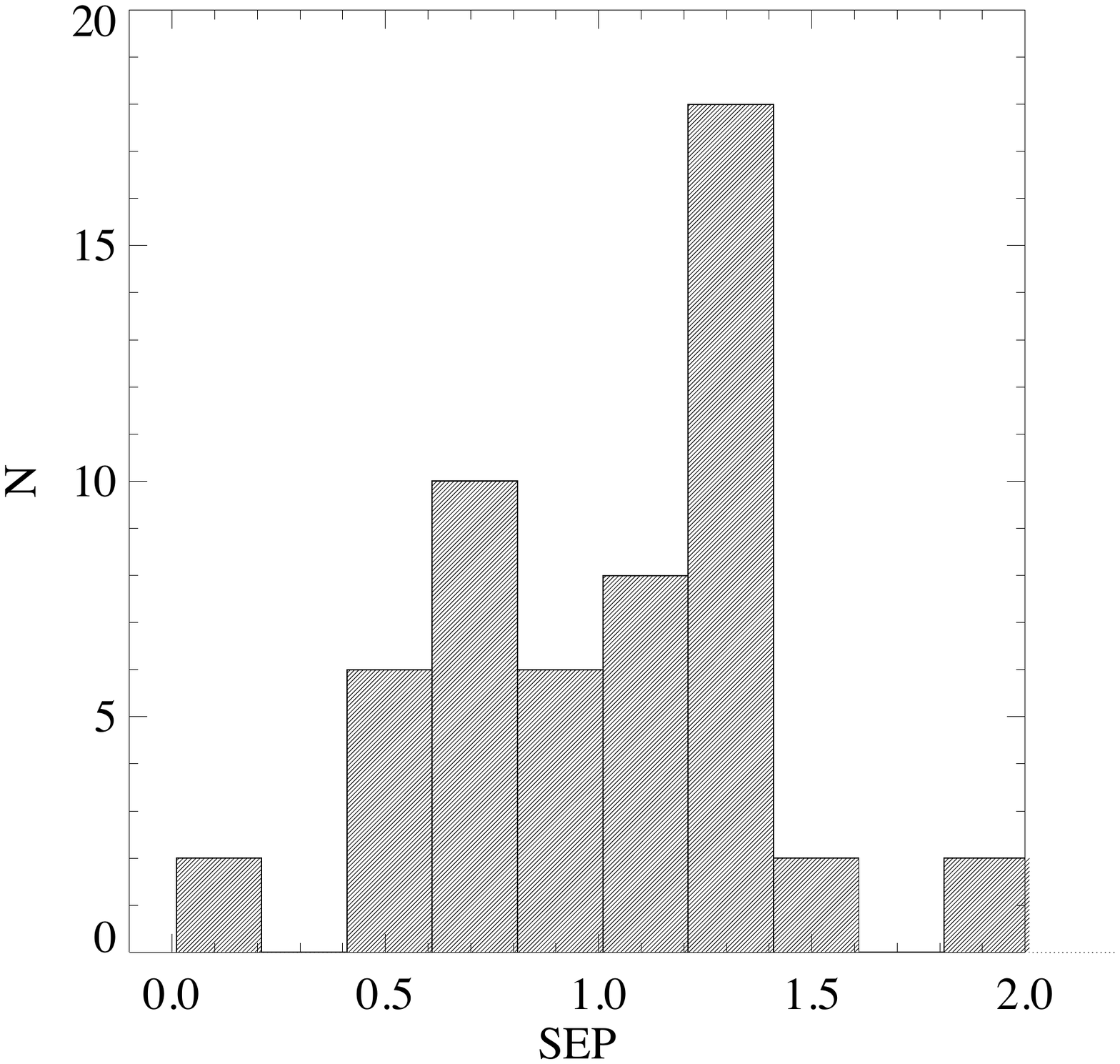}
\caption{Histogram of the 
SEP distribution of non-AGN spirals in the KPAIR sample.
}
\end{figure}

\begin{figure}
\plotone{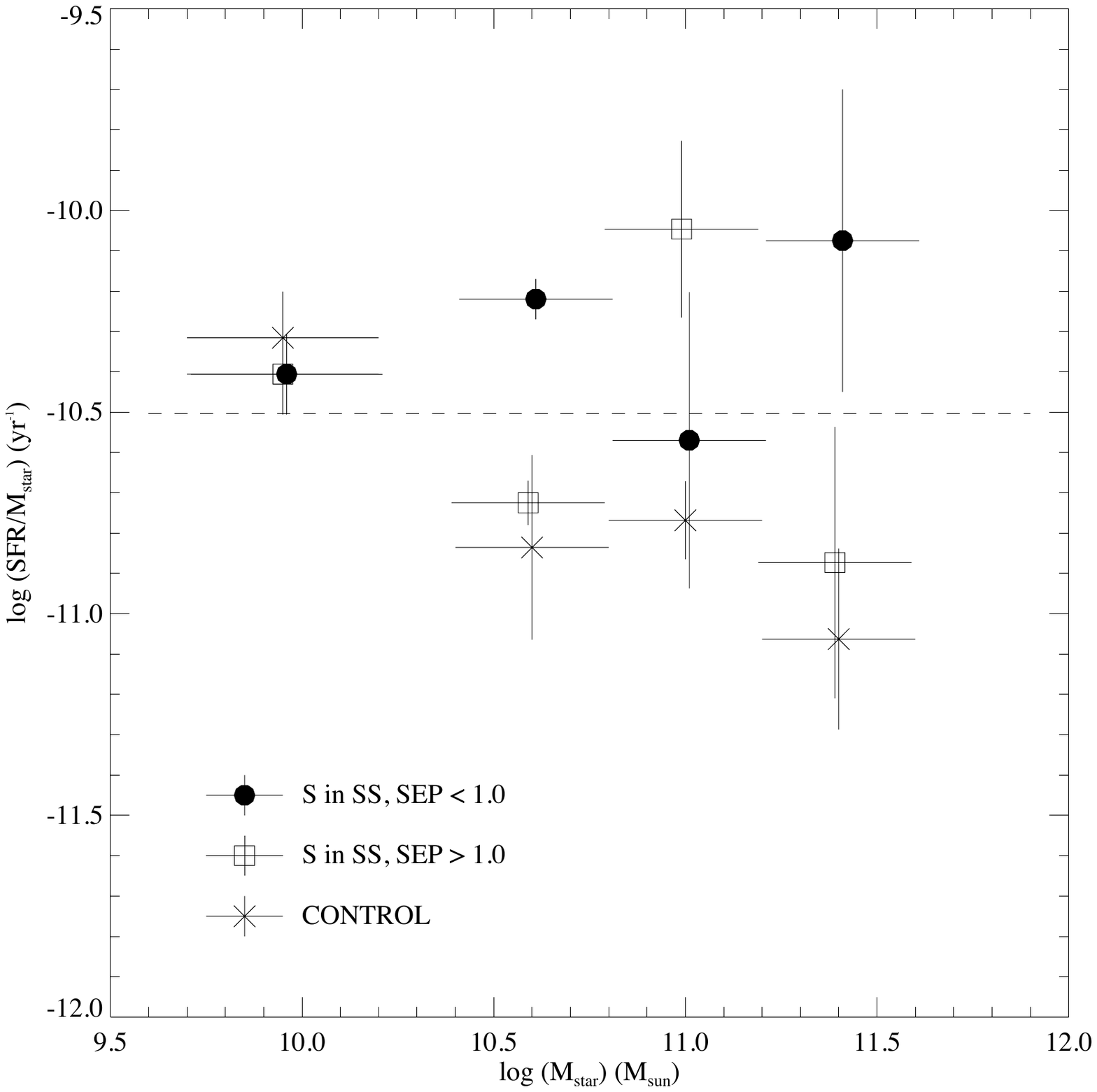}
\caption{Average
sSFR's of non-AGN spirals in S+S pairs,
 separated into two subsamples of $\rm SEP <1$
and $\rm SEP \geq 1$, respectively. The mass bins are the same as in Fig.5.
The dotted line marks the mean $SFR/M$ of the non-AGN spirals in
the KPAIR sample.
}
\end{figure}

\subsection{Contribution of KPAIR galaxies to the Cosmic SFR Density}
In this subsection we estimate the contribution of galaxies
in close major-merger pairs, as defined by the pair selection
criteria in Section 2, to the total cosmic SFR density in the local universe.
Assuming the contribution for E galaxies is negligible,
this can be estimated as follows:
\begin{equation}
\rm \rho^{.}_{KPAIR} = \int (SFR/M)_{KPAIR-S} \times \psi (M)\times f_{s}\times M\times dM
\end{equation}
where $\rm \psi (M)$ is the mass function of KPAIR galaxies and
$\rm f_{s}$ the S fraction ($\rm N_{S}/N$). Because
the mass dependence of $\rm (SFR/M)_{KPAIR-S}$ is rather flat,
we assume it is constant and equal to the 
mean SFR/M of the KPAIR-S sample: 
$\rm (SFR/M)_{KPAIR-S} =  10^{-10.50}\; yr^{-1}$.
The mass function $\psi$ and the S fraction $\rm f_{s}$ are taken
from Domingue et al. (2009). The result of the
integration is $\rm \rho^{.}_{KPAIR} = 2.54 \times 10^{-4} \; M_\sun \;
yr^{-1}\; Mpc^{-3}$, which is 1.7\% of total 
cosmic SFR density in the local universe 
($\rm \rho^{.} = 0.015 \; M_\sun \; yr^{-1}\; Mpc^{-3}$; Yun et al. 2001).

It should be pointed out that, because of the criterion on pair
separation (${\rm r \geq 5 h^{-1}}$ kpc), mergers already coalesced
are missing in the KPAIR sample. This population includes the 
majority of ULIRGs
and many LIRGs.  Assuming that all ULIRGs and 70$\%$ LIRGs are
coalesced mergers, and estimating their densities using the IR
luminosity function of Yun et al. (2001), we found that the total SFR
in these sources is $\rm \rho^{.} = 2.77 \times 10^{-4} \; (M_\sun \;
yr^{-1}\; Mpc^{-3})$, i.e. nearly the same as that in the KPAIR
galaxies. Therefore, in the z=0 universe, the total SFR in close
major mergers is $\rm \rho^{.} = 5.31 \times 10^{-4} \; (M_\sun \;
yr^{-1}\; Mpc^{-3})$. This is only 3.5\% of the local cosmic SFR density, a
truly negligible contribution indeed.

\section{Discussion}
\subsection{Dependence of SFR Enhancement on Companion's Morphological Type}
The star formation enhancement found in interacting galaxies is often
explained in terms of gas inflow caused by gravitational torques of
interaction-induced bars (Hernquist \& Barnes 1991; Barnes \&
Hernquist 1996). However, this theory cannot explain our result that the
star formation enhancement depends on the morphological type of the companion
galaxy: while spirals in S+S pairs show significant enhancement,
those in S+E pairs have star formation activity comparable
to that of single spirals. It appears that, 
in addition to pure gravitational effects, some other factors related
to the companion must play important roles in the interaction induced 
star formation. 

A related result was reported by Park \& Choi (2009) in a study of the
dependence of physical parameters of SDSS galaxies on small scale and
large scale environments. These authors found that the SFR of
late-type galaxies is enhanced when the nearest neighbor is also a
late-type, but reduced when the neighbor is an early-type. They
suggested that the hot gas halo of an early-type companion can
suppress the SFR of a late-type galaxy, through hydrodynamic effects
such as ram pressure stripping, viscous stripping and thermal
evaporation, analogous to what is being encountered by late-type
galaxies in clusters (Boselli \& Gavazzi 2006). This interpretation
can be applied to our result of low sSFR enhancement of spirals in the
S+E pairs.  Verdes-Montenegro et al. (2001) argued that a similar
mechanism might be responsible for the depressed SFR of galaxies in
compact groups (but see Rasmussen et al. 2008). The X-ray observations
of four S+E pairs by Gr\"untzbauch et al. (2007) indeed showed
evidence for extended X-ray halos in the E components, lending support
to this hypothesis.

It should be noticed, however, that our result of no SFR enhancement
for spirals in S+E pairs is based on a small sample of 11 S+E pairs,
and therefore should be confirmed by future studies using larger
samples.  Furthermore, there are known exceptions of strong starbursts
in S+E pairs in the literature, such as NGC~3561A (a LIRG) in Arp~105
(Duc et al. 1997).  It will be worthwhile to investigate why galaxies
like NGC~3561A behave differently. One noticeable feature of NGC~3561A
is the segregation of atomic and molecular gas: the HI gas is
completely displaced out of the disk, and the nuclear starburst is
supported by pure molecular gas (Duc et al. 1997).  Another spiral
galaxy in an S+E pair that has a similar atomic-molecular gas
segregation is NGC~1144 in Arp~118 (``Yin-Yang Galaxy'', Appleton et
al. 2003), also a LIRG, though in this case the nucleus is an AGN
rather than a starburst.

\subsection{``Holmberg Effect''}
The ``Holmberg effect'' on SFR/M of 
massive S+S pairs (Fig.14)
is in agreement with the result of Kennicutt et al. (1987) derived from
the integrated H$_\alpha$ fluxes, and that of Hern\'andez-Toledo 
et al. (2001) based on the (B-V) colors. Apparently, the effect
is present in star formation indicators of very different
time scales ($\sim 10^7$ yrs for the H$_\alpha$ emission,
$\sim 10^8$ yrs for the IR emission, and $\sim 10^9$ yrs for the 
(B-V) color). On the other hand, it has been found only in global
star formation indicators for entire galaxies, but not in those for the nuclear
star formation activity (Joseph et al. 1984). Interestingly,  
the SFR dependence on interaction parameters has been invoked to explain
both the presence of the
correlation between the SFR of the two components
(Kennicutt e al. 1987), and the absence of it (Joseph et al. 1984).
However, we have shown that for spirals in close major-merger pairs,
interaction parameters such as the separation are not important
factors in determining whether a galaxy has enhanced SFR/M or not
(Fig.17). It is possible that the concordant star formation behavior
of galaxies in a close major-merger pair is dictated by the local environment 
within/around the dark matter halo (DMH) surrounding the pair.
But, as shown in Fig.15, the level of star formation activity
of these pairs depends very little on the local density.
It is possible that 
the SFR is suppressed in those quiescent S+S pairs because there
is diffuse hot IGM gas in the DMH, in a similar way as what may be
happening in the S+E pairs (see Section 6.1). Or, in a related scenario,
it might be because the DMH's of
these quiescent S+S pairs have no ``cold streams'' of IGM gas
(Keres et al. 2009) to fuel the star formation in the component galaxies.
It will be worthwhile to confirm or refute these speculations in
future studies.

\subsection{Dependence of SFR Enhancement on Mass}

Our result indicating a 
lack of SFR enhancement in low mass late-type interacting 
galaxies is in agreement with observations of Brosch et al. (2004) and
Telles \& Maddox (2000). In the theory proposed by Mihos et al. (1997),
this is due to the fact that these galaxies do not have sufficient disk 
self-gravity to amplify dynamical instabilities, 
and this disk stability in turn inhibits 
interaction-driven gas inflow and starburst activity.

On the other hand, studies including minor-mergers (Woods \& Geller 2007;
Ellison et al. 2008; Li et al. 2008) found significant
SFR enhancement in low mass ($\rm M \lsim 10^{10} M_\sun$)
interacting galaxies. It appears that, in a minor-merger pair,
a low mass galaxy can have much stronger SFR enhancement
when its companion is much more massive (Woods \& Geller 2007).

\subsection{Overall SFR Enhancement in Major-Merger Pairs}
We have found that, for close major mergers, only massive ($\rm M
\gsim 10^{10.5} M_\sun$) galaxies in S+S pairs have significant star
formation enhancement.  These galaxies are less than $30\%$ in a
K-band selected sample of close major mergers (Domingue et
al. 2009), and some of them are locked in low sSFR
pairs (Fig.13). Therefore, even for spiral galaxies in close
major-merger pairs which harbor most merger-induced star
bursts in the universe, the star formation enhancement due to
galaxy-galaxy interaction is still confined to a small
sub-population. This is consistent with the observations of 
Bergvall et al. (2003) and the simulations of Di Matteo et
al. (2008), both argued that mergers are not very efficient in
triggering significantly enhanced star formation. This is also consistent
with the low contribution of major-merger galaxies to the cosmic
SFR density in the local universe (Section 5.8) and in the uninverse
of intermediate redshift (z$\sim  0.24$---0.80, Jogee et al. 2009).

\section{Summary}

We present Spitzer observations for a sample of close major-merger
pairs of galaxies, selected from a 2MASS/SDSS-DR3 cross-match.  The
scientific goals are (1) studying the star formation activity in these
galaxies and (2) setting a local bench mark for the cosmic evolution
of close major mergers.  The Spitzer KPAIR sample (27 pairs, 54
galaxies) includes all spectroscopically confirmed spiral-spiral (S+S)
pairs and spiral-elliptical (S+E) pairs in a parent sample that is
complete for primaries brighter than K=12.5 mag, projected separations
of $\rm 5 \leq s \leq 20$h$^{-1}$ kpc, and mass ratios $\leq 2.5$.
There are 42 spiral galaxies and 12 elliptical galaxies in the
sample. These galaxies harbor 6 known AGNs, 3 in spirals and 3 in
ellipticals.

Spitzer observations include images in the four IRAC bands at 3.6,
4.5, 5.8 and 8.0 $\mu m$, and the three MIPS bands at 24, 70 and 160
$\mu m$.  They show very diversified IR emission properties among
KPAIR galaxies. Among the paired spirals, the majority have
rather moderate IR luminosity ($\sim 10^{10} L_\sun$). There are four
LIRGs ($\sim 10\%$ of KPAIR-S subsample), but no ULIRGs.
The SFR, estimated using the IR luminosity, 
and the sSFR of non-AGN spirals (39 of them) in KPAIR are compared to those
of single spirals in a control sample. Each of the 39 galaxies
in the control sample matches a non-AGN spiral in KPAIR with the same
mass (estimated from the K-band luminosity). The following
results are found:
\begin{description}
\item{(1)} The mean SFR of non-AGN spirals in the KPAIR sample (KPAIR-S) 
is significantly enhanced 
compared to that of the single spirals in the control sample.
The means of $\rm log(SFR)$ of the KPAIR-S galaxies and of the galaxies in
the control sample are $0.36 \pm 0.12$ and $0.07 \pm 0.08$, respectively.
And 
the means of $\rm log(SFR/M)$ of the KPAIR-S galaxies and of the galaxies in
the control sample are $-10.50 \pm 0.10$ and $-10.78 \pm 0.08$, respectively.
The K-S test rejects at the 96.1\% confidence level 
the null hypotheses that the two samples are drawn from the same population.
\item{(2)}
When separating the non-AGN paired spirals into those in S+S pairs (28) 
and in S+E pairs (11),
only the former show SFR/M enhancement whereas the latter do not.
The means of $\rm log(SFR/M)$ of the spirals in S+S pairs and of  
those in S+E pairs are $-10.36 \pm 0.11$ and $-10.88 \pm 0.19$, respectively.
\item{(3)} The SFR/M enhancement
of spirals in S+S pairs is highly mass-dependent: only those with
$\rm M \geq 10^{10.5} M_\sun$ show significant enhancement, whereas relatively
low mass ($\rm M \sim 10^{10} M_\sun$) spirals in S+S pairs have
about the same SFR/M as their counterparts in the control sample.
\item{(4)} We define $\epsilon$ as the SFR enhancement parameter,
$\rm \epsilon = \log( (SFR/M)_{KPAIR-S}) - \log((SFR/M)_{control})$,
where $\rm (SFR/M)_{KPAIR-S}$ and $\rm (SFR/M)_{control}$ are
the SFR/M of a paired spiral
and that of its match in the control sample, respectively.
For spirals in the S+S subsample, there is a strong linear dependence 
of $\epsilon$ on $\rm \log(M)$, specified as
$\rm <\epsilon>_{S+S} = 0.03(\pm 0.14) + 0.47(\pm 0.15) \times 
(\log(M/10^{10} M_\sun))$. The relation is valid for the mass range of
$10.0 \lsim \log(M/M_\sun) \lsim 11.5$.
\item{(5)} For spirals in the KPAIR sample, which includes only close
  major mergers, there is no systematic difference between the $\rm
  log(SFR/M)$ of spirals with $\rm SEP <1$ and those with $\rm SEP \geq
  1$, $\rm SEP$ being the normalized separation parameter ($\rm SEP =
  s/(r_1 + r_2)$).  Also, there is no significant difference between
  the means of $\rm log(SFR/M)$ for the primaries and the
  secondaries.
\item{(6)} There is evidence for a correlation between the global star
  formation activities (but not the nuclear activities) of the
  component galaxies in massive S+S major-merger pairs (the ``Holmberg
  effect'').
\item{(7)}
The contribution of KPAIR galaxies to the cosmic SFR in the local universe
is $\rm \rho^{.}_{KPAIR} = 2.54 \times 10^{-3} \; M_\sun \;
yr^{-1}\; Mpc^{-3}$. This is 1.7\% of total 
cosmic SFR density in the local universe. Adding the
SFR in mergers already coalesced, which are missed by the KPAIR sample and
may include many ULIRGs and LIRGs,
the total SFR in close major mergers is
$\rm \rho^{.} = 5.31 \times 10^{-3} \; M_\sun \;
yr^{-1}\; Mpc^{-3}$. This is only 3.5\% of the local cosmic SFR density.
\end{description}

\noindent{\it Acknowledgments}:

CKX thanks David Shupe for useful discussions on the SWIRE data
handling. YG's research is partly supported by grants
\#10833006 \& \#10621303 of NSF of China. 
W.-H. Sun and Y.-W. Cheng acknowledge 
the support of National Science Council in 
Taiwan under the grant NSC 97-2112-M-002-014. 
 Constructive comments by an
anonymous referee are acknowledged.  This work is based on
observations made with the Spitzer Space Telescope, which is operated
by the Jet Propulsion Laboratory, California Institute of Technology
under a contract with NASA. Support for this work was provided by
NASA.  This research has made use of the NASA/IPAC Extragalactic
Database (NED) which is operated by the Jet Propulsion Laboratory,
California Institute of Technology, under contract with the National
Aeronautics and Space Administration.



\appendix
\renewcommand{\thefigure}{A-\arabic{figure}}
\setcounter{figure}{0}

\begin{center}
    {\bf APPENDIX}
\end{center}

\section{Notes on Individual Pairs}

\nid{\bf J0020+0049} The FIR emission of this S+E pair is dominated
by the S component. The E component was only barely detected in the MIPS
24 $\mu m$ band, and undetected in the 70 and 160 $\mu m$ bands.

\nid{\bf J0109+0020} Neither galaxy in this S+E pair is detected in any
of the MIPS bands, a unique case in the sample. The S component shows
an arm-like feature in the optical images, but was classified
as an E galaxy by the automatic classification routine based on
the optical color and light concentration. Apparently it is a
'red and dead' galaxy, perhaps of early S type.

\nid{\bf J0118-0013} Both components of the S+S pair 
are well detected by IRAC and MIPS. 
The western component has a narrow line AGN (Hao et al. 2005).
In its optical image there is a blue, jet-like feature pointing to the 
companion galaxy. It is a ``luminous IR galaxy'' (LIRG), with
$\log L_{TIR}/L_\sun = 11.41$, 
dominating the total dust emission of the pair.
From aperture photometry of the IRAC bands, 
the AGN contributes $\sim 40\%$ of the dust emission of the galaxy.
The eastern component looks like a normal late-type spiral,
contributing only $\sim 12\%$ of the L$_{TIR}$ of the pair.

\nid{\bf J0211-0039} This S+S pair is KPG~058 (Karachentsev 1972).
Both galaxies are seen edge-on.
The western component (an Sbc galaxy) has
a narrow line AGN (Hao et al. 2005). It dominates the dust emission of
the pair.

\nid{\bf J0906+5144} Another KPG pair, KPG~185.
The western component is classified as E type in our scheme,
though according to NED it is an Sa galaxy.
The eastern component is an S type with
a narrow-line AGN (Hao et al. 2005). The IRAC 8 $\mu m$ band image shows a 
nucleus$+$ring morphology, with most of the emission from the ring.

\nid{\bf J0937+0245} This S+E pair is Arp~142 (= VV~316), classified as
`ring galaxy' by RC2 (de Vaucouleurs et al. 1976). The S component (NGC~2936)
is very disrupted. The optical/NIR morphology looks like a bird head,
with the nucleus being the eye. The tidal tails, which form the 'neck' of
the bird, contain several star forming regions bright in the IR
(see the 8 and 24 $\mu m$ image). However, most of the dust emission is
still confined within the nucleus$+$disk (i.e. the bird head) region.
The E component (NGC~2937) is undetected in the 70 and 160$\mu m$ bands. 

\nid{\bf J0949+0037} This S+S pair is KPG~216.
It is one of the three pairs in our sample larger than $2'.5$.
The two relatively low mass galaxies ($\rm \sim 10^{10} M_\sun$) are 
well separated from each other. 
The Spitzer observations were carried out for the two components separately,
and the final maps are coadds of these separate observations.
Both components are well detected by IRAC and MIPS. And both show
extended IR emission through out the entire discs. There is a strong
outer disc starburst on the east side of the eastern component (NGC~3023),
whose FIR luminosity is comparable to that of the nucleus.

\nid{\bf J1020+4831} The eastern component (E type)
is a strong radio source 
(4C~+48.29), classified as an AGN in the literature 
according to NED. It is a rather weak FIR
source, only marginally detected in the 24 $\mu m$ band, and undetected
in the 70 and 160 $\mu m$ bands. The western component (S type),
detected in all 7 Spitzer bands, dominates the dust emission of the pair. 

\nid{\bf J1027+0114} This close S+E pair is actually in a triplet 
    (Karachentseva et al. 1988). The third galaxy,
 west of the pair, is about 2$'$ away
    from the pair center.
    The southern component (S type) is a strong IR source. The northern
    component (E type) is detected marginally by MIPS in the 24 $\mu m$
    band.

\nid{\bf J1043+0645} Both components of this S+S pair
are detected by IRAC and MIPS.
The two components have nearly equal mass (the mass ratio derived
from the K band lumiosities is $\rm M_1/M_2 =1.2$), but rather
uneven dust emissions. The $\rm L_{TIR}$ of the 
western component is more than 5 times of that of the eastern component.

\nid{\bf J1051+5101} This is KPG~253, which appears to be
a very close S+E pair. The projected separation of the two components
($\rm s = 4.7\; h^{-1} kpc$) is actually less than the 5 $h^{-1}$ kpc lower 
boundary of the separation criterion. However, since the separation is only
$6\%$ off the boundary, we chose to keep the pair in this
work. It is in the center of a
cluster. Both galaxies are massive ($\rm > 10^{11} M_\sun$). The
westhern galaxy is classified as E while the eastern galaxy as S.
At low surface brightness levels, a ring like structure shows up
around the nucleus of the S component both in the optical and in
the 8$\mu m$ images.
The pair is detected by IRAS with $\rm f_{60} = 0.78$ Jy.
Interestingly, both the 24$\mu m$ and 70$\mu m$ band emission contours
coincide with the nucleus of the E component while there is little
emission detected in the nuclear region of the S component.  This
indicates that most of the star formation in the pair 
is occuring in the nucleus of the more massive E
component, perhaps due to an ISM transfer from the S component.  The
total SFR derived from the IR emission is quite low, at the level of
$\sim 1 M_\sun/yr$, very different from IR selected mergers.  No
AGN or any detectable radio source has been found in the pair (van Driel
et al. 2000).

\nid{\bf J1202+5342} The weastern component (S type) dominates the 
dust emission the S+E pair.

\nid{\bf J1308+0422} $=$UGC~8217. Both galaxies in this S+S pair
are well detected in all 7
Spitzer bands with nearly equal $\rm L_{TIR}$ ($\rm \simeq 4\; 10^9 L_\sun$).

\nid{\bf J1332-0301} Both spiral galaxies are detected by IRAC and MIPS.
The western component, with the lower mass and more compact
morphololgy among the two,
has warmer (i.e. higher $\rm f_{70}/f_{160}$ ratio)
and stronger dust emission than the eastern component.

\nid{\bf J1346-0325} The two galaxies of this S+E pair are well separated from
each other. Both are detected by IRAC and MIPS, with comparable, moderate
dust emission. The eastern component (an E type) has a Sy2 nucleus 
(Maia et al. 2003).

\nid{\bf J1400+4251} This S+S
pair is a bright IRAS source ($\rm f_{60} = 2.32$ Jy).
The combined $\rm L_{TIR}$ ($ = 1.3 \times 10^{11} L_\sun$) makes the pair a LIRG.
The two spiral galaxis, both detected by IRAC and MIPS, 
are about equally bright in the dust emission. When
separated, each is slightly fainter than a LIRG.

\nid{\bf J1425+0313} This S+E pair is VIII Zw~415. The western 
component (E type), a moderate 
IR source, contains a broad line AGN (Hao et al. 2005).

\nid{\bf J1433+4004} $=$KPG~426. These are two IR bright spiral galaxies,
both are detected by IRAC and MIPS.
There are bright hot-spots in the 8 $\mu m$ image of the southern galaxy.
As a close-interacting S+S pair, it has been studied by Xu et al. (2001)
using ISOCAM.

\nid{\bf J1453+0317} One of the three pairs in our sample that are 
larger than $2'.5$. These are 
two well separated S-type galaxies, both are detected by IRAC and MIPS.
The eastern galaxy displays a ring formation in both 8 and 24$\mu m$ images.
The western galaxy was detected in the ISOPHOT 170 micron Serendipity
Survey (Stickel et al. 2004), with 
$\rm f_{170}=1.94\pm 0.58$ Jy. This is significantly lower than
the MIPS 160 $\mu m$ flux: $\rm f_{160}=3.03\pm 0.32$ Jy.

\nid{\bf J1506+0346} An S+S pair. 
The western component (IC~1087), classified as
S0-a (via NED), is a weak IR source. Most of the dust emission
is from the eastern component (UGC~09710).

\nid{\bf J1510+5810} These are two very close S-types. The source
about 1$'$ west of the pair is a star. Most of the dust emission is
due to the eastern component.  It was detected in the ISOPHOT 170
micron Serendipity Survey (Stickel et al. 2004), with $\rm
f_{170}=0.87\pm 0.26$ Jy. This is slightly lower than the MIPS 160
$\mu m$ flux: $\rm f_{160}=1.16\pm 0.12$ Jy.

\nid{\bf J1528+4255} $=$I~Zw~113. These two spiral galaxies show
signs of interaction. Most of the dust emission is found in the western
component (NGC~5934).

\nid{\bf J1556+4757} The western component (S type) 
dominates the dust emission in the S+E pair.
The source between the two galaxies is a star.

\nid{\bf 1602+4111} $=$KPG~479. Both spiral galaxies are IR bright.
They show signs of interactions.

\nid{\bf 1704+3448} These are two closely-interacting, star-forming spiral
galaxies. The northern component is a LIRG.

\nid{\bf J2047+0019}  $=$KPG~548. The S+E pair is one of the 3 pairs in the 
KPAIR
sample with size $>2'.5$. The western component (S type), well resolved
in both the IRAC and the MIPS bands, shows a nucleus+ring structure. 
In the MIPS bands, the ring is rather faint. The eastern component (E type)
is only marginally detected in the 24$\mu m$ band,
undetected in the 70 and 160 $\mu m$ band.  This pair has been studied by
Domingue et al. (2003) with ISOPHOT.

\nid{\bf J1315+6207} $=$UGC08335$=$Arp~238$=$KPG~369$=$VV~250.  This S+S
pair is not displayed in Fig.1. It was observed successfully
by Mazzarella et
al. (2009) in the IRAC bands, but the MIPS observations failed.  The IRAC
images (Mazzarella et al. 2009) and the KAO image at 100$\mu m$
(Bushouse et al. 1998) show that the eastern component dominates the IR
emission of this LIRG pair. The west component was undetected in the
KAO observation (Bushouse et al. 1998).  The radio continuum
observations at 4.85 GHz (Condon et al. 1991) revealed a flux ratio
between the two components of 4.4 (22mJy/5mJy), close to the 8$\mu m$
flux ratio of 3.5 (187mJy/54mJy) measured by Mazzarella et al.
(2009). Compared to the IRAS data, Condon et al. (1991) found a
FIR/radio ratio index q = 2.74 for the pair, very close to the mean
($<q> = 2.64 \pm 0.16$) for star forming galaxies. Assuming that both
components have the same FIR/radio ratio (e.g. Hattori et al. 2004),
we divided the total IR luminosity of the pair, $\log (L_{TIR}/L_\sun)
= 11.74$ (Mazzarella et al. 2009), and found $\log (L_{TIR}/L_\sun) =
11.65$ for the east component and $\log (L_{TIR}/L_\sun) = 11.01$ for
the west component. Therefore, both galaxies qualify as LIRGs.

\section{MIPS Point Response Functions (PRFs)}

The 24$\mu m$ PRF was taken from Engelbracht et al. (2007).
It was generated by smoothing the standard Spitzer TinyTim PSF 
(Krist 1993) with a $4''.41$ (`1.8-pixel') boxcar.
Similarly, the 70$\mu m$ PRF was taken from Gordon et al. (2007),
which is a TinyTim PSF smoothed by a $13''.30$ (`1.35-pixel') boxcar.
As shown in Fig.A-1, good agreements were found between our data and the above 
model PRFs.
\begin{figure}
\plottwo{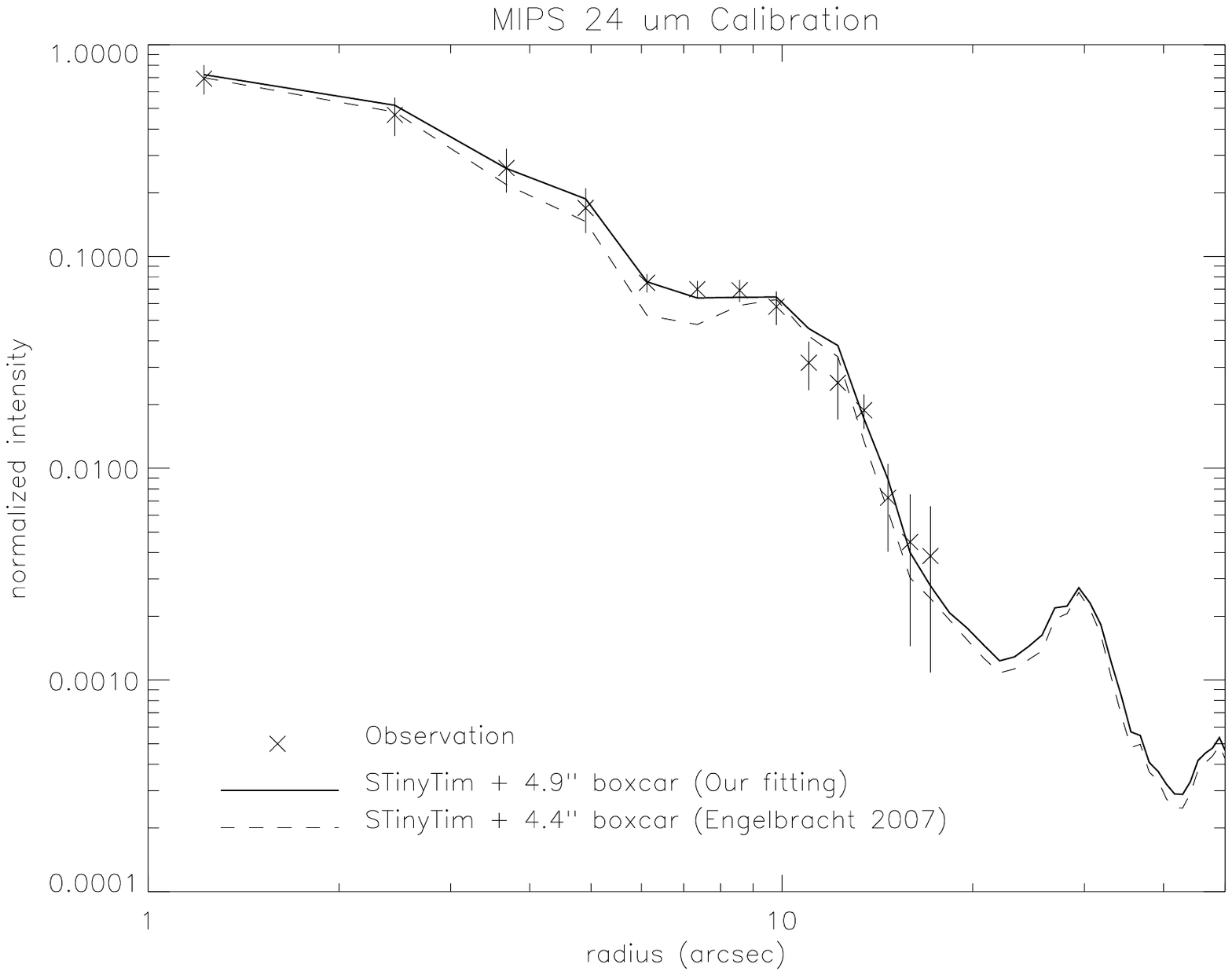}{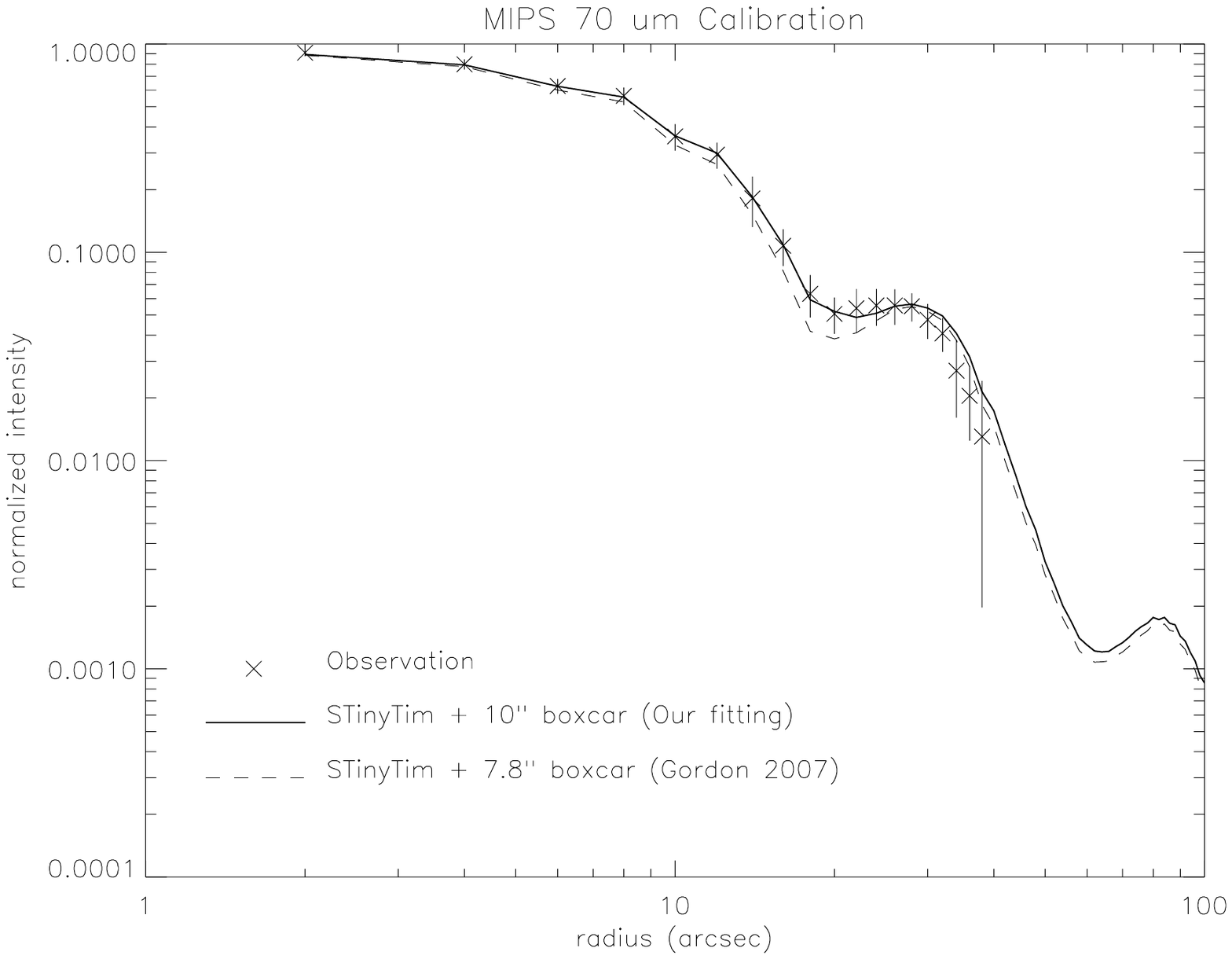}
\caption{{\it Left}: The 24$\mu m$ point response functions (PRFs).
{\it right}: The 70$\mu m$ PRFs.
}
\end{figure}

On the other hand, the comparison between our data in the 160$\mu m$ band
and the standard PRF (``STinyTim+1.6-pixel boxcar'') of the MIPS team 
(Standsberry et al. 2007) showed an excess in the first Airy-ring in
our data (Fig.A-2). In order to better fit the data, we adopted
an empirical PRF. It is the standard PRF plus a ring,
which has a truncated Gaussian profile: 
\begin{eqnarray}
 S & = & S_1 \times \exp{{(r-r_1)^2\over 2\sigma_1^2}}\hskip3truecm (r \leq r_2) \\
  & = & 0 \hskip6truecm (r > r_2);
\end{eqnarray}

where $S_1= 0.05$, $r_1= 5$ pixel, $r_2= 8$ pixel and $\sigma_1 = 4$ pixel.
The pixel size is 8$''$ for the 160$\mu m$ map.
The comparisons between 160$\mu m$ PRFs and the data can be found in 
Fig.A-2.
\begin{figure}
\plotone{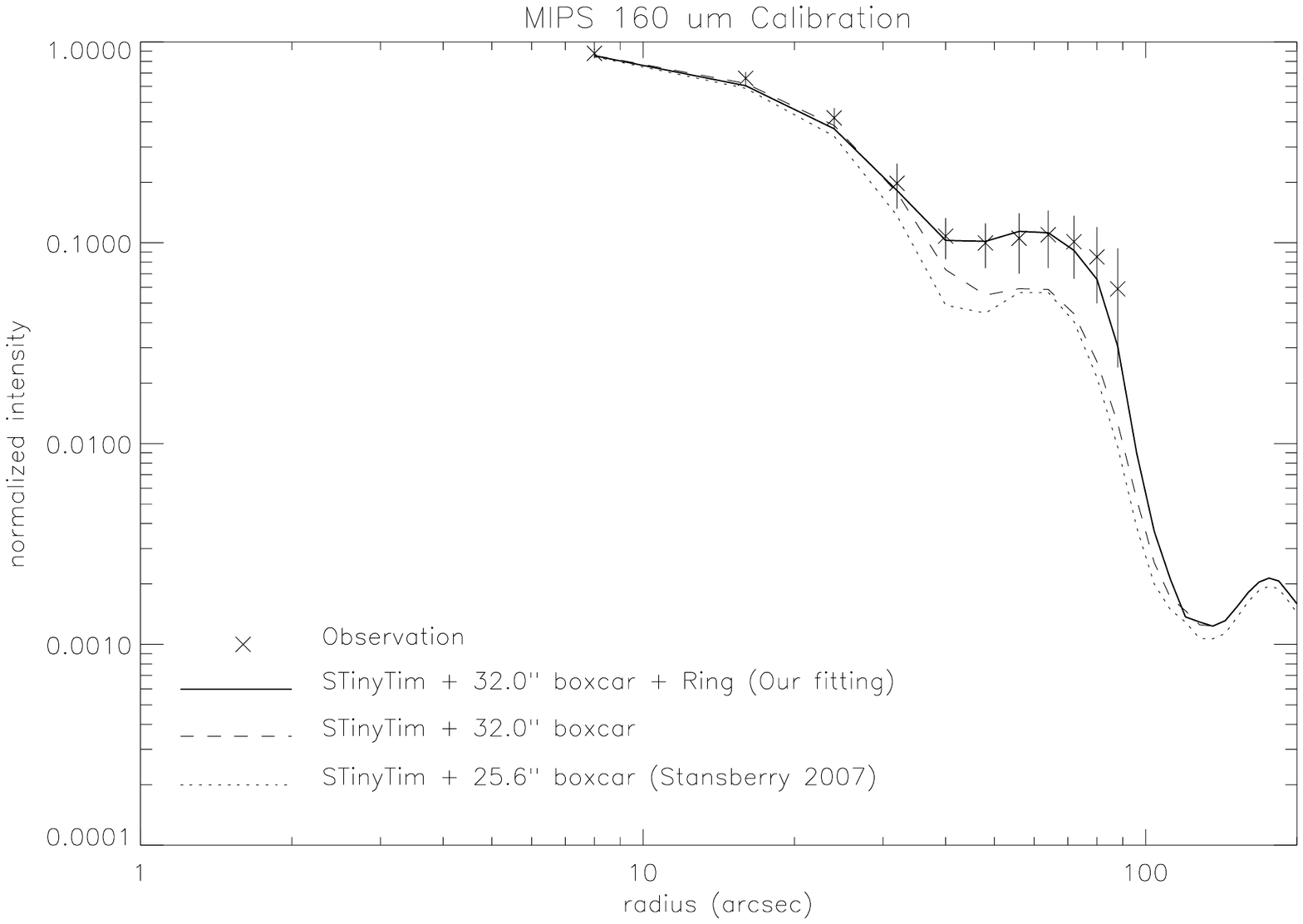}
\caption{The 160$\mu m$ PRFs.
}
\end{figure}

\end{document}